\documentclass[11pt,a4paper]{article} 
\usepackage{jcappub} 
\usepackage{amsmath,amsfonts,amssymb}
\usepackage{txfonts}
\usepackage{graphicx}
\usepackage{fixltx2e}
\usepackage{natbib}
\usepackage{multirow}

\bibpunct{(}{)}{;}{a}{}{,}

\def\reff@jnl#1{{\rm#1\/}}

\def\apj{\reff@jnl{ApJ}}                
\def\apjl{\reff@jnl{ApJ}}               
\def\apjs{\reff@jnl{ApJS}}              
\def\aap{\reff@jnl{A\&A}}               
\def\mnras{\reff@jnl{MNRAS}}            
\def\prd{\reff@jnl{Phys.Rev.D}}         
\def\nar{\reff@jnl{New~Astr.~Rev.}}     

\newcommand{\thedate}{\today}

\def\muK{\ifmmode \,\mu$K$\else \,$\mu$\hbox{K}\fi}
 
\newcommand{\vt}[1]{\boldsymbol{\rm {#1}}}
\newcommand{\tn}[1]{\rm {#1}}

\newcommand{\CORE}{\textit{CORE}}
\newcommand{\miniCORE}{\textit{miniCORE}}
\newcommand{\WMAP}{\textit{WMAP}}
\newcommand{\Planck}{\textit{Planck}}
\newcommand{\LiteBIRD}{\textit{LiteBIRD}}
\newcommand{\Spider}{\textit{Spider}}

\begin{document}

\title{Exploring cosmic origins with CORE: mitigation of systematic effects}

\author[1,2,a]{P.~Natoli,\note[$a$]{Corresponding author}}
\affiliation[1]{Dipartimento di Fisica e Scienze della Terra, Universit\`a di
Ferrara, Via Saragat 1, 44122 Ferrara, Italy}
\affiliation[2]{INFN, Sezione di Ferrara, Via Saragat 1, 44122 Ferrara, Italy}

\author[3,4]{M.~Ashdown,}

\affiliation[3]{Astrophysics Group, Cavendish Laboratory, University
  of Cambridge, J.\ J.\ Thomson Avenue, Cambridge, CB3 0HE, UK}
\affiliation[4]{ Kavli Institute for Cosmology, Univerisity of
  Cambridge, Madingley Road, Cambridge, CB3 0HA, UK}

\author[5]{R.~Banerji,}
 \affiliation[5]{APC, AstroParticule et Cosmologie, Universit{\'e}
Paris Diderot, CNRS/IN2P3, CEA/lrfu, Observatoire de Paris, Sorbonne
Paris Cit{\'e}, 10, rue Alice Domon et L{\'e}onie Duquet, 75205 Paris
Cedex 13, France}

\author[6,7]{J.~Borrill,}
\affiliation[6]{ Computational Cosmology Center, Lawrence Berkeley
National Laboratory, Berkeley, California, U.S.A. }
\affiliation[7]{Space Sciences Laboratory, University of California, Berkeley, CA, 94720, USA}

\author[8,9,10]{A.~Buzzelli,}
\affiliation[8]{ Dipartimento di Fisica, Universit\`a di Roma  La
Sapienza , P.le A. Moro 2, 00185 Roma, Italy }
\affiliation[9]{ Dipartimento di Fisica, Universit\`a  di Roma  Tor
Vergata,  Via della Ricerca Scientifica 1, I-00133, Roma, Italy }
\affiliation[10]{ INFN, Sezione di Tor Vergata, Via della Ricerca
Scientifica 1, I-00133, Roma, Italy }

\author[9,10]{G.~de Gasperis,}

\author[5]{J.~Delabrouille,}

\author[11]{E.~Hivon,}
\affiliation[11]{Institut d' Astrophysique de Paris (UMR7095: CNRS \&
UPMC-Sorbonne Universities), F-75014, Paris, France}

\author[1,2,12]{D.~Molinari,}
\affiliation[12]{ INAF/IASF Bologna, via Gobetti 101, I-40129 Bologna, Italy}

\author[5]{G.~Patanchon,}

\author[1,2]{L.~Polastri,}

\author[13,14]{M.~Tomasi,}
\affiliation[13]{Dipartimento di Fisica, Universit\`a degli Studi di Milano,
Via Celoria, 16, Milano, Italy}
\affiliation[14]{INAF/IASF Milano, Via E. Bassini 15, Milano, Italy}

% ...

\emailAdd{paolo.natoli@unife.it}

\arxivnumber{...}


\abstract{We present an analysis of the main systematic effects that
  could impact the measurement of CMB polarization with the proposed
  \CORE\ space mission. We employ timeline-to-map simulations to
  verify that the \CORE\ instrumental set-up and scanning strategy
  allow us to measure sky polarization to a level of accuracy adequate
  to the mission science goals.  We also show how the
  \CORE\ observations can be processed to mitigate the level of
  contamination by potentially worrying systematics, including
  intensity-to-polarization leakage due to bandpass mismatch,
  asymmetric main beams, pointing errors and correlated noise.
  We use analysis techniques that are well validated on data from
  current missions such as \Planck\ to demonstrate how the residual
  contamination of the measurements by these effects can be brought to
  a level low enough not to hamper the scientific capability of the
  mission, nor significantly increase the overall error budget.
  We also present a prototype of the \CORE\ photometric calibration
  pipeline, based on that used for \Planck, and discuss its robustness
  to systematics, showing how \CORE\ can achieve its calibration
  requirements. While a fine-grained assessment of the impact of
  systematics requires a level of knowledge of the system that can
  only be achieved in a future study phase, the analysis presented
  here strongly suggests that the main areas of concern for the
  \CORE\ mission can be addressed using existing knowledge, techniques
  and algorithms.}

\thedate
\maketitle
\flushbottom


\section{Introduction}
\label{sec:intro}

The Standard Model of Cosmology owes its emergence to increasingly
accurate observations as much as to theoretical advancement. As new
experiments are designed and deployed, the quest for precision and
accuracy is becoming more demanding than ever. In the field of cosmic
microwave background (CMB) observations, the forefront of research has
shifted in recent years from the temperature anisotropies to
polarization, a much weaker signal, which has increased scientific
expectations and concerns about the analysis. Accurate measurements of
CMB polarization pose significant challenges to observational
strategies \citep{ECO.instrument.paper} as well as to the analysis of
data. The \Planck\ results have set a milestone by reaching a level
where systematic errors, arising either in the instrumental chain or
from contamination by spurious emission, surpass those from stochastic
noise in the detectors, both for the CMB temperature power spectrum
\citep{2014A&A...571A..15P,2016A&A...594A..11P} and for polarization
on large angular scales
\citep{2016A&A...594A..11P,2016A&A...596A.107P}. The error budget of
future experiments, whose focal plane arrays will contain thousands of
polarization sensitive detectors, will be dominated by systematics
even for small scale polarization. It is therefore critical to ensure
that these contaminants can be controlled to a level that does not
jeopardize the science goals of the mission.

The impact of systematic effects plays a central role in the analysis
of modern CMB experiments
\citep{2016A&A...594A...2P,2016A&A...594A...7P,2013ApJS..208...20B,2016arXiv161002360L,2015ApJ...806..247B,2016PhRvL.116c1302B}
and is the main subject of several papers as well (e.g.,
\cite{2013ApJS..207...14K,2016A&A...594A...3P,2009PhRvD..79f3008M,2004ApJ...603..371G}),
many of them focusing specifically on polarization specific
systematics and their treatment
\citep{2002AIPC..609..209K,2008PhRvD..77h3003S,2009PhRvD..79j3002M,2009PhRvD..80d3522P}. The
definition of a systematic effect is somewhat dependent on the
context. Strictly speaking, any contamination which is not the signal
of interest and does not exhibit a purely stochastic behavior may be
regarded as a systematic. The CMB community has traditionally used the
term in a wider sense, considering any contamination that deviates
from ideal, white noise as a systemetic. In this sense, long time
scale (i.e., correlated or `$1/f$') noise may be considered as a
systematic contribution, while being from another point of view a
purely random component with a zero expectation value.

This paper is part of a set describing the scientific performance of
the proposed \CORE\ satellite, which is designed to map CMB
polarization to an accuracy only limited by cosmic variance over a
broad range of scales. It explores several aspects related to the
expected quality of \CORE's polarization measurements. We employ a
realistic simulation pipeline to produce time ordered data for a
year's worth of observations, which we then reduce to maps of
intensity and polarization using a state of the art map-making
code. We analyse these maps to assess the overall quality of the
\CORE\ full sky polarization measurements, in view of the proposed
scanning strategy and instrumental design. We include in the
simulations a number of realistic effects that may impact the accuracy
of the observations, and show that they are either under control or
can be kept under control by employing analysis techniques already
used by the CMB community. The approach we follow consists in studying
one effect at a time, which allows us to evaluate each contribution in
isolation and carefully assess its impact. The obvious drawback is
that we may miss potential interactions between different effects, a
situation that may be addressed by employing full end-to-end
simulations (see, e.g., \cite{2016A&A...594A..12P}). We defer this
very demanding analysis to future studies.

The plan of this paper is as follows. We provide in
Sect.~\ref{sec:map-making} a brief introduction to the CMB map-making
methodology, which we use throughout this work. In
Sect.~\ref{sec:simulations} we describe the timeline-to-map simulation
engine that was used in this work, based on the publicly available
TOAST software package. We produce noise-only maps based on a
realistic noise model, which are analyzed in Sect.~\ref{Analysis_maps}
to infer results about the purity of the Stokes parameter maps and
show how polarization can be resolved by modulating observations using
only the scanning strategy, as opposed to adopting specific hardware
such as a rotating half wave plate. In this Section, we also explore
possible ways of optimizing the scanning strategy and study the noise
properties of detectors in several positions in the focal plane. In
Sect.~\ref{sec:CorrelatedNoise} we address the case of dealing with
noise that is correlated between detectors. We begin addressing signal
related simulations with Sect.~\ref{sec:BandpassMismatch} where we
show how a bandpass mismatch between detectors can be effectively
mitigated for \CORE. Temperature-to-polarization leakage arising from
beam non-idealities is discussed in Sect.~\ref{sec:AsymmetricBeam}
where we present correction schemes to be applied either in making the
map or in harmonic space afterwards, the latter being supported by a
specific semi-analytical approach whose performance is compared to
simulations. Sect.~\ref{sec:Calibration} presents a prototype
in-flight calibration pipeline for \CORE\ and discusses its robustness
to selected systematic contamination. We discuss in
Sect.~\ref{sec:pointing_accuracy} the impact of effects not considered
earlier and finally draw our conclusions in
Sect.~\ref{sec:conclusions}.

\section{Map-making for CMB experiments}
\label{sec:map-making}

This paper deals extensively with the propagation of \CORE\ simulated
data from time-ordered observations (also called `timelines') to maps
of the sky. To provide some context, we briefly review map-making
algorithms for CMB experiments. We begin by considering a simple
model, which only accounts for ideal sky signal and stochastic
instrumental noise, and discuss the standard approaches and their
computational implications. This model will be elaborated in the
following sections to include systematics contributions and to discuss
specific procedures to mitigate their impact.

Map-making deals with estimation of maps from timelines containing
redundant observations of the sky. This subject has closely followed
experimental progress in the field. Map-making schemes devised for
COBE \citep{1994ApJ...436..452L}, whose differential measuring
technique proved effective in reducing correlated noise, were extended
to maps containing millions of pixels for
\WMAP\ \citep{1996ApJ...458L..53W}. More recent CMB experiments
(including \Planck) adopt a direct measurement scheme, as opposed to a
differential one, in order to gain sensitivity and reduce the
complexity of the optical system. This approach, also adopted for
\CORE, faces higher levels of $1/f$ noise, which has to be kept under
control by employing suitable analysis methods (see, e.g.,
\citet{2001A&A...374..358D,2001A&A...372..346N,2002PhRvD..65b2003S,Tristram_2011,2005A&A...436.1159D,2008ApJ...681..708P})

A widely employed model assumes that the timeline $\mathbf{d}$ depends
linearly on the map $\mathbf{m}$ by means of a `pointing' operator
$\mathbf{A}$:
\begin{equation}\label{eq:data_model_mm}
\mathbf{d} = \mathbf{A}\, \mathbf{m} + \mathbf{n},
\end{equation} 
where the time-ordered vector $\mathbf{n}$ is a stochastic noise
component with zero mean and (usually non-diagonal) covariance matrix
$N_{tt^\prime} \equiv \langle n_t n_{t^\prime}\rangle$ ($t$ labels
time samples) and the vector $\mathbf{m}$ is a discretized image of
the sky\footnote{We shall employ the HEALPix pixelization scheme in
  what follows \citep{2005ApJ...622..759G}.}, containing maps of the
Stokes parameters for intensity $I$ and linear polarization $Q$ and
$U$\footnote{Circular polarization $V$ is seldom considered for CMB,
  since it cannot be produced by Thomson scattering over electrons by
  an unpolarized and anisotropic radiation field
  \citep{1999NewAR..43..157K}. Instruments employed for CMB
  measurements do not normally possess the capability to measure
  circular polarization.}. The simplest possible model for
$\mathbf{A}$ is the so-called `pencil beam' approximation, which
ignores the convolution of the signal by the instrumental beam. In
this limit, the projection from the sky to the timeline of
Eq.~\ref{eq:data_model_mm} reads:
\begin{equation}\label{eq:signal_model_mm}
d_t = I + Q \cos2\psi_t +  U \sin2\psi_t +n_t,  
\end{equation} 
where $(I,Q,U)$ are the value of the Stokes parameters of the sky for
a given instrumental pointing and $\psi_t$ is the instantaneous
detector orientation with respect to a chosen celestial frame. Hence,
the pencil beam pointing matrix has only three non-zero entries in
each row, equal to $[1, \cos2\psi_t, \sin2\psi_t]$. If the
instrumental beam is azimuthally symmetric, the pointing and beam
convolution operations commute. If this is the case, we may retain the
pencil beam approximation and look for an estimate of the
beam-convolved map. On the other hand, if the beam is asymmetric the
model \ref{eq:signal_model_mm} leads to a biased estimate of the map
unless proper treatment is included. This situation is addressed in
Section~\ref{sec:AsymmetricBeam} below.

An estimate of the map, $\widetilde{\mathbf{m}}$, can be obtained by
applying the generalized least squares (GLS) procedure to
Eq.~\ref{eq:data_model_mm}:
\begin{equation}\label{eq:gls_mm}
\widetilde{\mathbf{m}} = (\mathbf{A}^T \mathbf{N}^{-1} \mathbf{A})^{-1}\mathbf{A}^T\mathbf{N}^{-1}\mathbf{d},
\end{equation} 
where $\mathbf{A}^T$ denotes the transpose of the pointing
operator. The quantity $(\mathbf{A}^T \mathbf{N}^{-1}
\mathbf{A})^{-1}$ is the covariance matrix of
$\widetilde{\mathbf{m}}$. The GLS estimate enjoys a number of
desirable properties: provided the noise matrix $\mathbf{N}$ is
correct (in practice, it must be estimated from the data) it is the
minimum variance estimator. Furthermore, if the noise is drawn from a
multivariate Gaussian distribution, the GLS estimate becomes the
maximum likelihood solution. It is, however, intractable to compute
the matrix for a real world situation with trillions of time samples
and millions of map pixels. The problem can be effectively solved by
resorting to iterative techniques, typically employing a conjugate
gradient solver \citep{2001A&A...372..346N}.

For the moment, we restrict the model to a single detector. Multiple
detector maps can be trivially accounted for in the absence of noise
that is correlated between detectors. If this is not the case, an
optimal solution can still be obtained by taking the correlations in
account. We discuss an application to \CORE\ of this scenario is
Section \ref{sec:CorrelatedNoise} below.

In the case of large datasets, it may be desirable to further reduce
the computational burden. This can be achieved by using approximate
versions of Eq.~\ref{eq:gls_mm}. A straightforward way to obtain such
an approximation is to model the correlated component of the noise
using a set of basis functions (typically piecewise constant offsets
of given constant length, although more complicated bases can be used)
superimposed on white noise. The problem then reduces to finding a
suitable estimate of the coefficients of the basis functions. This
class of map-making codes is called destripers
\citep[see, for example,][]{burigana.1997.destriping,Delabrouille_1998_destriping,Maino1999,keihanen2004,2005MNRAS.360..390K,2009A&A...506.1511K}.
Sophisticated implementations of these algorithms can produce results which are
statistically indistinguishable from GLS map-making while requiring
significantly less computational resources. In this scenario, prior
information on the correlated noise properties may be needed
\citep{2009A&A...506.1511K}. The most desirable feature of destriping
algorithms is that they can be tuned to the desired precision while
still controlling their computational cost. The latter of course
scales unfavourably with precision, but in real-world applications an
advantageous compromise can be usually found by tweaking the offset
length. In the following we will make extensive use of a public domain
implementation of a generalized destriper, MADAM
\citep{2005MNRAS.360..390K,2010A&A...510A..57K}.

\section{Simulations}
\label{sec:simulations}

\begin{figure}[htbp]
\begin{center}
\includegraphics[scale=0.15]{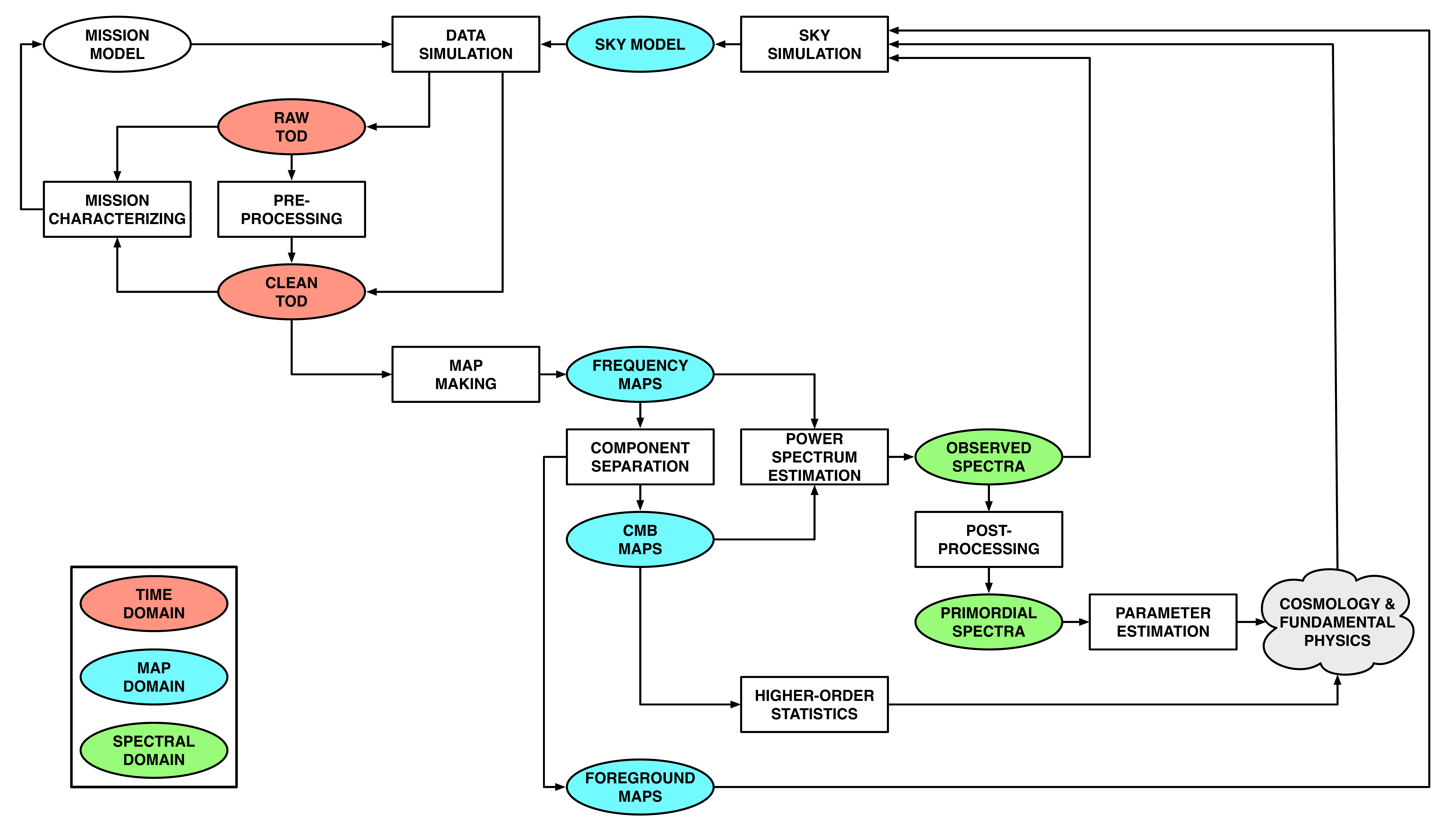}
\caption{A schematic CMB simulation and analysis pipeline, with
  rectangular operators acting on oval data objects, which may be time
  samples (red), map pixels (blue) or spectral multipoles
  (green). Note the many loops, implying iterative processing.}
\label{simda}
\end{center}
\end{figure}

Simulations play a number of critical roles in CMB missions:
\begin{enumerate}
\item Optimization of the design of the mission (both the instrument
  and the observation) to ensure that the dataset obtained will be
  sufficient to meet the science goals;
\item Validation and verification of the data analysis pipeline to
  ensure that the science can be extracted from the mission dataset;
\item Uncertainty quantification and debiasing of the data analysis
  results using Monte Carlo methods in lieu of the computationally
  intractable full data covariance matrix.
\item Encapsulation of knowledge on the data taking and processing,
  allowing e.g.\ for novel analyses outside of the team.

\end{enumerate}
All of these require a joint simulation and analysis pipeline capable
of generating a detailed realization of the full mission dataset and
reducing it to the science results; figure \ref{simda} provides a
schematic overview of such a pipeline. The model of the mission
includes both the instrument (including the detector properties, focal
plane layout and optical path) and the observations (including the
scanning strategy and data-flagging), while the sky model includes the
CMB together with all foregrounds (and their impact on the CMB through
lensing and scattering). The data simulation operator then applies the
mission model to the sky model to generate synthetic time-domain
data. The steps of the analysis pipeline alternate between mitigation
of the systematic effects in the current data domain (pre-processing,
component separation, post-processing) and reduction of the
statistical uncertainties by transforming the data to a new domain
with higher signal-to-noise (map-making, power spectrum
estimation). The map- and spectral-domain products are then used to
constrain the parameters of any given model of cosmology and
fundamental physics, typically in conjunction with other cosmological
datasets. Finally the various data representations can be used to
provide feedback to refine the mission and sky models.

In this work we particularly focus on the simulation and mitigation of
systematic effects to address all three questions, the optimization of
the mission design, the validation and verification of the mitigation
algorithms and implementations, and the quantification of the
residuals after mitigation and their impact on the science results.

In the absence of the explicit data covariance matrix, the most
computationally challenging elements of this pipeline are those that
manipulate the full time-domain data, and in particular the generation
and analysis of Monte Carlo realizations used in lieu of this matrix
for uncertainty quantification and debiasing. Given the volume of data
to be processed, we require highly optimized massively parallel
implementations of the simulation, pre-processing and map-making
algorithms and significant high performance computing
resources. Moreover, since data movement -- whether between disk and
memory or across distributed memory -- is expensive, these steps must
be tightly-coupled within an overall time-domain data framework. One
such framework, developed for the \Planck\ satellite mission
\citep{2016A&A...594A..12P} but with broad applicability for both
satellite and suborbital CMB missions, is the Time-Ordered
Astrophysics Scalable Tools (TOAST)
package\footnote{http://github.com/hpc4cmb/toast}.

As well as being highly computationally efficient, any such framework
must also be readily adaptable, allowing the rapid prototyping of new
algorithms. TOAST is therefore implemented as a python wrapper and
data management layer into which new modules can easily be dropped,
coupled with compiled libraries (both internal and external) which can
be called wherever computational efficiency is a limiting
factor. TOAST has been extensively validated and verfied, primarily in
conjunction with its use in the \Planck\ full focal plane simulations
\citep{2016A&A...594A..12P} but also through simulations of the
\CORE\ and \LiteBIRD\ satellite missions, and in stand-alone
comparisons with both analytic calculations and other computational
tools.

In this work the TOAST framework calls four main libraries, two
internal and two external to the TOAST package:
\begin{enumerate}
\item the TOAST pointing library, which generates the dense-sampled
  pointings for each detector from the sparse-sampled satellite
  boresight pointing.
\item the TOAST noise simulation library, which generates timelines of
  noise from each detector's piecewise stationary noise power spectral
  density functions, provided either as a set of arrays of explicit
  frequency/power pairs, or as the parameters of an analytic function
  (typically a white noise level and correlated noise knee frequency
  and spectral index).
\item the libCONVIQT beam convolution
  library\footnote{http://github.com/hpc4cmb/libconviqt}, a
  TOAST-compatible implementation of the CONVIQT beam convolution
  algorithm \citep{2010ApJS..190..267P}, which generates timelines of
  sky signals from each detector's full asymmetric beam and pointings
  and the simulated sky being observed.
\item the libMADAM map-making
  library\footnote{http://github.com/hpc4cmb/libmadam}, a
  TOAST-compatible implementation of the MADAM map-making algorithm
  \citep{2005MNRAS.360..390K,2010A&A...510A..57K}, which makes a
  destriped map of the sky given some set of time-ordered data and
  pointings, for some set of detectors.
\end{enumerate}
Using 1+2+4 we generate coverage and noise maps to evaluate scanning
strategies and correlated noise performance, while using 1+3+4 we
generate sky signal maps to evaluate the impact of asymmetric
beams. In general, the parameters used, and the analyses of the
resulting maps, are discussed in detail in the following sections. For
consistency though we employ the same MADAM destriping parameters
throughout, in particular setting the destriping offset length and the
prior on the correlated noise to maximize statistical efficiency.

We have carried out several tests, considering a variety of offset
lengths with and without a noise prior. For simulations, where we know
the noise properties, it can be taken to be a priori known and exact;
for real data it would be necessary to estimate the noise properties
from the timeline data, although the accuracy of this estimate does
not need to be especially high for typical applications
\citep{2002A&A...383.1100N}. For the \CORE\ scanning strategy and the
noise properties described in Table \ref{tab:toast_parameters}, we
found that the best MADAM performance is achieved for a offset of $1$
s and using the exact noise prior (i.e. the description provided to
the TOAST noise simulation tool).

\section{Analysis of simulated noise maps}
\label{Analysis_maps}

In this section we describe the noise maps produced with MADAM from
timelines simulated with the TOAST pipeline described above.  We
analyse these maps to assess the robustness of the \CORE\ scanning
strategy in measuring the sky Stokes parameters with adequate
purity. We also explore possible tweaks to the scanning parameters to
verify if they lead to increase robustness. Finally, we analyse the
properties of the noise maps to find requirements on the detector knee
frequency that ensure that residual contributions to the map on large
angular scales are kept under control.

\begin{table}
\centering
\begin{tabular}{|l|l|c|}
\hline
\multicolumn{2}{|l|}{Parameter} & Value \\
\hline
\hline
\multicolumn{2}{|l|}{Precession angle [$^\circ$]} & 30 \\
\hline
\multicolumn{2}{|l|}{Spin angle [$^\circ$]} & 65 \\
\hline
\multicolumn{2}{|l|}{Precession period [days]} & 4 \\
\hline
\multicolumn{2}{|l|}{Spin period [s]} & 120 \\
\hline
\multicolumn{2}{|l|}{Hours of observation per day [h]} & 24 \\
\hline
\multicolumn{2}{|l|}{Length of a single chunk of data [h]} & 24 \\
\hline
\multicolumn{2}{|l|}{Observation duration [days]} & 366 \\
\hline
\multicolumn{2}{|l|}{Number of detectors} & 2 \\
\hline
\multicolumn{2}{|l|}{Frequency [GHz]} & 145 \\
\hline
\multicolumn{2}{|l|}{FWHM [arcmin]} & 7.68 \\
\hline
\multicolumn{2}{|l|}{Sampling rate [Hz]} & 84.97 \\
\hline
\multicolumn{2}{|l|}{Polarization orientation detector 1 [$^\circ$] }& -22.5 \\
\hline
\multicolumn{2}{|l|}{Polarization orientation detector 2 [$^\circ$]} & 67.5 \\
\hline
\multicolumn{2}{|l|}{Knee frequency $f_k$ [mHz]} & 0, 10, 20, 50 \\
\hline
\multicolumn{2}{|l|}{Noise slope $\alpha$} & 1.0 \\
\hline
\multicolumn{2}{|l|}{NET [$\mathrm{\mu K\!\sqrt{\mathrm{s}}}$]} & 52.3 \\
\hline
\multirow{2}{3 cm}{Deviation from boresight [$^\circ$]} & `high' & +4.7 \\
\cline{2-3}
 & `low' & -4.7 \\
\hline
\multicolumn{2}{|l|}{$\mathrm{N_{side}}$} & 1024 \\
\hline
\multicolumn{2}{|l|}{Offset length (with noise prior) [s]} & 1.0  \\
\hline
\end{tabular}
\caption{\footnotesize{Parameters supplied to TOAST to generate the baseline simulations. See text for details. The sampling rate is chosen to ensure four samples per beam FWHM.}}\label{tab:toast_parameters}
\end{table}

In Table \ref{tab:toast_parameters} we summarize the parameters we
selected as input to TOAST to produce the maps we analyse in this
Section. More detail about the parameters in this Table and on \CORE's
scanning strategy is given in \cite{ECO.instrument.paper}). We
consider the \CORE\ baseline scanning strategy with spin and
precession angles of $65^\circ$ and $30^\circ$ respectively, with
corresponding periods of 120 s and 4 days respectively.  In this
Section we consider timelines containing only instrumental
noise. Signal contributions are examined in Sections
\ref{sec:BandpassMismatch}, \ref{sec:AsymmetricBeam} and
\ref{sec:Calibration} below.  We simulate an entire year of
observations divided into segments of 24 hours. These are then
combined to produce the final map. This segment size is a reasonable
compromise between the need to capture long timescale features in the
noise and the desire to minimize computational and memory
requirements. We assume a noise model with power spectrum density:
\begin{equation}\label{eq:noisemodel}
P(f)=A\left[\left(\frac{f_k}{f}\right)^\alpha+1\right]
\end{equation}
where $f$ is the frequency, $f_k$ the knee frequency which we will
vary below, $A$ the amplitude and $\alpha$ a slope equal to 1. A
reasonable choice for \CORE\ two detector system is $f_k=20$\,mHz and
an amplitude corresponding to a NET of $53.2\; \mathrm{\mu
  K\!\sqrt{\mathrm{s}}}$.  The impact of cross-correlation of noise
between detectors is discussed in Section \ref{sec:CorrelatedNoise}
below, were we employ, in place of MADAM a dedicated map-making code,
ROMA \citep{2016A&A...593A..15D}, capable of taking cross-correlation
information in account to deliver a lower noise solution.

In the remainder of this Section we consider a pair of polarization
sensitive detectors at 145\,GHz at the same position in the focal
plane, either at the boresight or at the edges of the focal plane,
oriented at $-22.5^\circ$ and $67.5^\circ$ with respect to the scan
direction. This choice equalizes the noise power in the $Q$ and $U$
Stokes parameters and produces EE and BB angular power spectra with
similar amplitude. In any case, any particular choice of orientation
becomes irrelevant when producing maps from a large number of
detectors, assuming their orientations are evenly spaced. We adopt
these particular values for the sole purpose of achieving balance in
the Stokes parameters in this minimal two-detector exercise.  We also
simulate the sky as observed by detectors at the edge of the focal
plane. These are modelled by considering two pairs of detectors at
$\pm 4.7^\circ$ with respect to the boresight along the direction
orthogonal to the scan direction. These have the same polarization
orientation as the boresight detectors and are labelled as `high' and
`low' detectors in Table \ref{tab:toast_parameters} and hereinafter.

\begin{figure}
\centering
\includegraphics[scale=0.6]{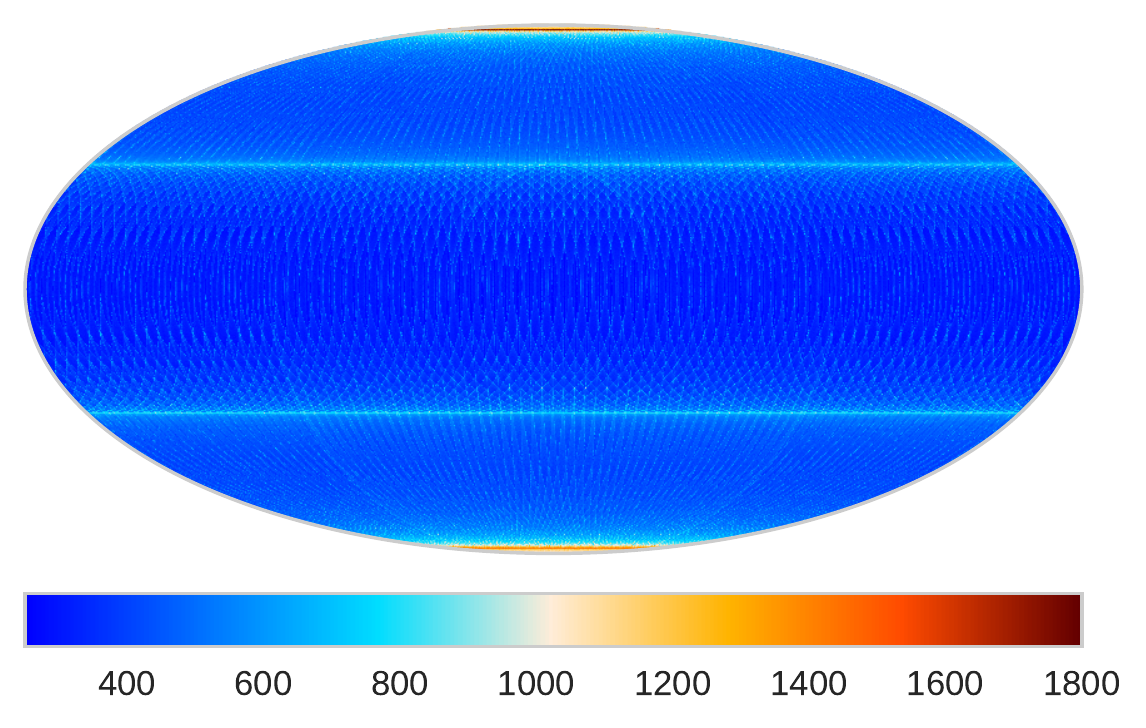}
\includegraphics[scale=0.6]{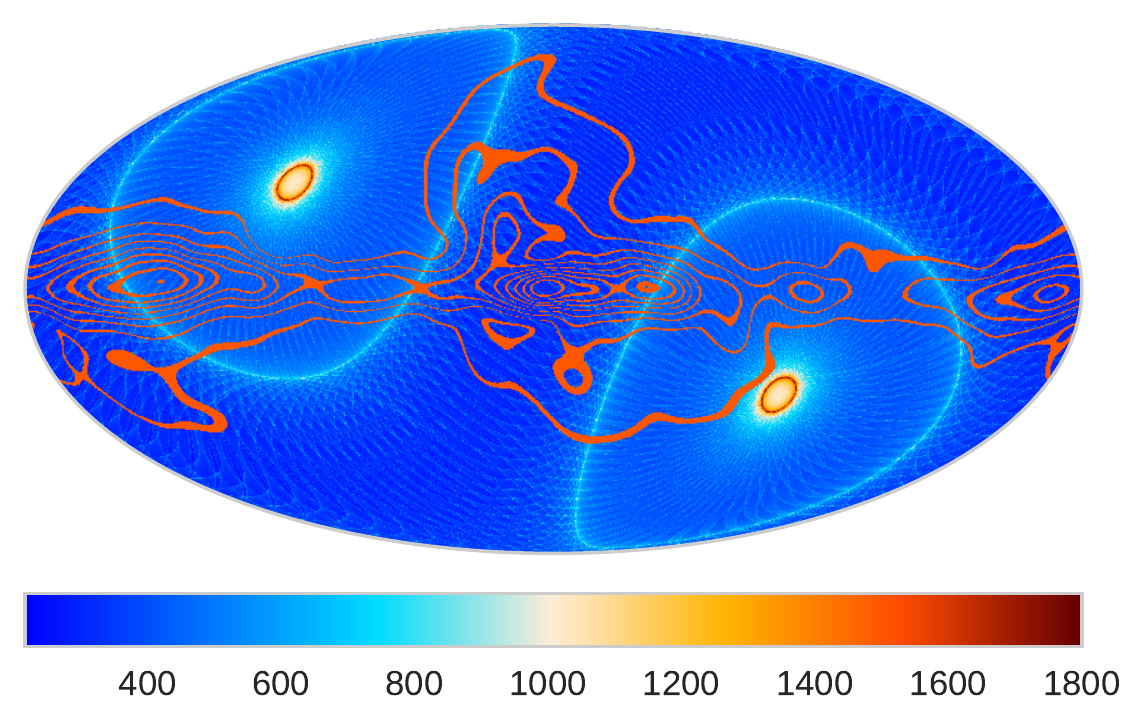}
\caption{\footnotesize{Hit map for a pair of detectors located at the
    center of the focal plane after one year of observation. The map
    is shown in Ecliptic coordinates (left) and in Galactic
    coordinates (right). We also show (orange contours) an estimate of
    the polarization amplitude of the foregrounds at 70\,GHz, a
    frequency close to the minimum emission of diffuse foregrounds.
    The outermost contour corresponds to $1.3$\muK\ in polarized
    intensity, and the subsequent contours to further steps of
    $1.3$\muK.}}
\label{fig:hitsmap}
\end{figure}

The hit map for the two-detector case described above is given in
Fig.~\ref{fig:hitsmap}, having chosen a boresight direction. The
irregular small-scale features, hardly visible at standard figure
size, would be diluted when considering a larger number of
detectors. From this simple exercise, we show that the \CORE\ scanning
strategy leads to a complete sky coverage in around 6 months, yet a
coverage of around 45\% of the full sky is achieved in just 4 days
thanks to the wide precession. After one year, all pixels in the sky
have been observed at least 200 times by a pair of detectors, assuming
3.4 arcmin pixels (HEALPix\ $\mathrm{N_{side}}=1024$). The hit map in
Galactic coordinates is overplotted with an estimate of the total
diffuse polarized foregrounds at 70\,GHz, where emission is close to a
minimum \citep{Planck_2015_X}. This estimate was obtained using the
\Planck\ 353\,GHz and 30\,GHz polarized maps respectively as dust and
synchrotron templates. This simple exercise shows how the
\CORE\ scanning strategy provides high signal-to-noise sampling of
regions that are remarkably clean of polarized foreground emission. Of
course, the use of specific component separation techniques will
reduce residual foreground emission considerably
\citep{ECO.foregrounds.paper}.

We compute the white noise covariance matrix for the chosen scanning
strategy. This gives a $3\times3$ symmetric positive definite matrix
for $(I,Q,U)$ in each pixel. In so doing, we ignore $1/f$
contributions that would generate correlations between pixels. For
these $3\times3$ matrices we compute the reciprocal condition number
(RCN), defined as the ratio of its smallest to largest eigenvalue. The
RCN is an useful indicator to decide whether a matrix is
ill-conditioned. We employ it here to verify the purity of the
map-making solution for the Stokes parameters. A RCN of $1/2$ is
achieved only in ideal cases, while values too low, even if still
adequate from a purely numerical standpoint, may leave the system
vulnerable to non-idealities, by amplifying the effects of systematic
contributions in the timeline. We set a limit of a minimum RCN of
$10^{-2}$ for the present analysis.  We also compute the angular power
spectra (APS) of the simulated noise maps. The noise APS allow to
assess the destriping efficiency of MADAM in controlling spurious
low-frequency contributions.

\subsection{Baseline scanning strategy}

\begin{figure}
\centering
\includegraphics[scale=0.3]{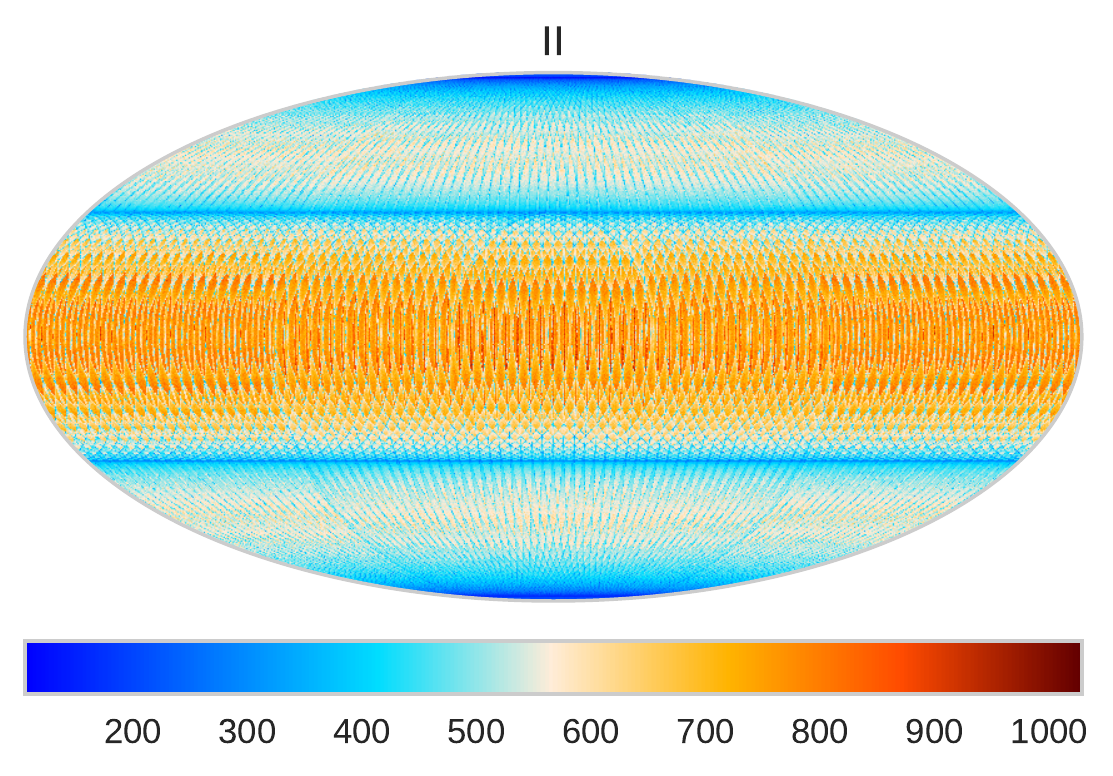}
\includegraphics[scale=0.3]{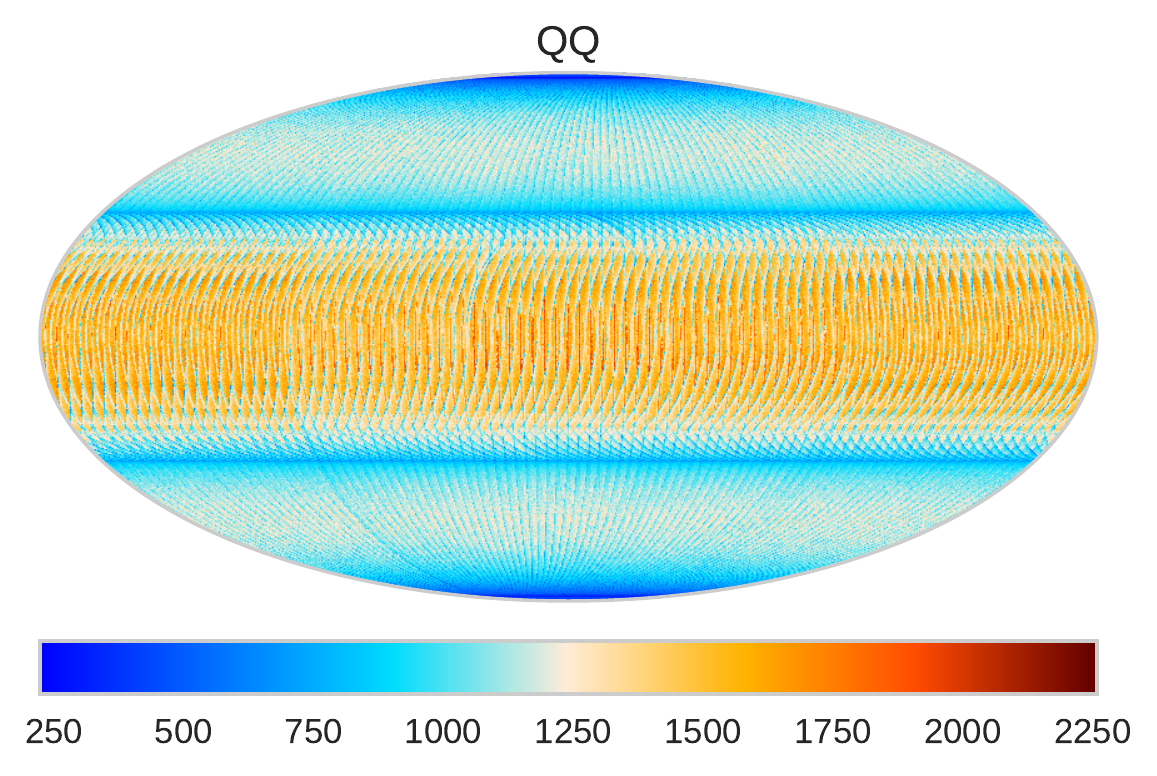}
\includegraphics[scale=0.3]{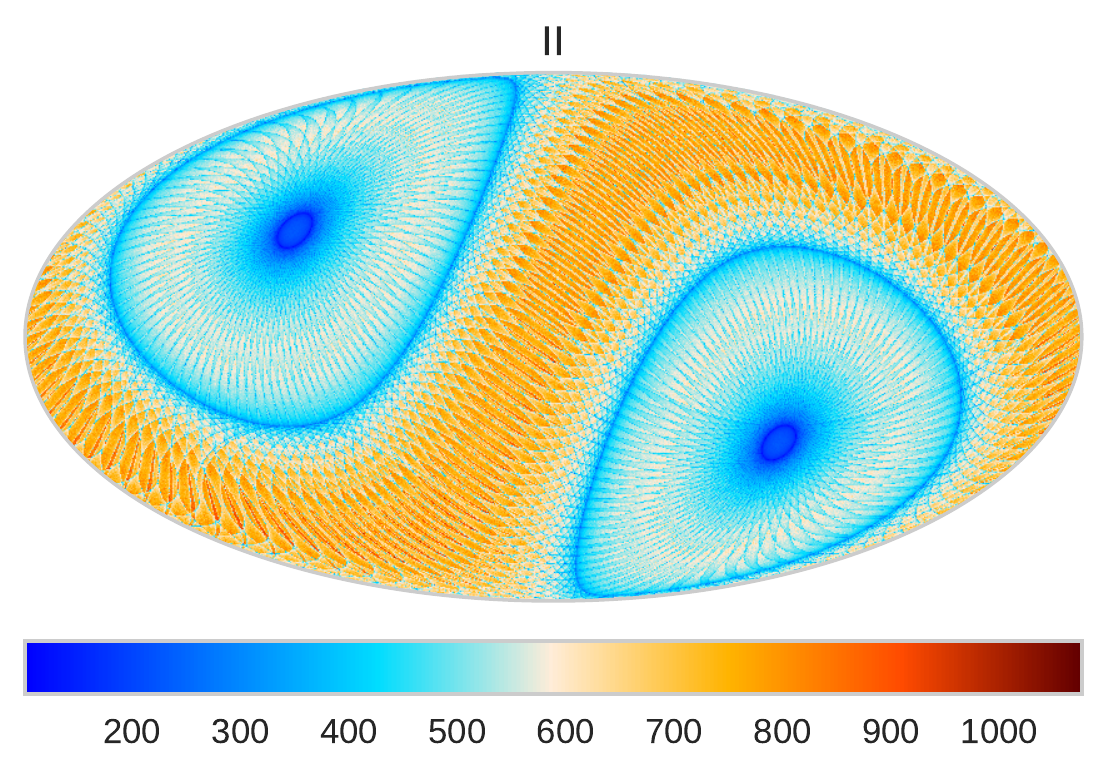}
\includegraphics[scale=0.3]{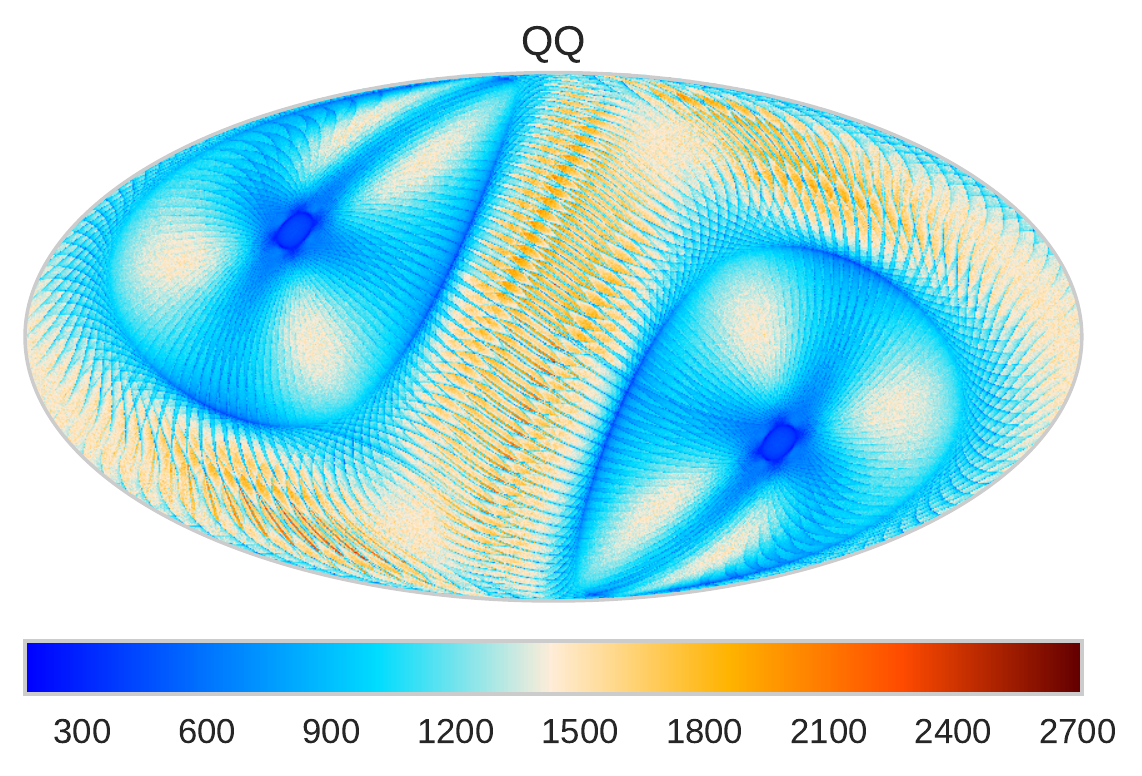}\\
\includegraphics[scale=0.3]{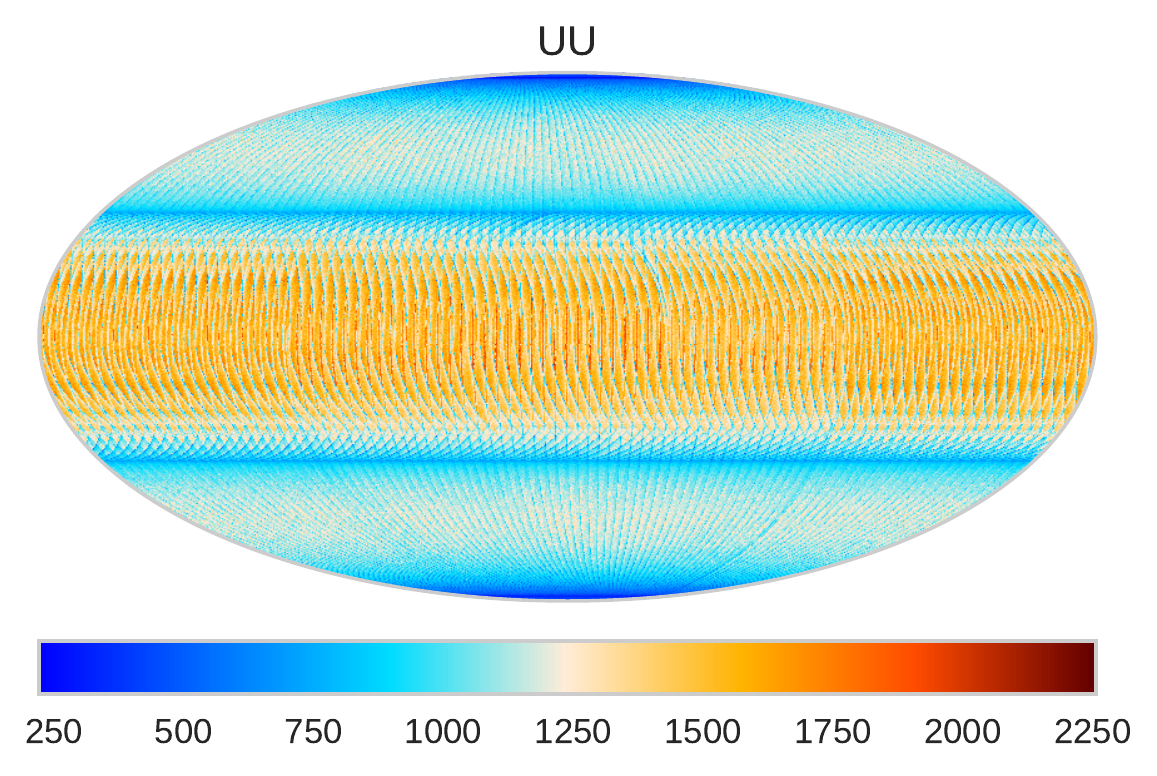}
\includegraphics[scale=0.3]{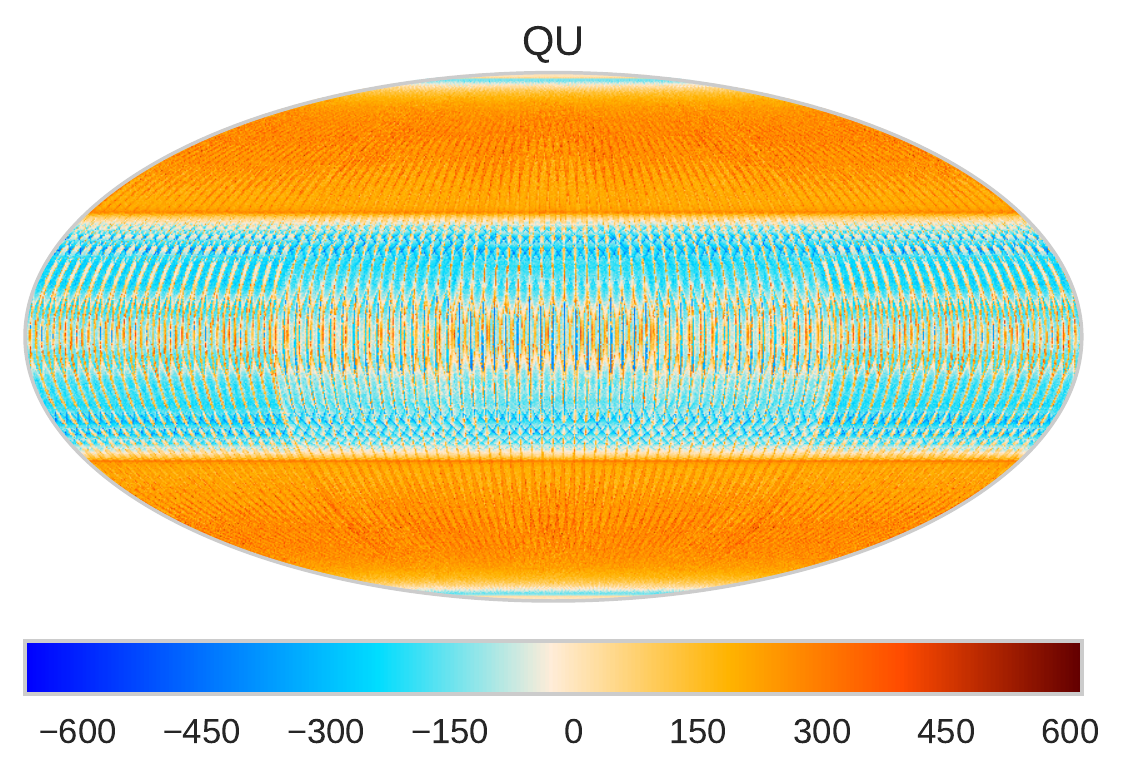}
\includegraphics[scale=0.3]{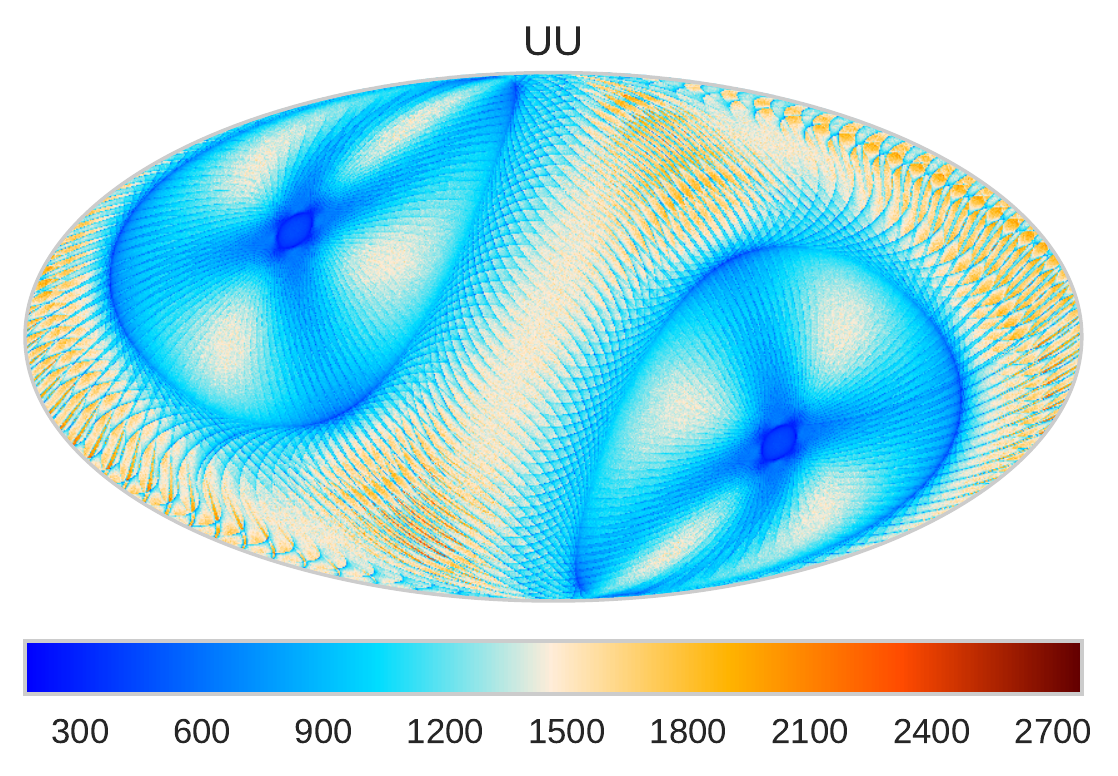}
\includegraphics[scale=0.3]{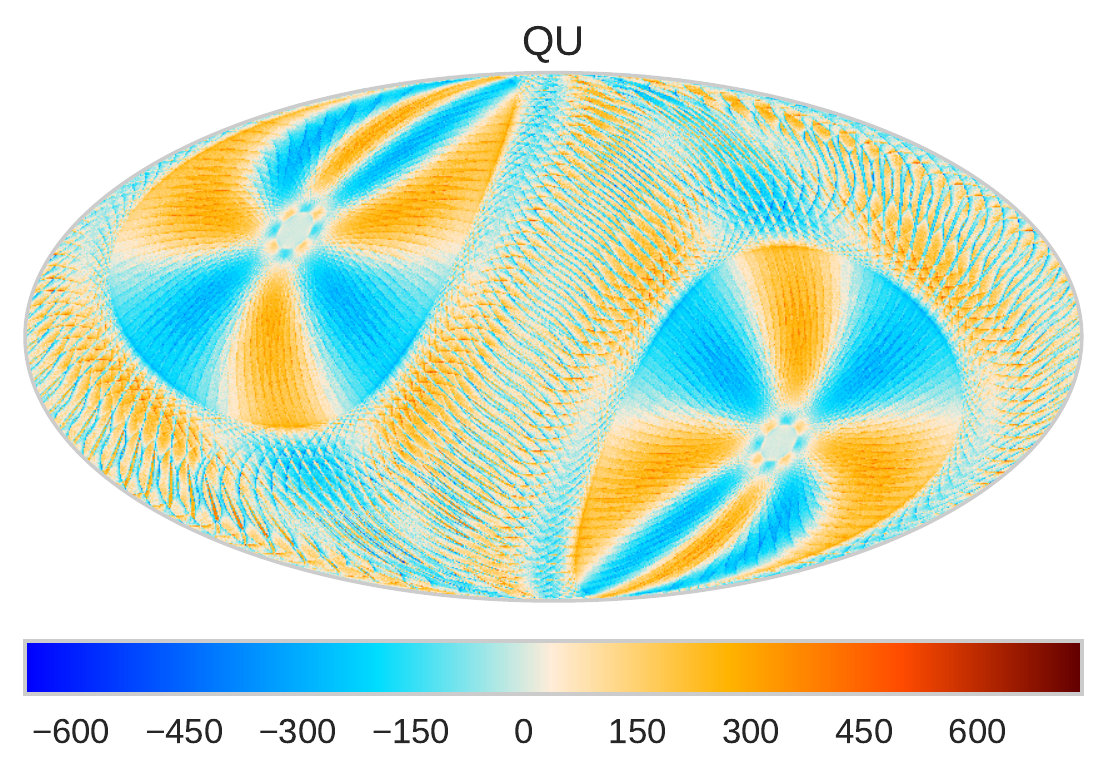}\\
\includegraphics[scale=0.3]{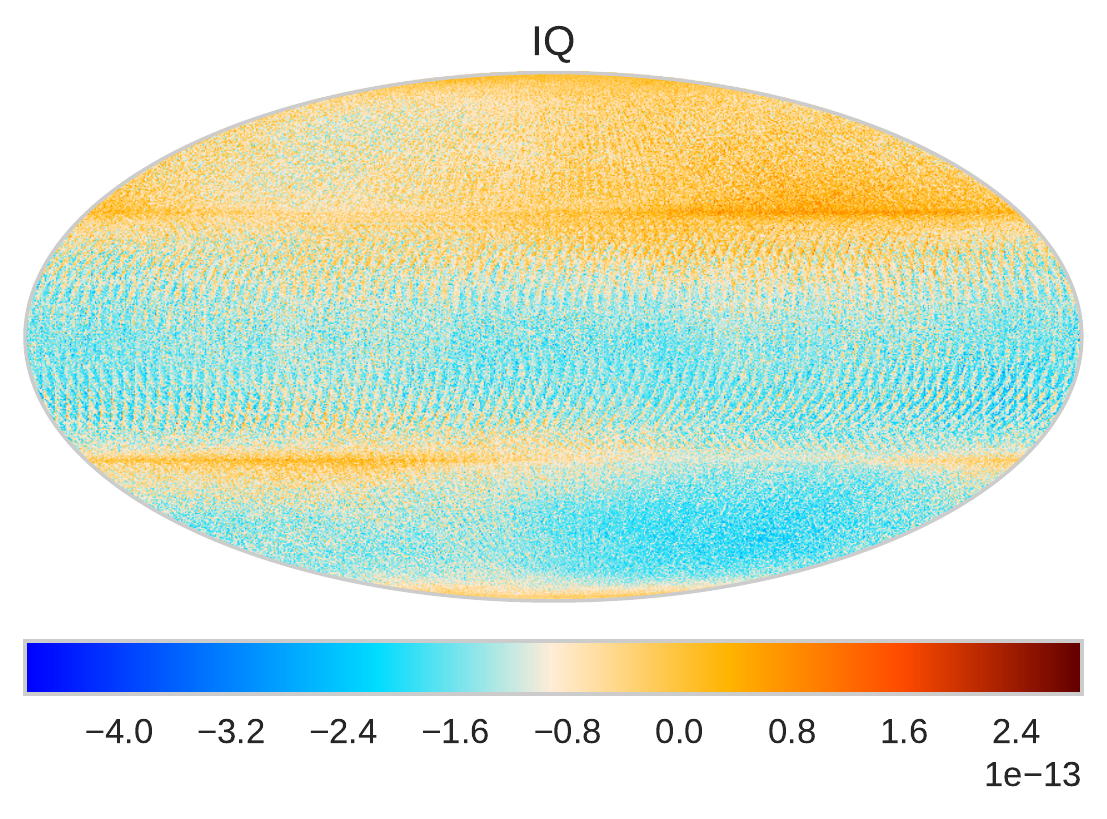}
\includegraphics[scale=0.3]{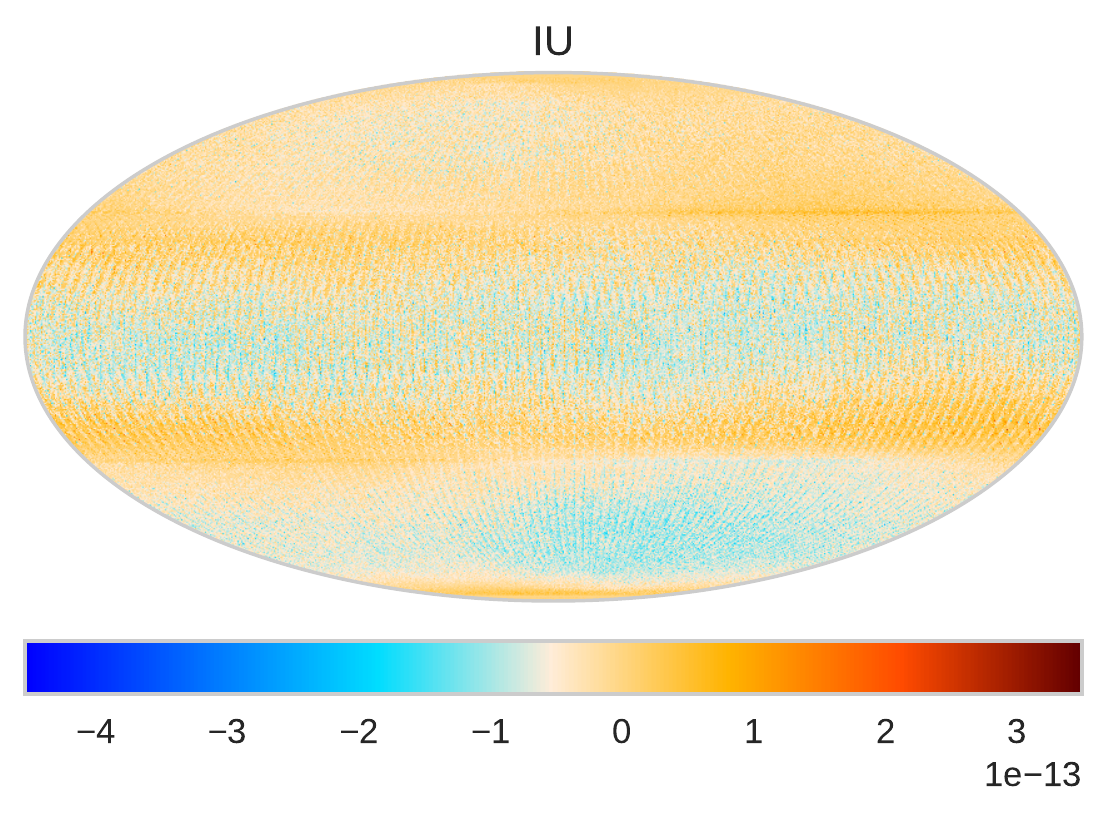}
\includegraphics[scale=0.3]{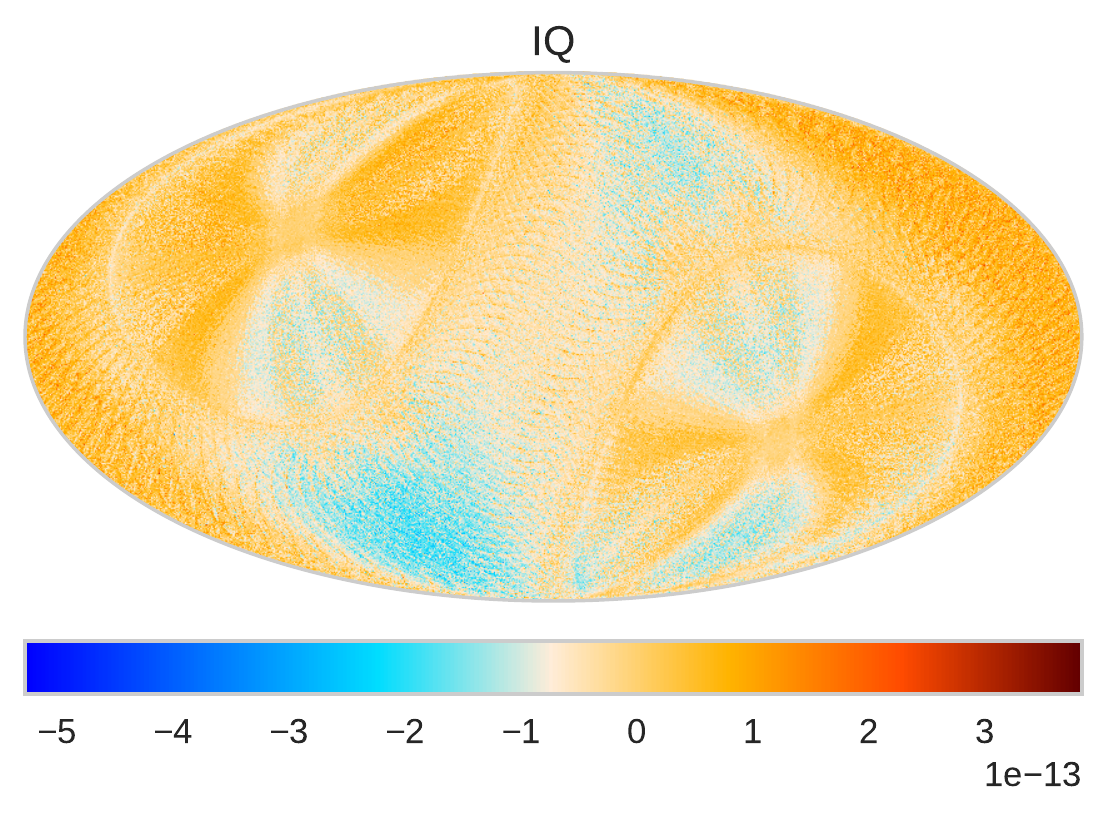}
\includegraphics[scale=0.3]{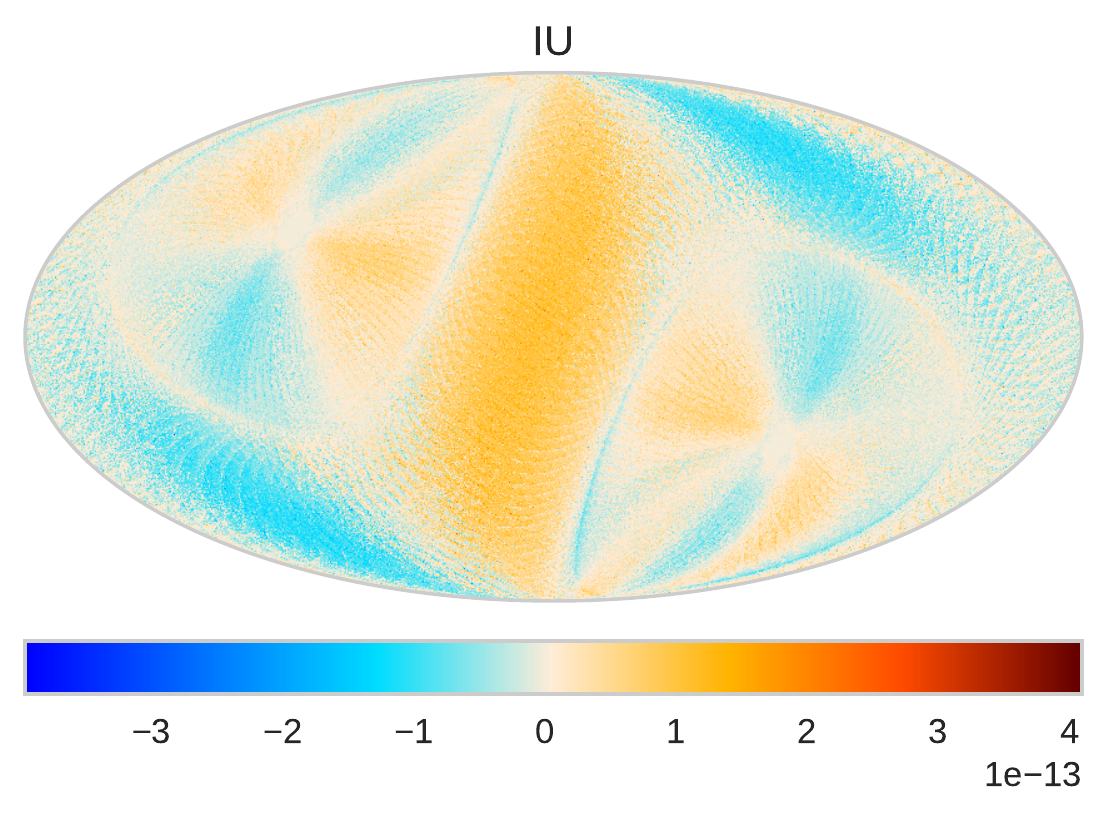}
\caption{\footnotesize{Elements of the white noise covariance matrix
    for a pair of boresight detectors displayed as maps in units of
    $\mu K^2$: $II$ (top left), $QQ$ (top right), $UU$ (center left),
    $QU$ (center right), $IU$ (bottom left), $IQ$ (bottom
    right). Coordinates are Ecliptic (left columns) and Galactic
    (right columns). Notice that $IQ$ and $IU$ correlations are very
    weak.}}\label{fig:noisemaps}
\end{figure}

\begin{figure}
\centering
\includegraphics[scale=0.39]{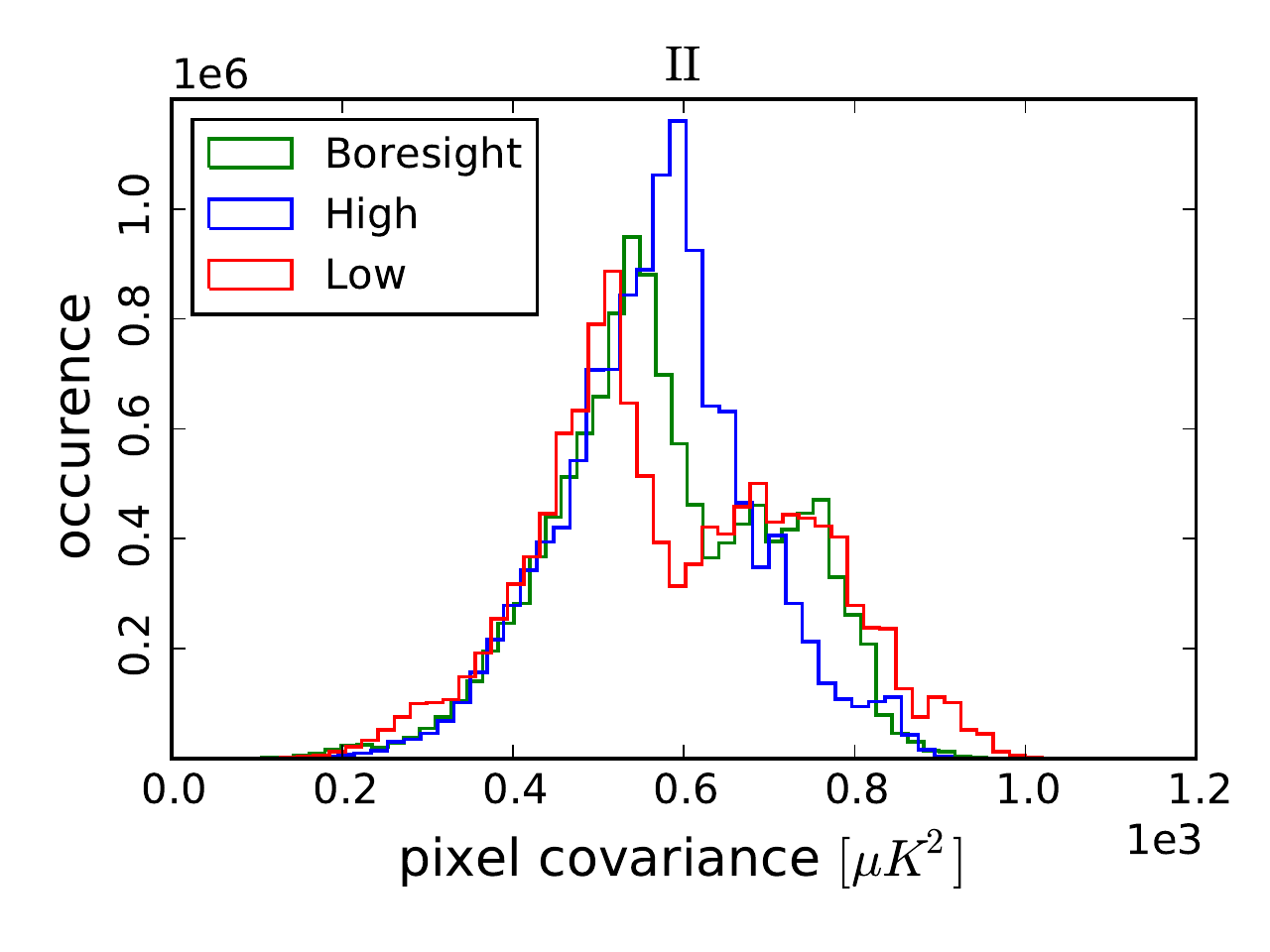}
\includegraphics[scale=0.39]{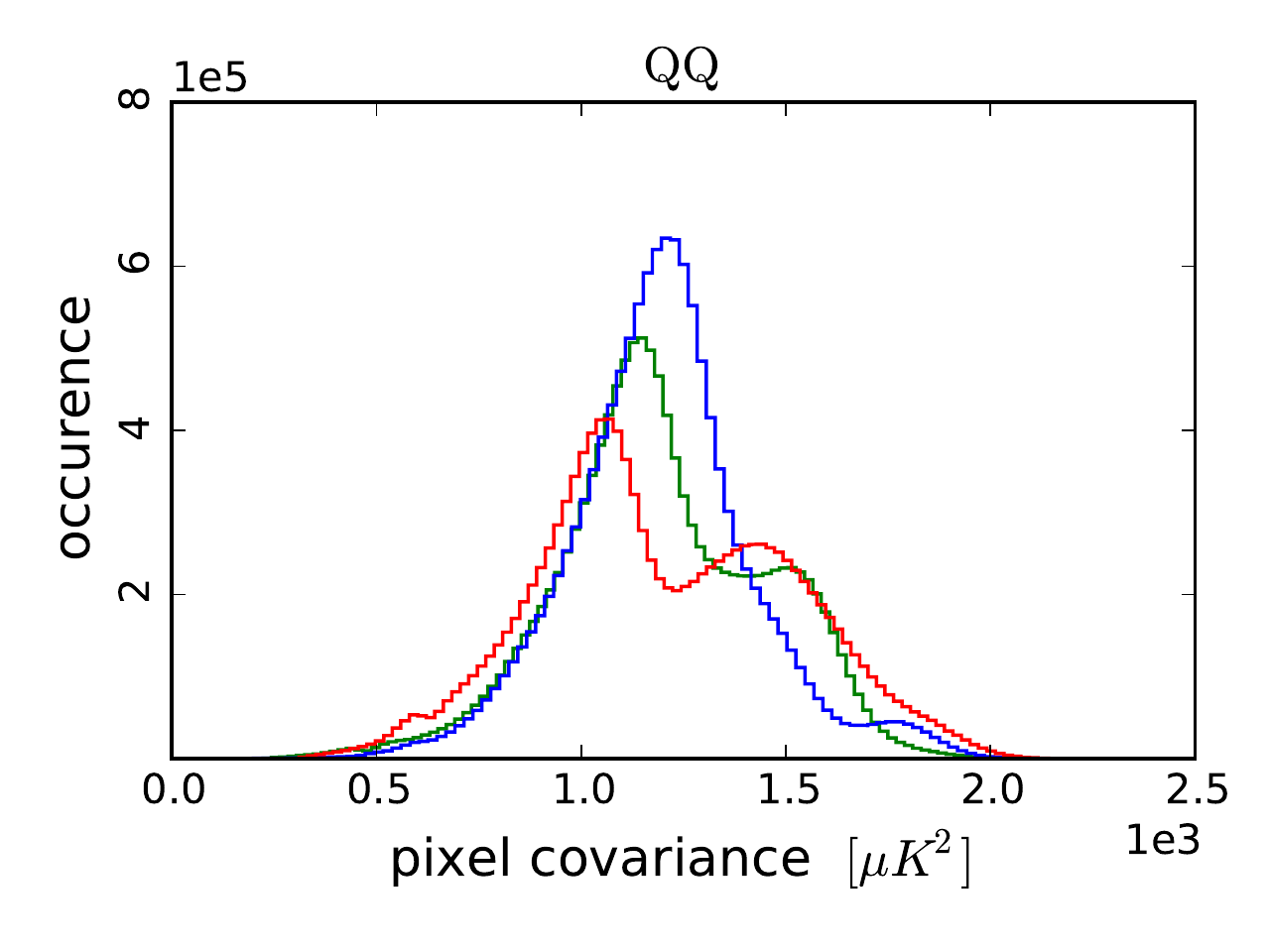}
\includegraphics[scale=0.39]{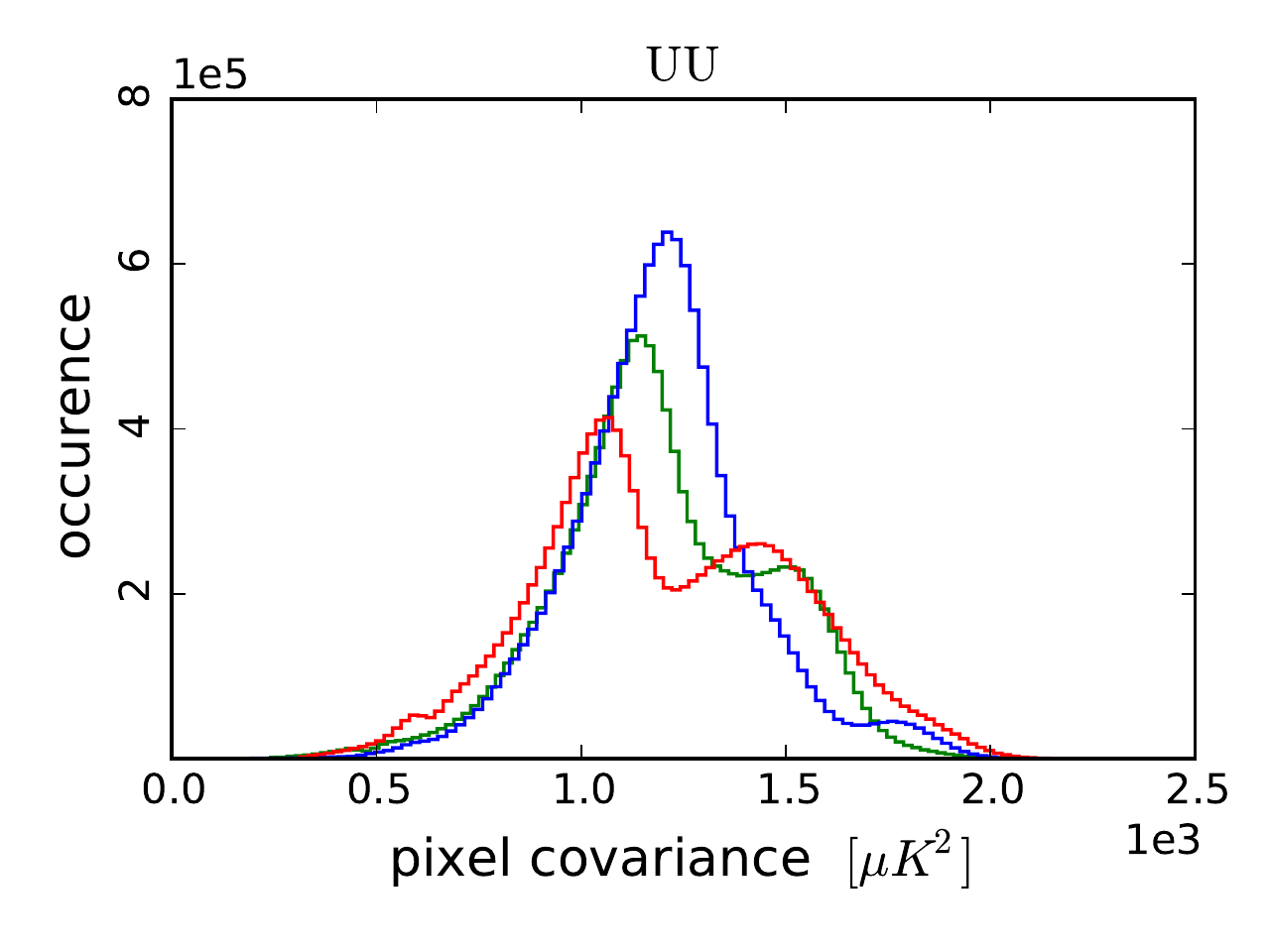}\\
\includegraphics[scale=0.39]{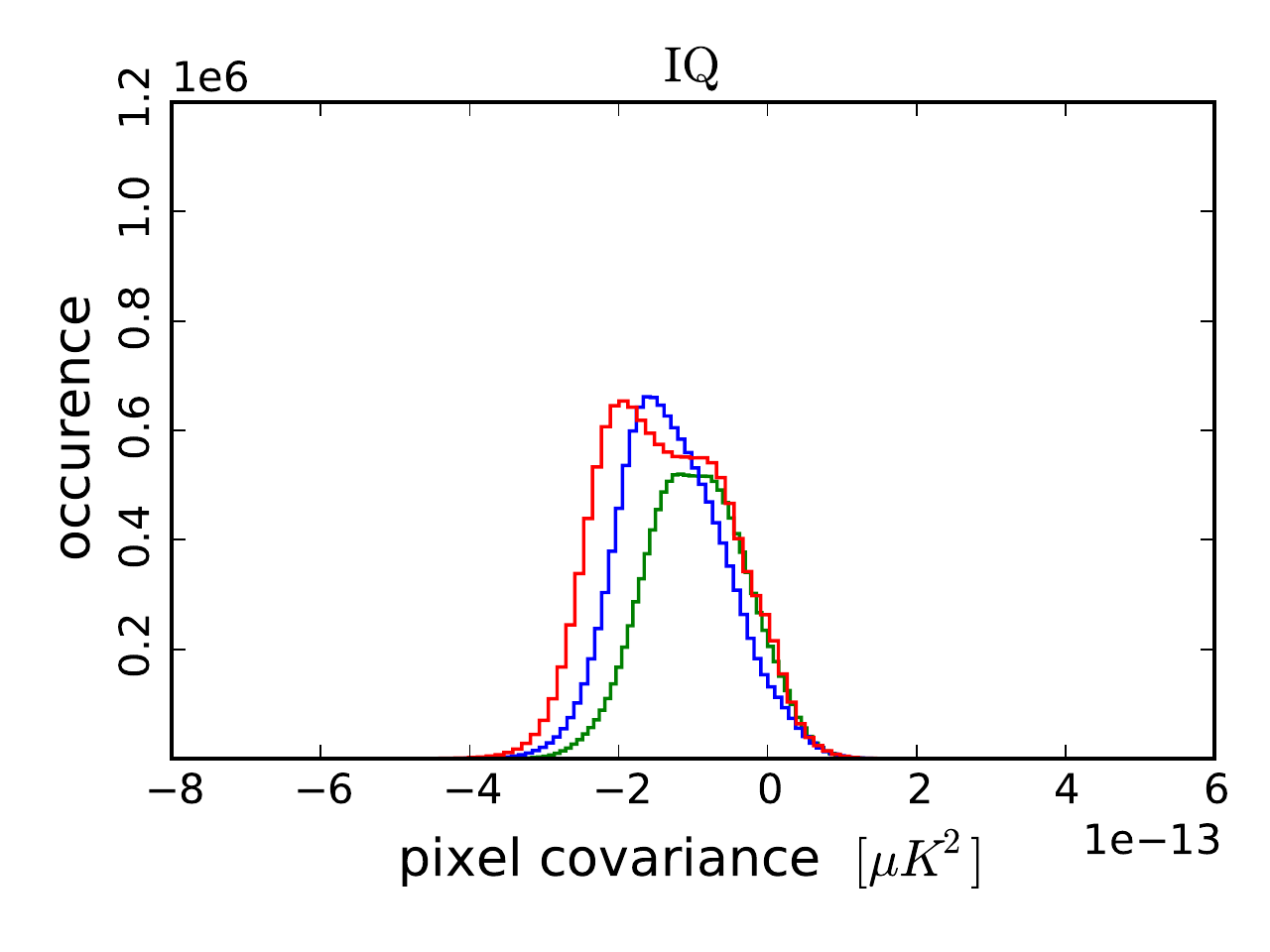}
\includegraphics[scale=0.39]{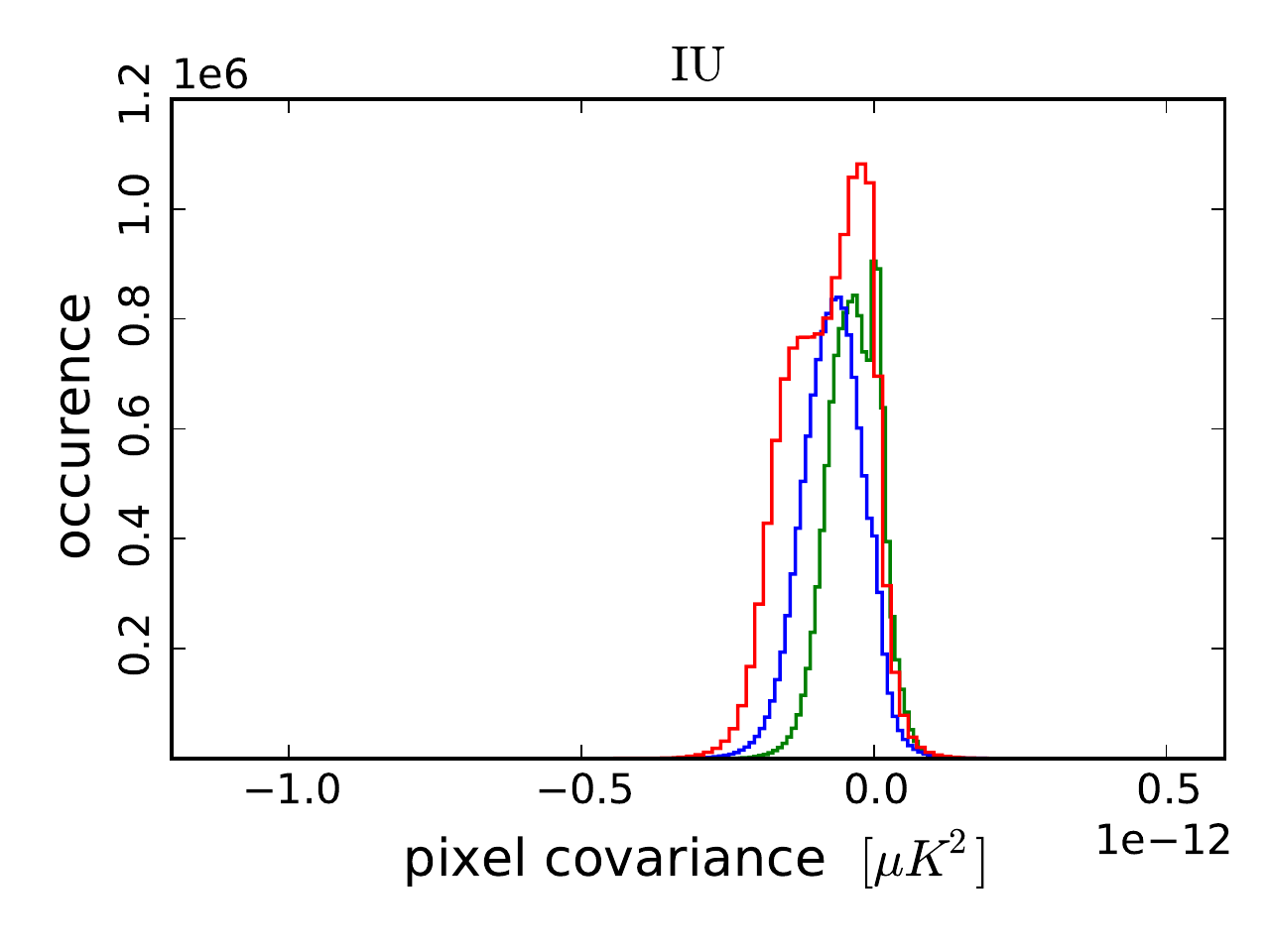}
\includegraphics[scale=0.39]{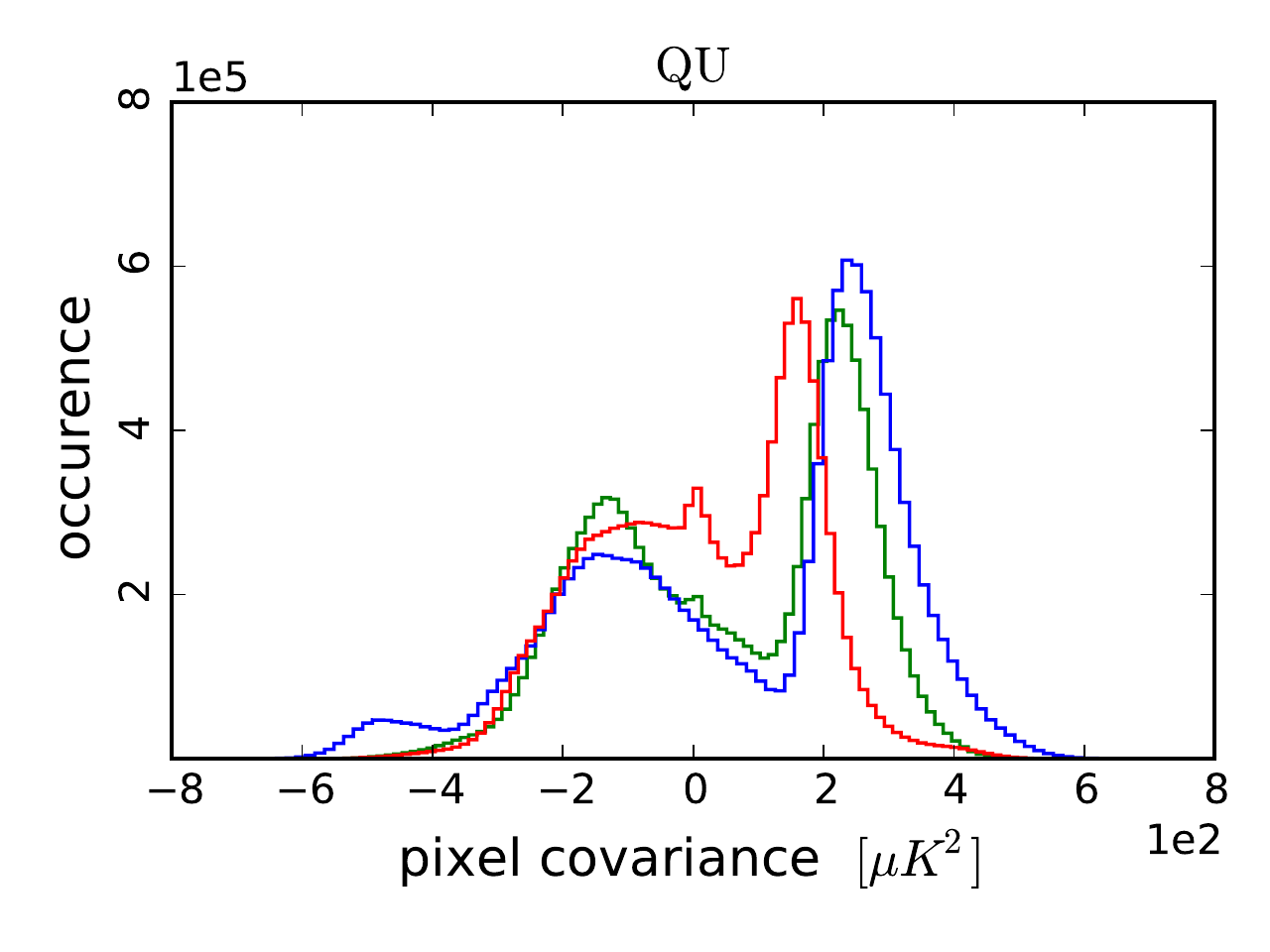}\\
\includegraphics[scale=0.39]{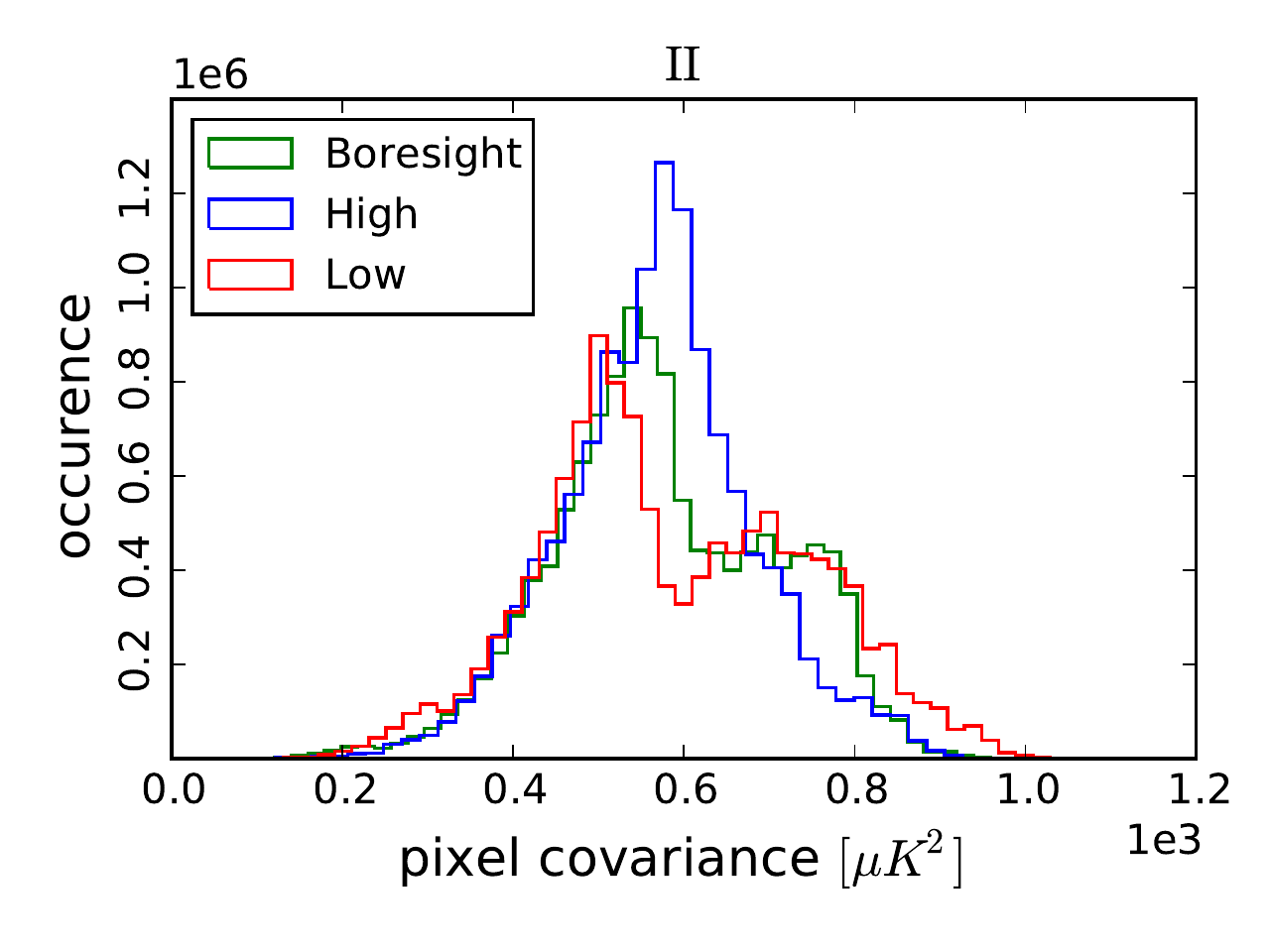}
\includegraphics[scale=0.39]{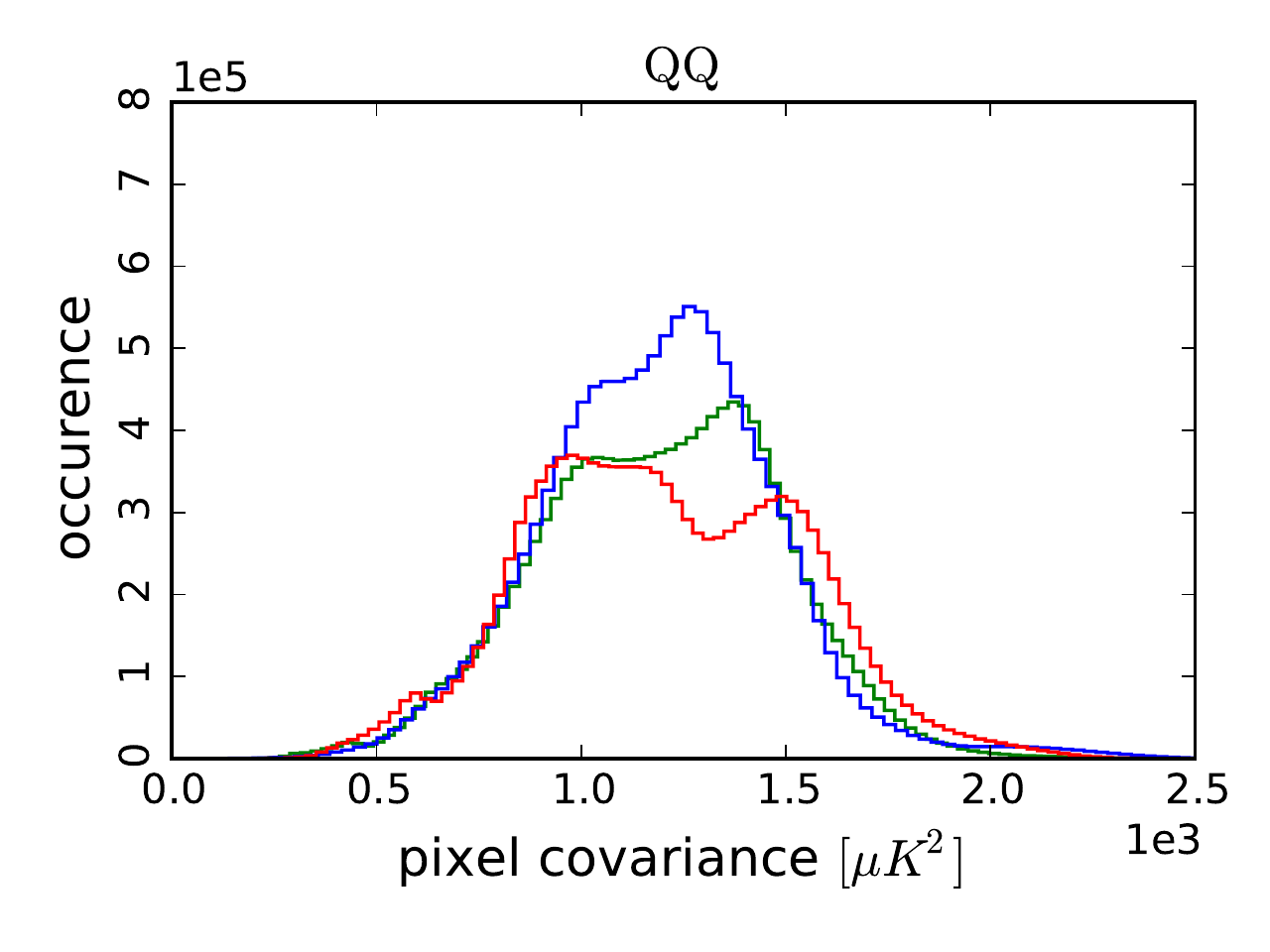}
\includegraphics[scale=0.39]{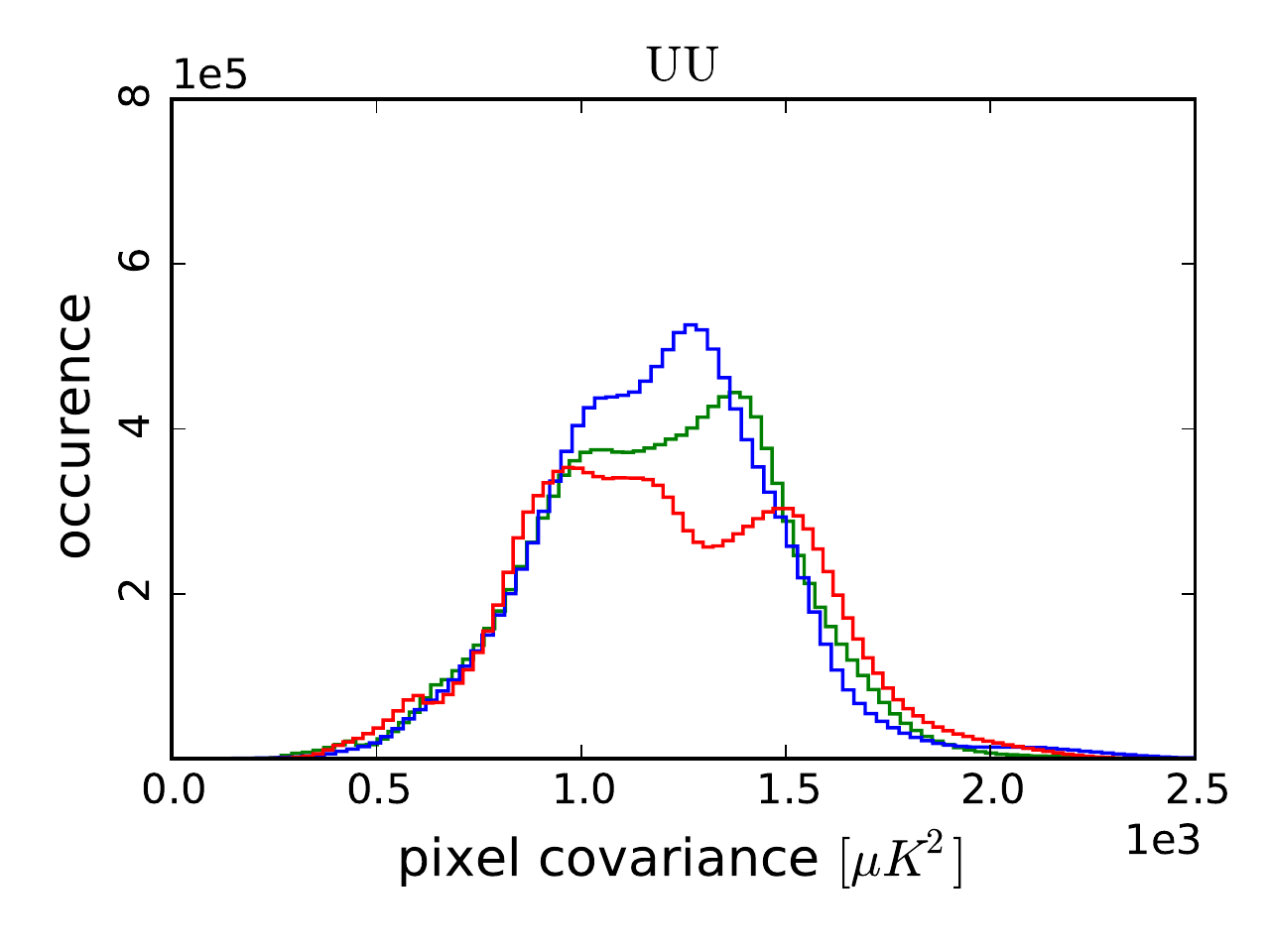}\\
\includegraphics[scale=0.39]{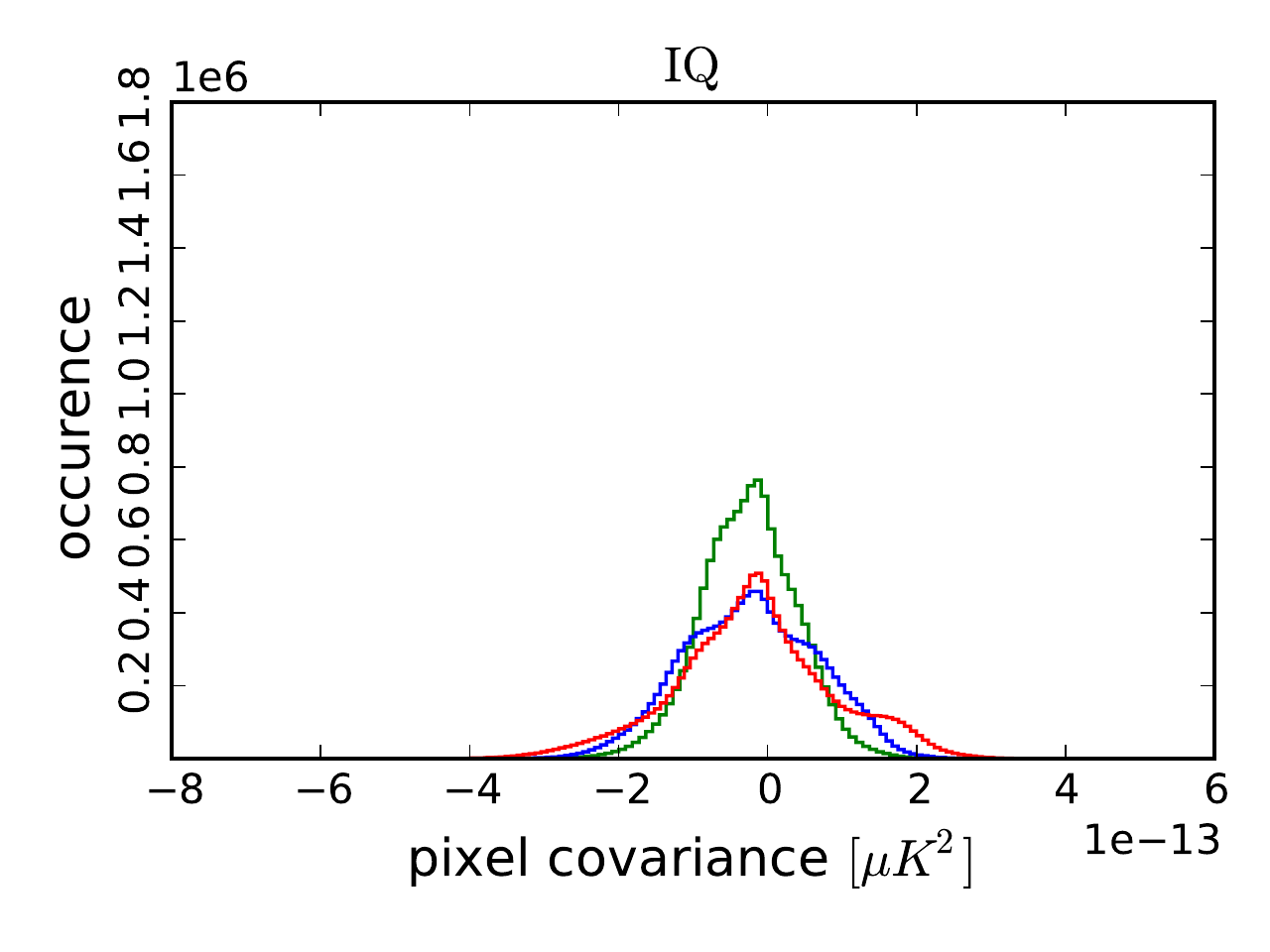}
\includegraphics[scale=0.39]{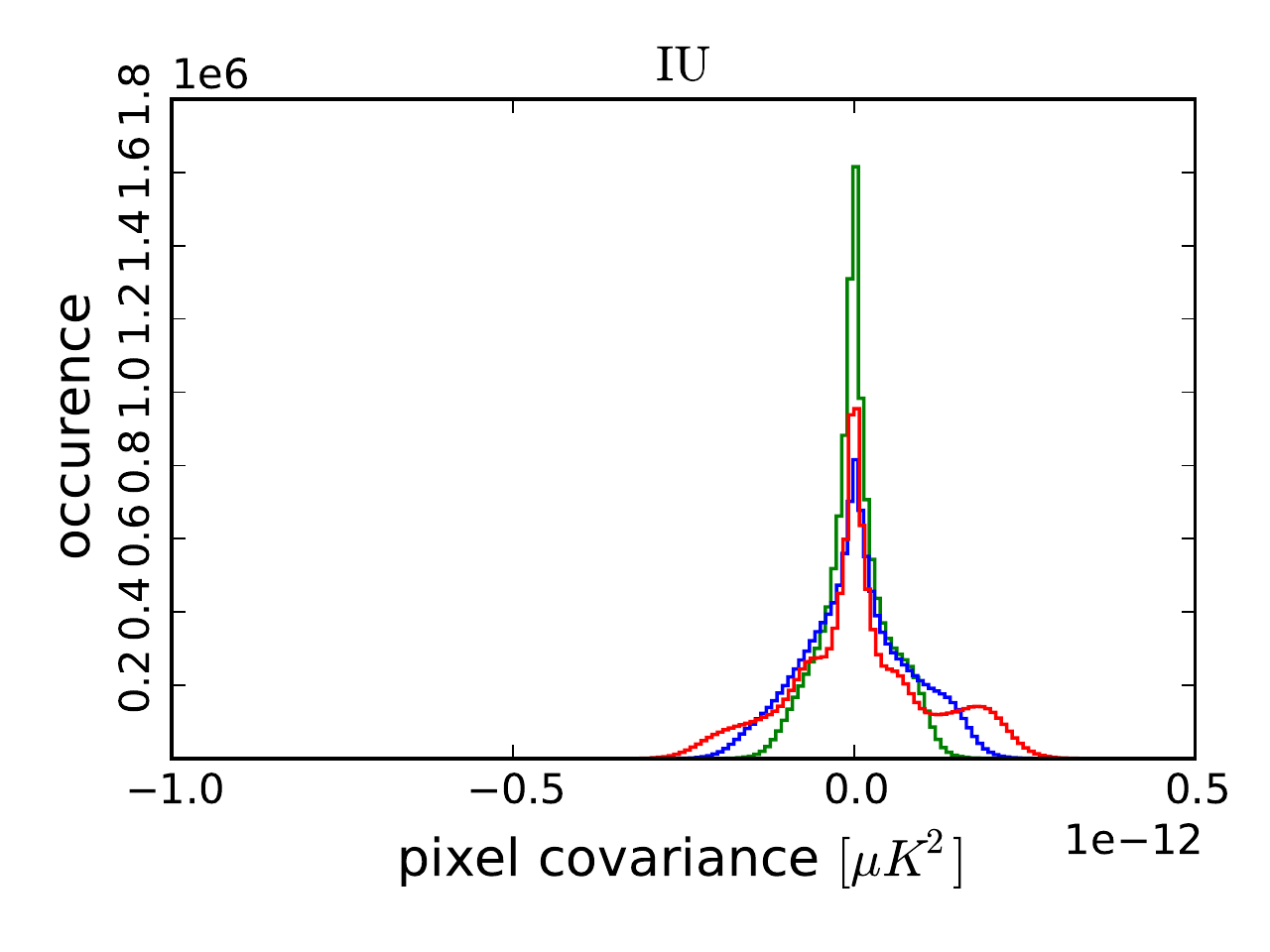}
\includegraphics[scale=0.39]{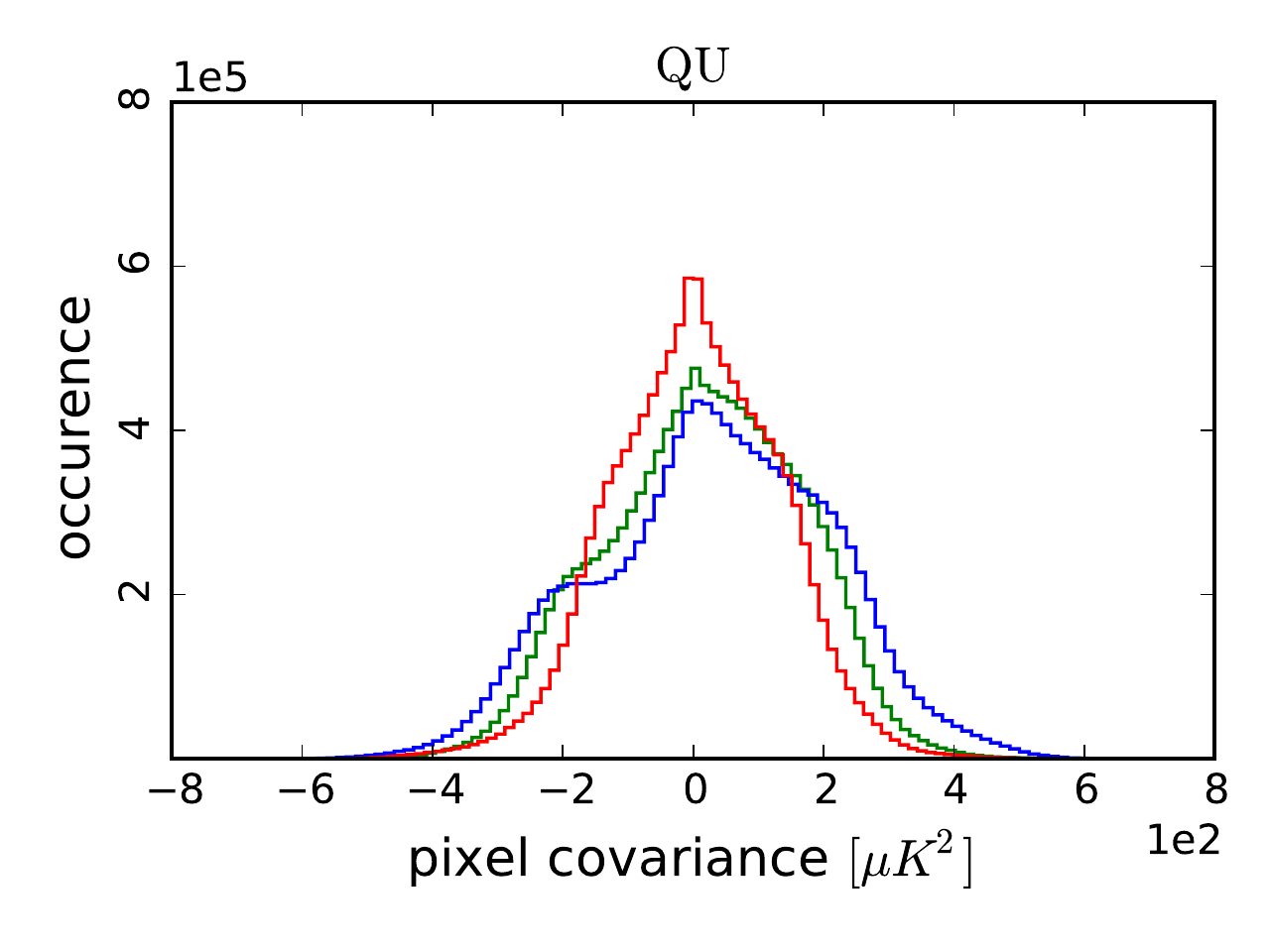}
\caption{\footnotesize{Histograms of the $3\times3$ pixel covariance matrix elements in Ecliptic coordinates (first and second rows) and in Galactic coordinates (third and fourth rows). There are minimal intensity-to-polarization couplings (notice the change of scale) but significant $QU$ residual correlation.}}
\label{fig:histcov}
\end{figure}

In Fig. \ref{fig:noisemaps} we show maps of the elements of the
$3\times3$ white noise covariance matrices produced by MADAM for the
case of boresight detectors for both Ecliptic and Galactic
coordinates, and histograms of these matrix elements are shown in
Fig. \ref{fig:histcov}. Larger values of these histograms reflect
larger pixel variance of the noise maps.  An effective scanning
strategy will achieve compact histograms with low mean values and
minimal tails. These requirements are reasonably satified by \CORE, as
observed in Fig. \ref{fig:histcov}: the histograms do not possess
large tails, total intensity has smaller values with respect to
polarization by a factor of 2, and the $QQ$ and $UU$ histograms are
very similar to one another. This last property is influenced by the
particular choice of the orientations of the detectors as explained
above. In addition, intensity and polarization show almost negligible
correlations, while $QU$ does show significant correlation
features. These however are expected to vanish when a multi-detector
map from an entire frequency channel is produced, in view of the large
number of detectors per frequency expected by \CORE, and the desirable
variation in mutual orientation (in fact, it would suffice to consider
only four detectors with polarization angles at exactly $45^\circ$ to
each other to have this correlation vanish).

\begin{figure}
\centering
\includegraphics[scale=0.5]{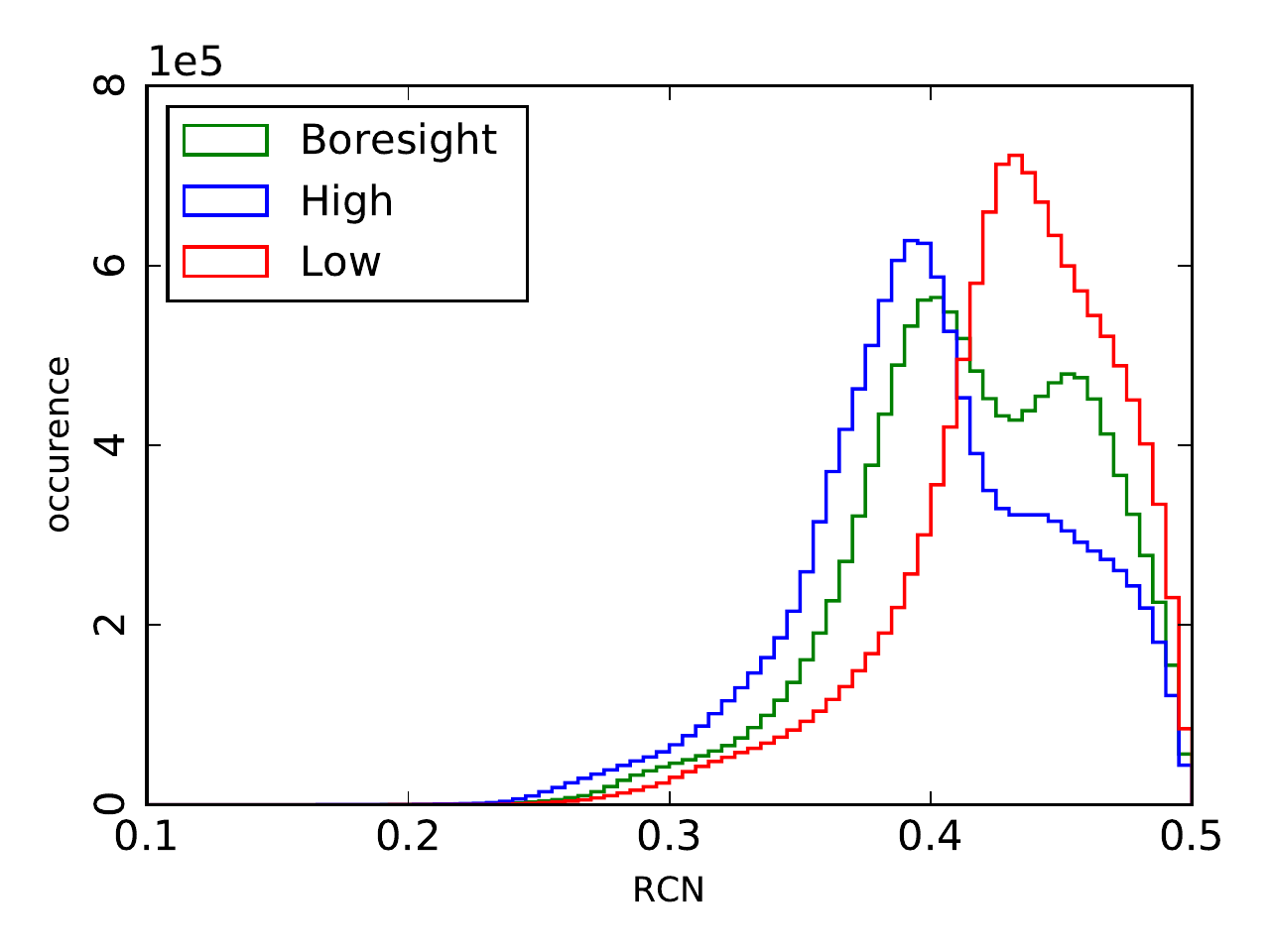}
\caption{\footnotesize{Histograms of the reciprocal condition numbers
    for the boresight, high and low detectors.}}
\label{fig:invcov_baseline}
\end{figure}

As described above, a useful quantity for assessing the map-making
inversion is the RCN. The RCN for the boresight solution is shown in
Fig \ref{fig:invcov_baseline}. Reasonable requirements for optimal
inversion are an average value across the histogram higher than 0.25
and no pixels with values lower than $10^{-2}$. With the
\CORE\ scanning strategy, we obtain an average value of about 0.41 and
no pixels with values lower than 0.2, hence the separation of the
Stokes parameters as allowed by the scanning strategy alone is very
good. This shows that from the point of view of map-making
effectiveness, \CORE\ can efficiently modulate polarization without
resorting to a rotating half-wave plate. It should be mentioned that
we have investigated the ideal performance of the scanning strategy
here, neglecting, for example, cross-polar leakage.

In Fig.~\ref{fig:histcov} we also show the histograms of the noise
covariance matrix for the high (blue) and low (red) detectors. One of
the risks for detectors at the edge of the focal plane is not
achieving complete sky coverage. This is avoided by imposing the
condition that the sum of the spin and precession angles is more than
$90^\circ$ for the entire focal plane. In \CORE, the sum of these
angles for the low detectors is $90.3^\circ$ allowing for the complete
sky coverage across the whole focal plane. We have used these
simulations to verify that this is indeed the case.  The histogram
shapes are similar to the boresight ones, and there are no anomalous
values of the noise covariance matrix elements. In
Fig. \ref{fig:invcov_baseline} we show the RCN for the high and low
detectors, which on average are quite similar to boresight. Low
detectors show slightly higher RCN, high detectors show slightly lower
RCN. This allows us to extend the above conclusions about the clean
separation of the Stokes parameters to the whole focal plane.

\begin{figure}
\centering
\includegraphics[scale=0.39]{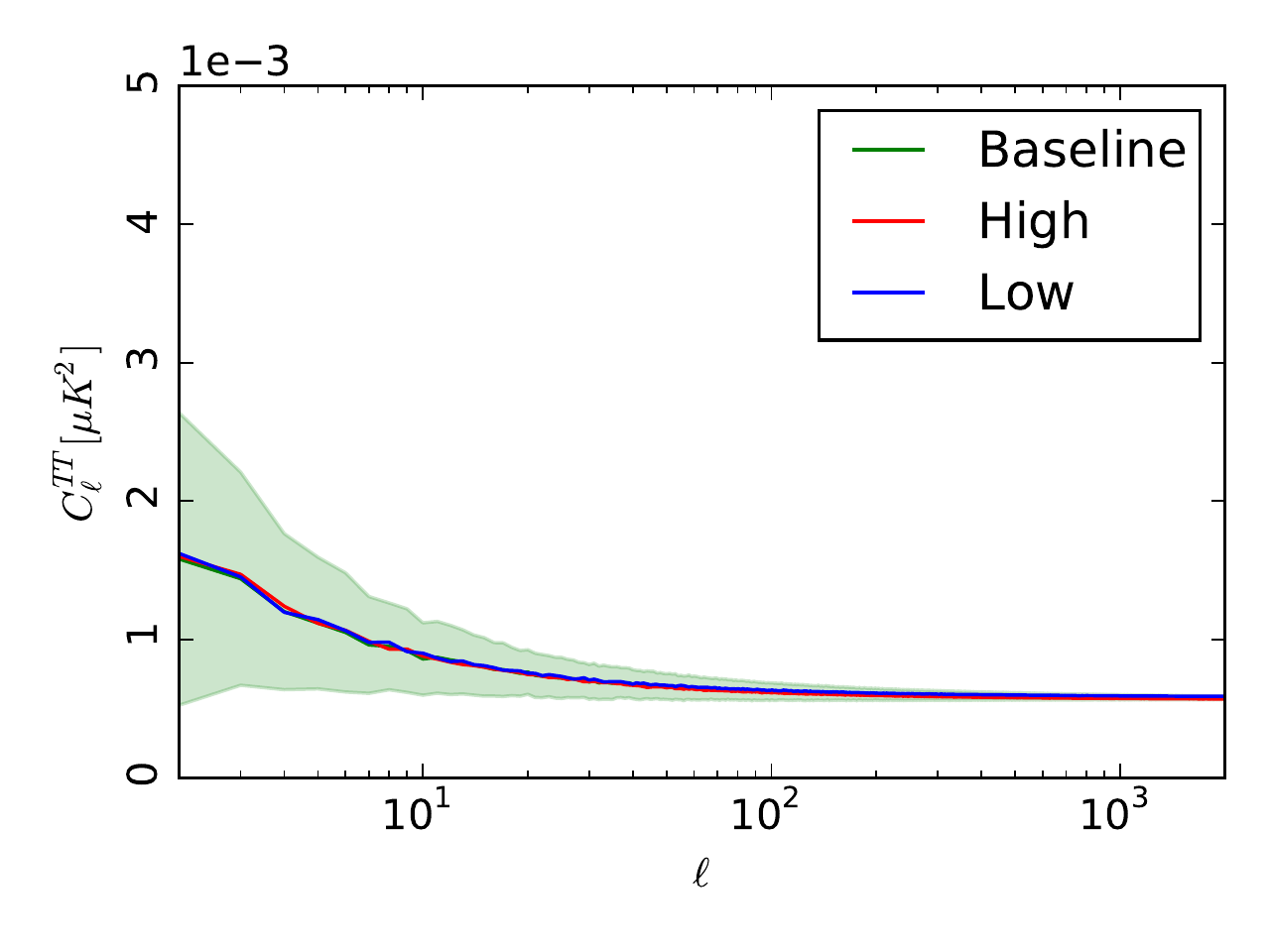}
\includegraphics[scale=0.39]{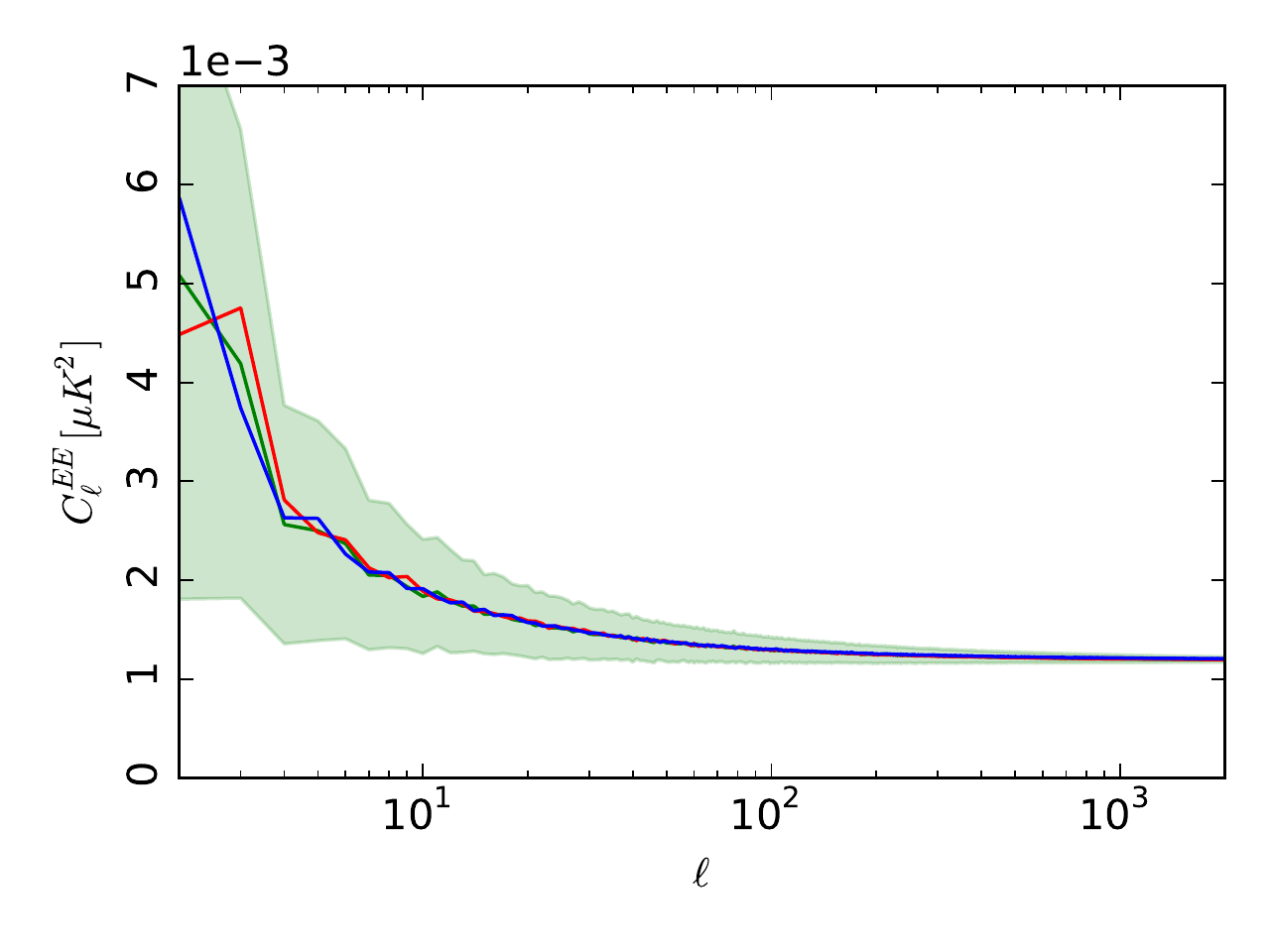}
\includegraphics[scale=0.39]{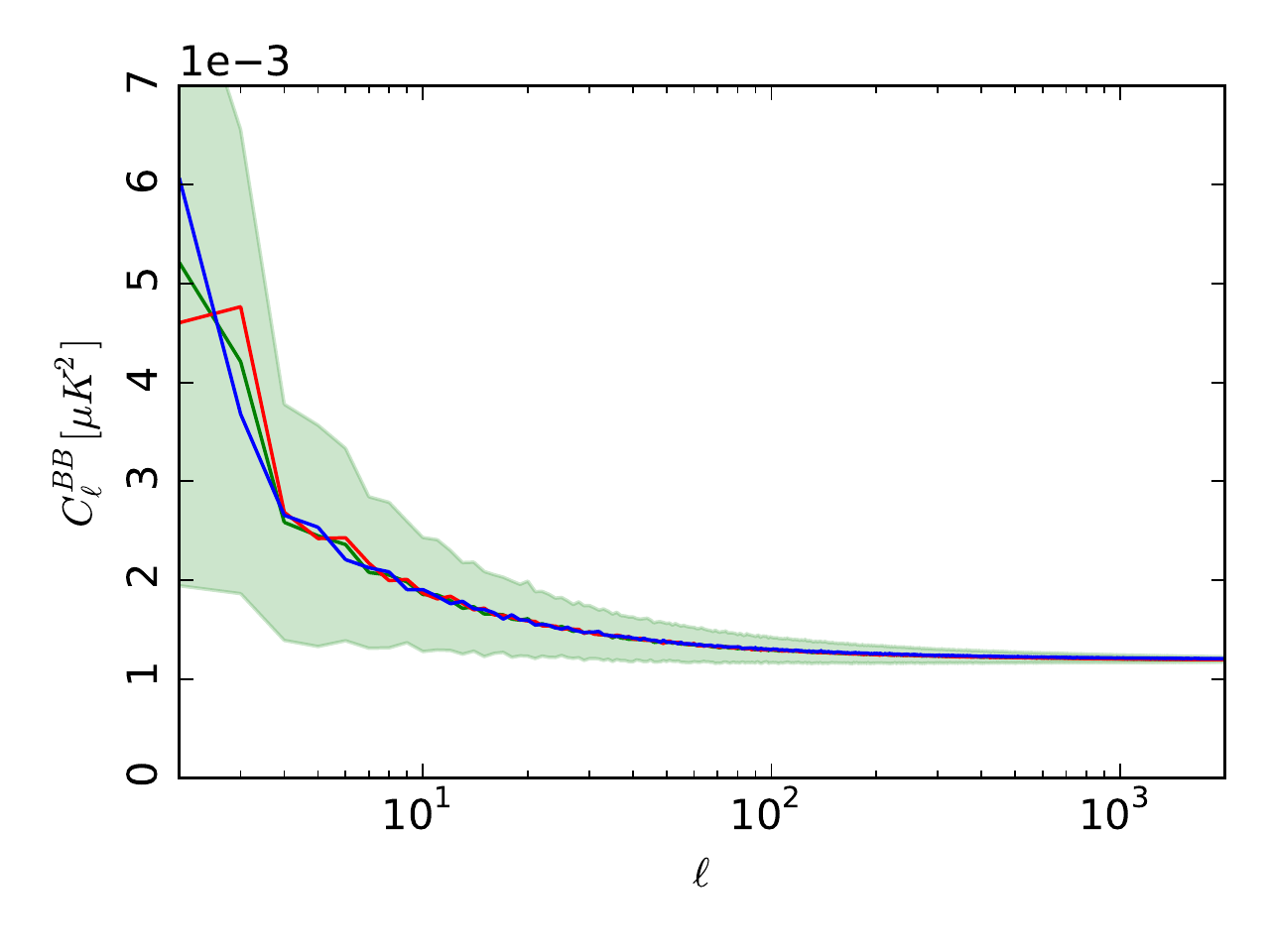}
\caption{\footnotesize{Angular power spectra for $TT$ (left), $EE$
    (centre) and $BB$ (right) of the baseline simulations for the
    boresight, high and low detectors. We show the average of 1000
    simulations and the $1\,\sigma$ dispersion for the boresight case
    (shaded regions).}}
\label{fig:APSbase}
\end{figure}

In Fig.~\ref{fig:APSbase} we show the average $TT$, $EE$ and $BB$ APS
from 1000 noise realizations for the boresight, high and low detectors
(details of our Monte Carlo pipeline are given in Appendix A). We also
show the $1\,\sigma$ dispersion of the boresight case. As already
noted for the RCN, the APS of different detectors are all similar. The
APS of low detectors show slightly lower amplitudes than the other
two. The $EE$ and $BB$ amplitudes are practically the same as a result
of the choice of the polarization orientations. All spectra show a
large scale (low multipole) excess, due to residual $1/f$ contribution
after destriping. The impact of different knee frequencies is
discussed in Section \ref{fknee}.

\subsection{Optimizing the scanning strategy}

We investigate possible optimizations of the \CORE\ scanning strategy
by analysing the effect of varying the spin angle and the precession
angle. We consider seven pairs of values keeping the sum of these
angles equal to $95^\circ$ for the boresight detectors in order to
preserve full sky coverage for the entire focal plane. In this way we
define seven `tweaked' cases to be compared to the baseline
\CORE\ scanning strategy (see Table \ref{tab:tweak_cases} for the
chosen values, all the other parameters are the same as in
Table~\ref{tab:toast_parameters}).

\begin{table}
{\footnotesize
\centering
\begin{tabular}{|l|c|c|c|c|c|c|c|c|}
\hline
Parameter & Baseline & Tweak 1 & Tweak 2 & Tweak 3 & Tweak 4 & Tweak 5 & Tweak 6 & Tweak 7 \\
\hline
\hline
Precession angle [$^\circ$] & 30 & 32 & 34 & 36 & 38 & 40 & 45 & 50 \\
\hline
Spin angle [$^\circ$] & 65 & 63 & 61 & 59 & 57 & 55 & 50 & 45 \\
\hline
\end{tabular}}
\caption{\footnotesize{Parameters modified with respect to
    Table~\ref{tab:toast_parameters} to obtain tweaked cases to
    evaluate a possible optimization of the \CORE\ scanning
    strategy. The first column gives the baseline
    parameters.}}\label{tab:tweak_cases}
\end{table}

\begin{figure}
\centering
\includegraphics[scale=0.39]{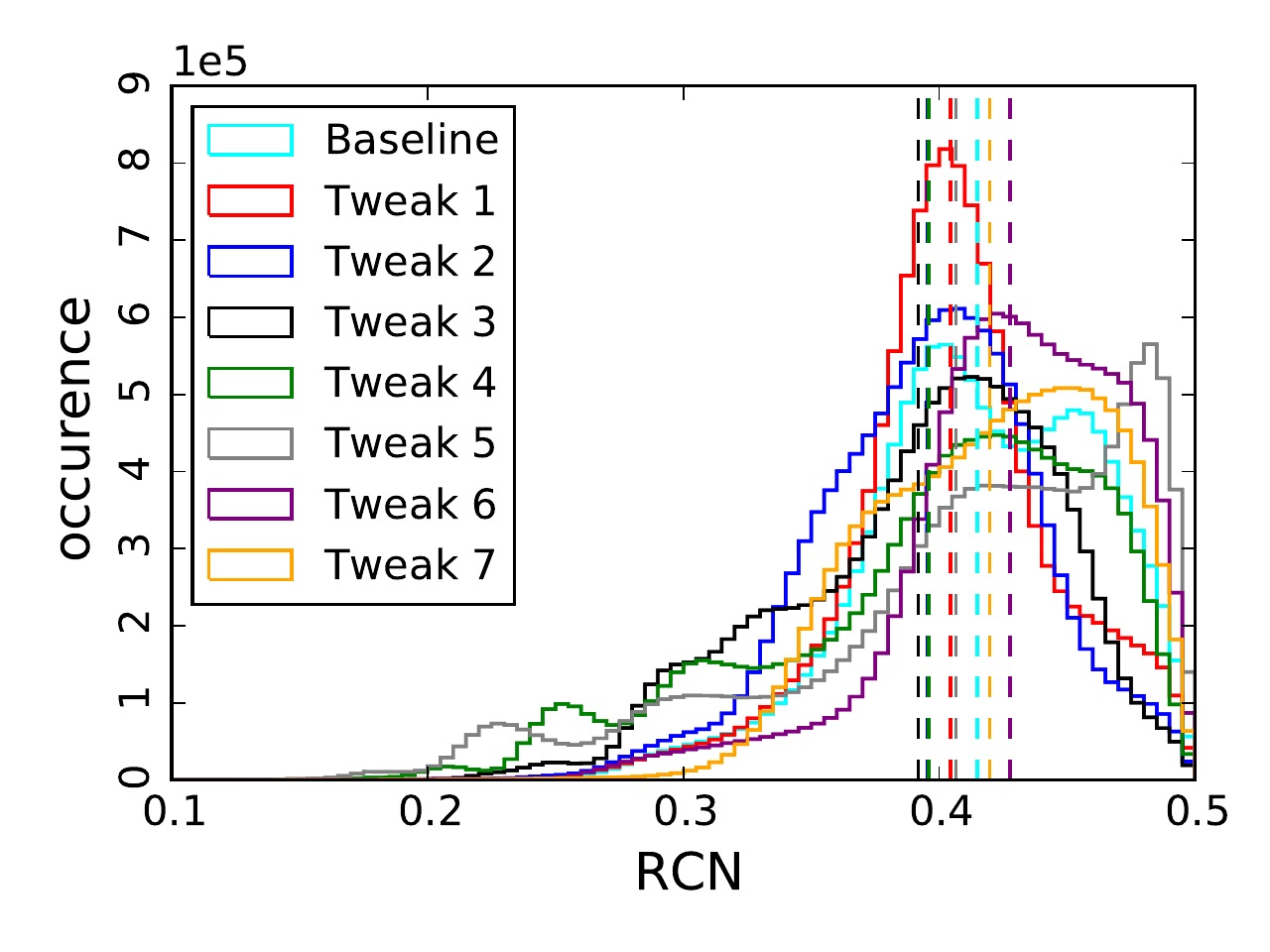}
\includegraphics[scale=0.39]{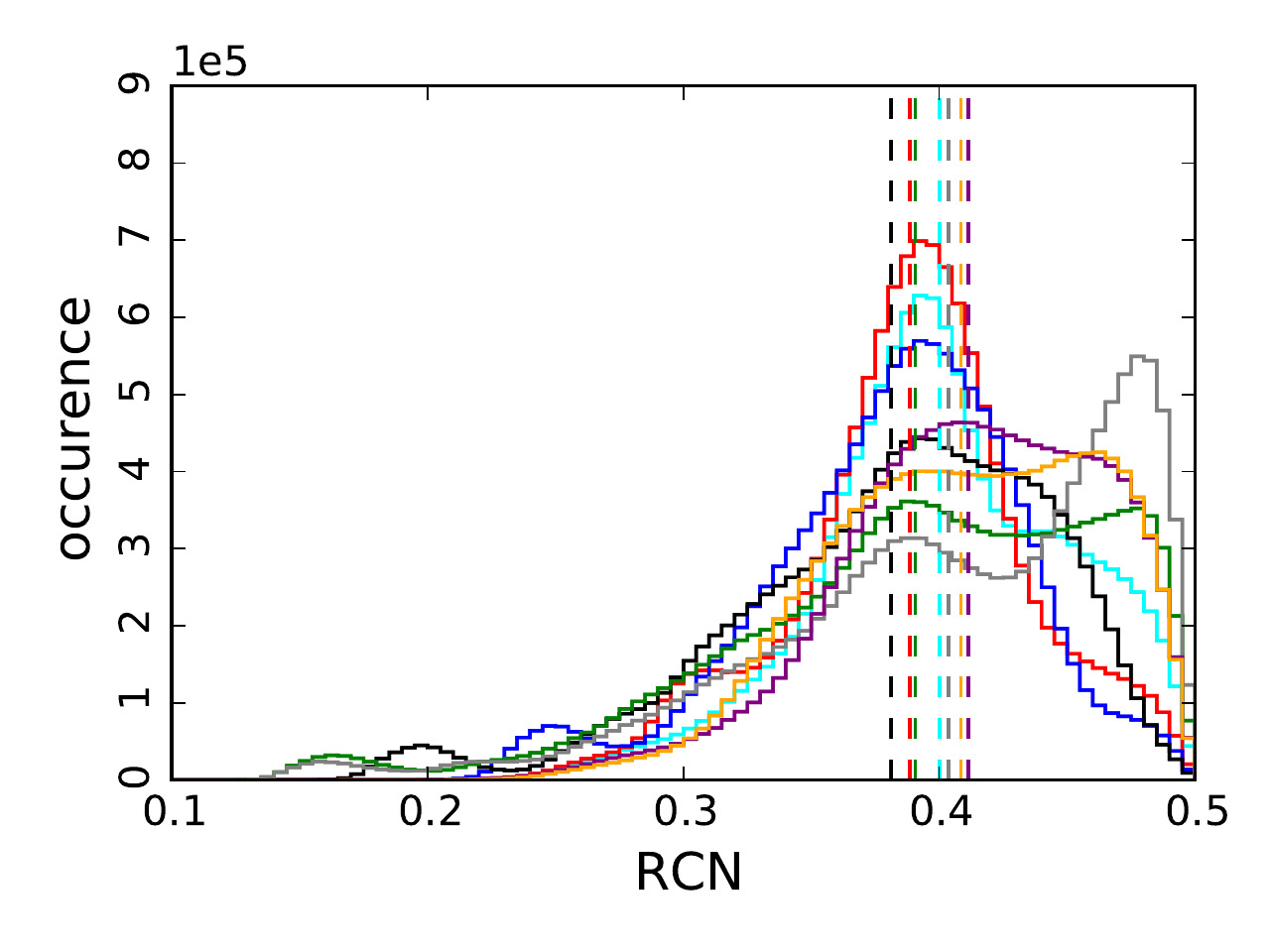}
\includegraphics[scale=0.39]{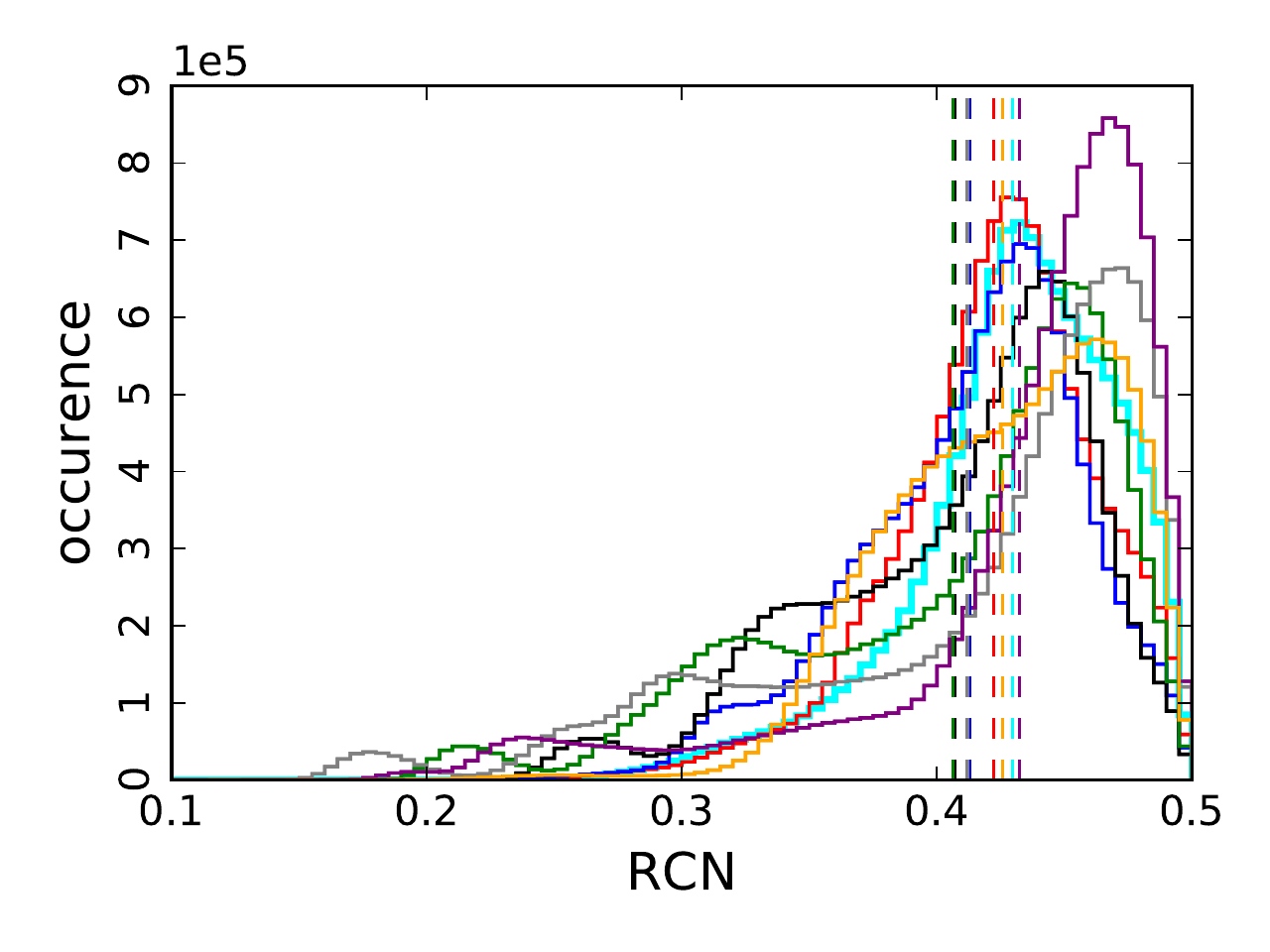}
\caption{\footnotesize{Histograms of the RCN for the boresight (left),
    high (centre) and low (right) detectors in the tweaked cases
    compared to the baseline (cyan). The vertical dotted lines show
    the mean values.}}
\label{fig:invcov_tweak}
\end{figure}

In Fig. \ref{fig:invcov_tweak} we show the RCN of the noise covariance
matrices for the tweaked cases considering the boresight, high and low
detectors. The RCN are all quite similar with average values around
$0.4$ for all cases. Cases from 1 to 5 show larger tails towards lower
RCN values and therefore their average RCN is slightly lower. Cases 6
and 7 show slightly improved RCN with respect to the baseline
especially for the boresight detectors. The improvements are less
evident when the high and low detectors are considered. The highest
mean RCN is achieved by case 6 with a value of about 0.42 for the
boresight detector which, given the dispersion of the RCN values shown
in Fig.~\ref{fig:invcov_tweak}, is not significantly different from
the 0.41 achieved by the baseline, in view of the generous spread of
RCN values. This is a small improvement that would require significant
changes in spin and precession angles, and would have negative impacts
on other subsystems of the spacecraft (for example, a lower power
supply due to the change in solar aspect angle).

\begin{figure}
\centering
\includegraphics[scale=0.39]{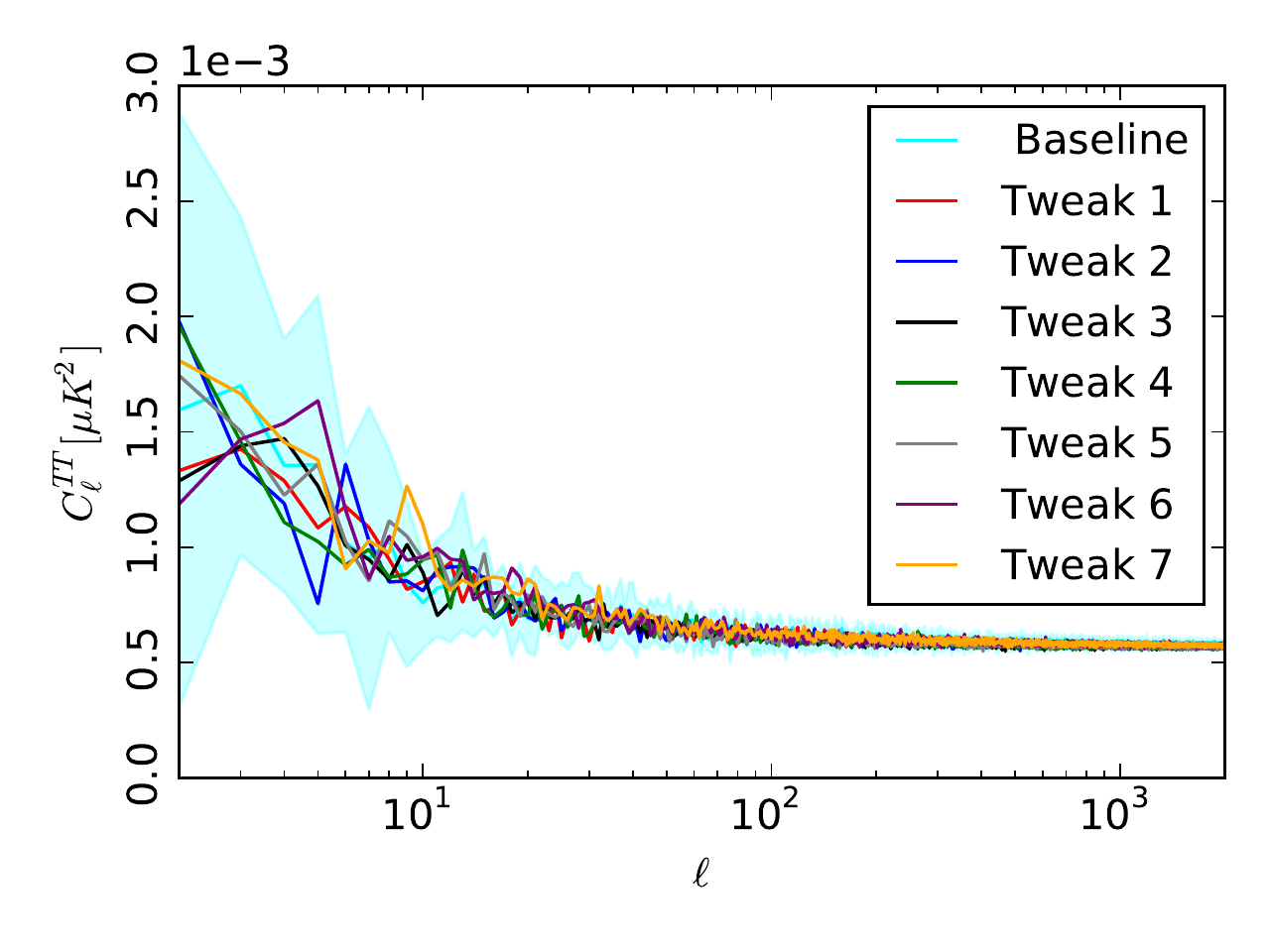}
\includegraphics[scale=0.39]{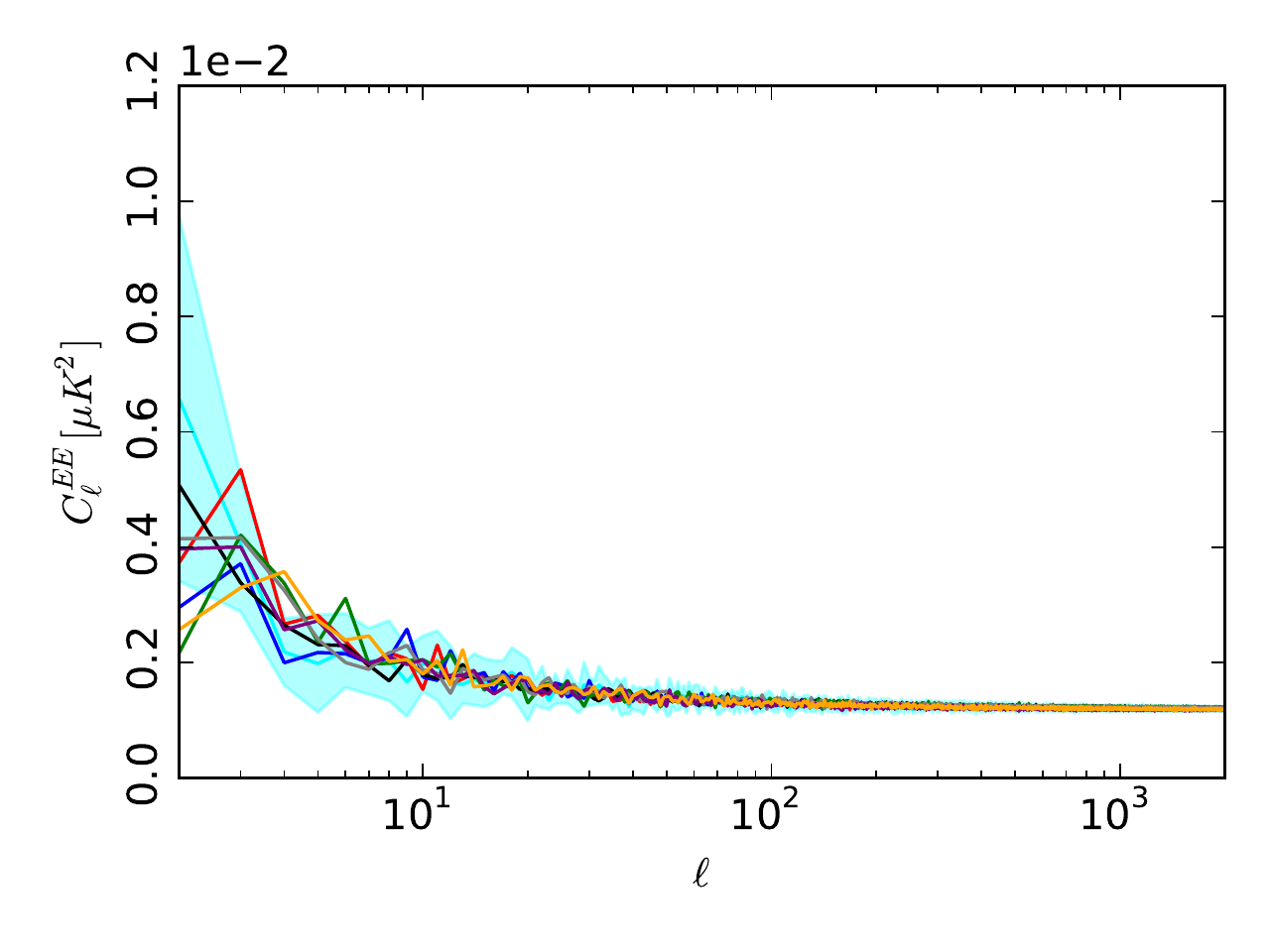}
\includegraphics[scale=0.39]{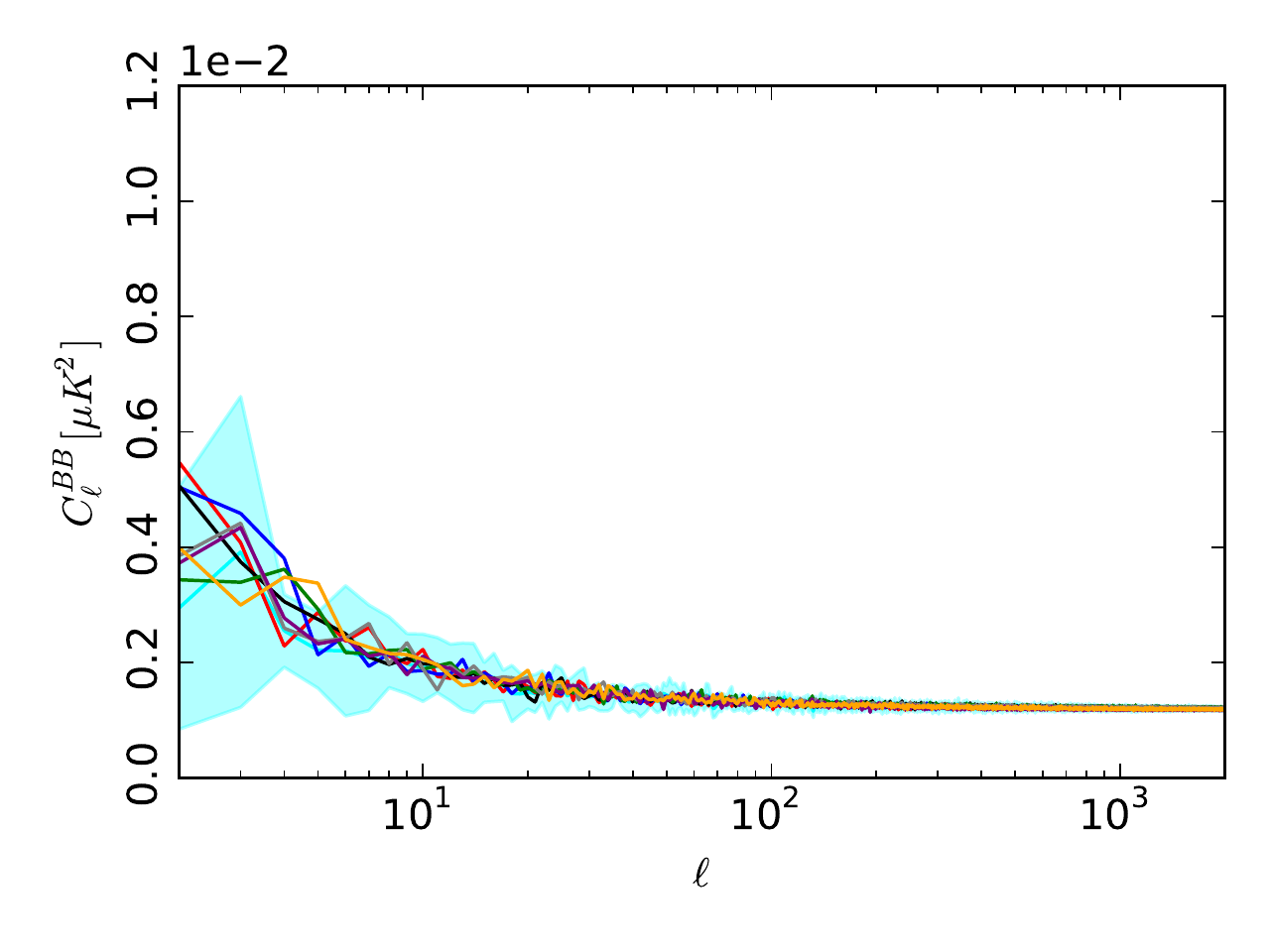}
\caption{\footnotesize{$TT$ (left), $EE$ (centre) and $BB$ (right) APS
    of the baseline simulations for the boresight detectors compared
    to the tweaked cases described in Table
    \ref{tab:tweak_cases}.}}\label{fig:APS_tweak}
\end{figure}

In Fig.~\ref{fig:APS_tweak} we show the APS of the noise maps for the
boresight, high and low detectors. All the APS here are the result of
the average over 10 noise realizations. The APS of the tweaked cases
are compared to the baseline and its $1\,\sigma$ dispersion delimited
by the cyan shaded region. At small scales the APS are all almost
identical. Larger differences are evident at large scales, but they
are well inside the $1\,\sigma$ dispersion.

Our conclusion from this exercise is that any gain in tweaking the
scanning strategy parameters is modest and probably is not worth
attempting, at least for the figures of merit considered above, but
does leave some flexibility to optimize others.

\subsection{$1/f$ noise performance}\label{fknee}

In this Section we investigate the effect of the low frequency noise
properties. We simulate a year of observations for a pair of
boresight, high or low detectors considering several knee frequencies
$f_k$ in the range between $0$-$50$\,mHz, a range that appears
reasonable in view of \CORE's planned detectors.  As mentioned above,
we make use of a noise prior in MADAM, which requires as input an
estimate of the noise power spectral density. We provide here the true
underlying power spectrum of the noise. Even if this may be considered
an optimistic choice, in practice the impact on the results of a
mismatch between the true and estimated noise properties is weak, as
noted in Sect.~\ref{sec:simulations} above. We always use 1\,s as the
MADAM offset length. We generate 1000 Monte Carlo (MC) realizations
(see Appendix A) and apply MADAM to produce noise-only maps. The
amplitude of residuals can be turned in a requirement on the maximum
acceptable knee frequency.

\begin{figure}
\centering
\includegraphics[scale=0.39]{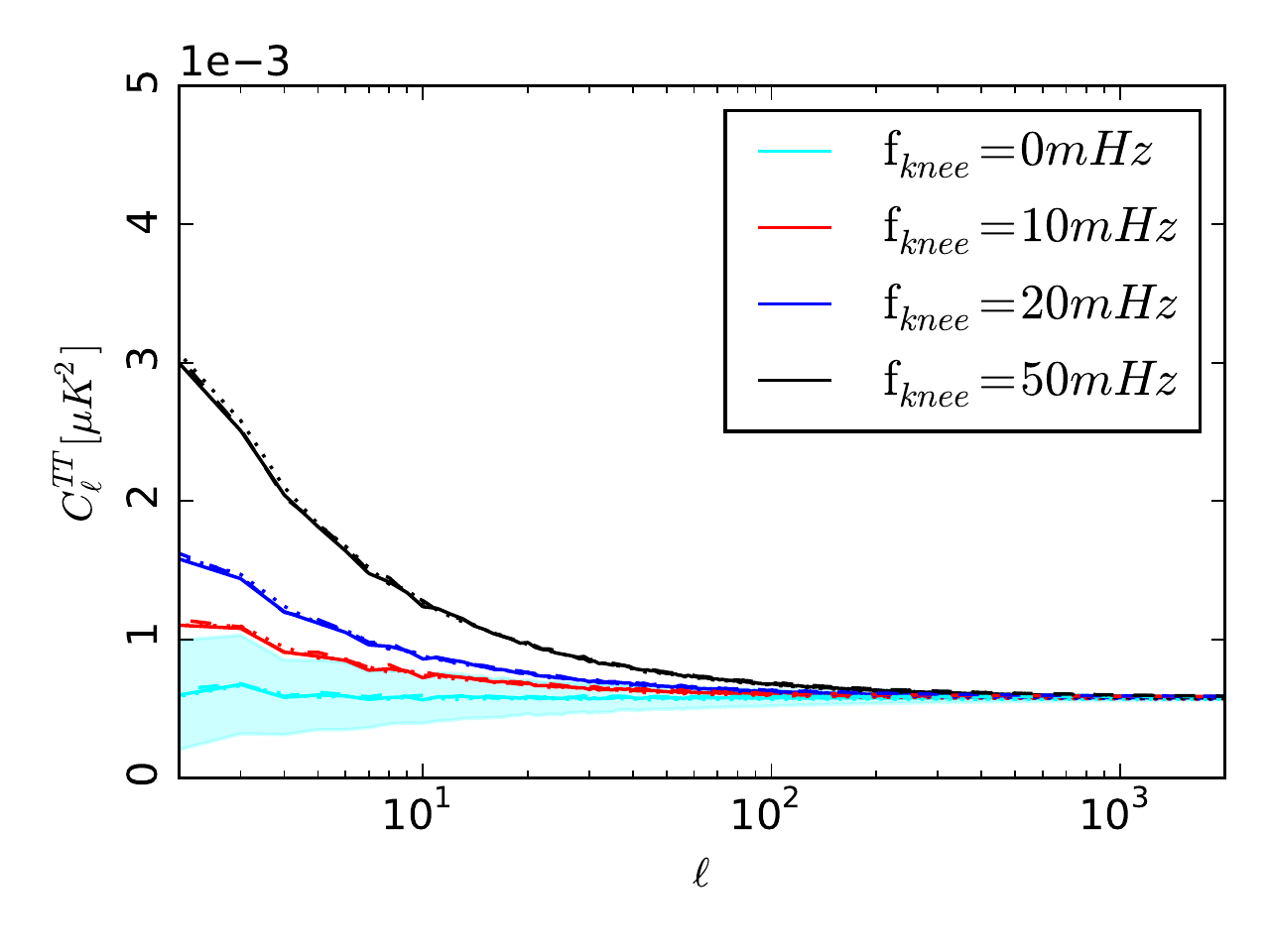}
\includegraphics[scale=0.39]{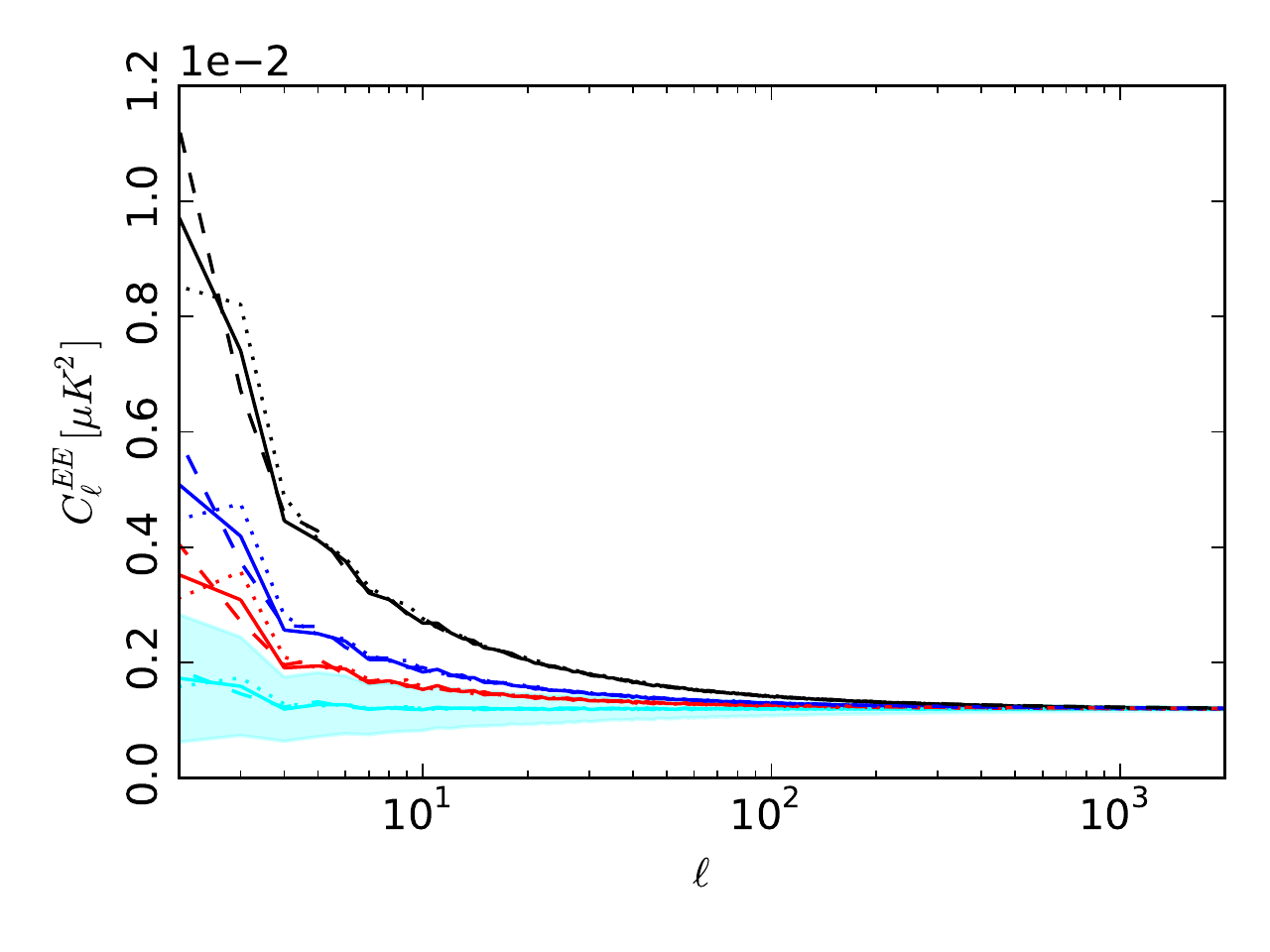}
\includegraphics[scale=0.39]{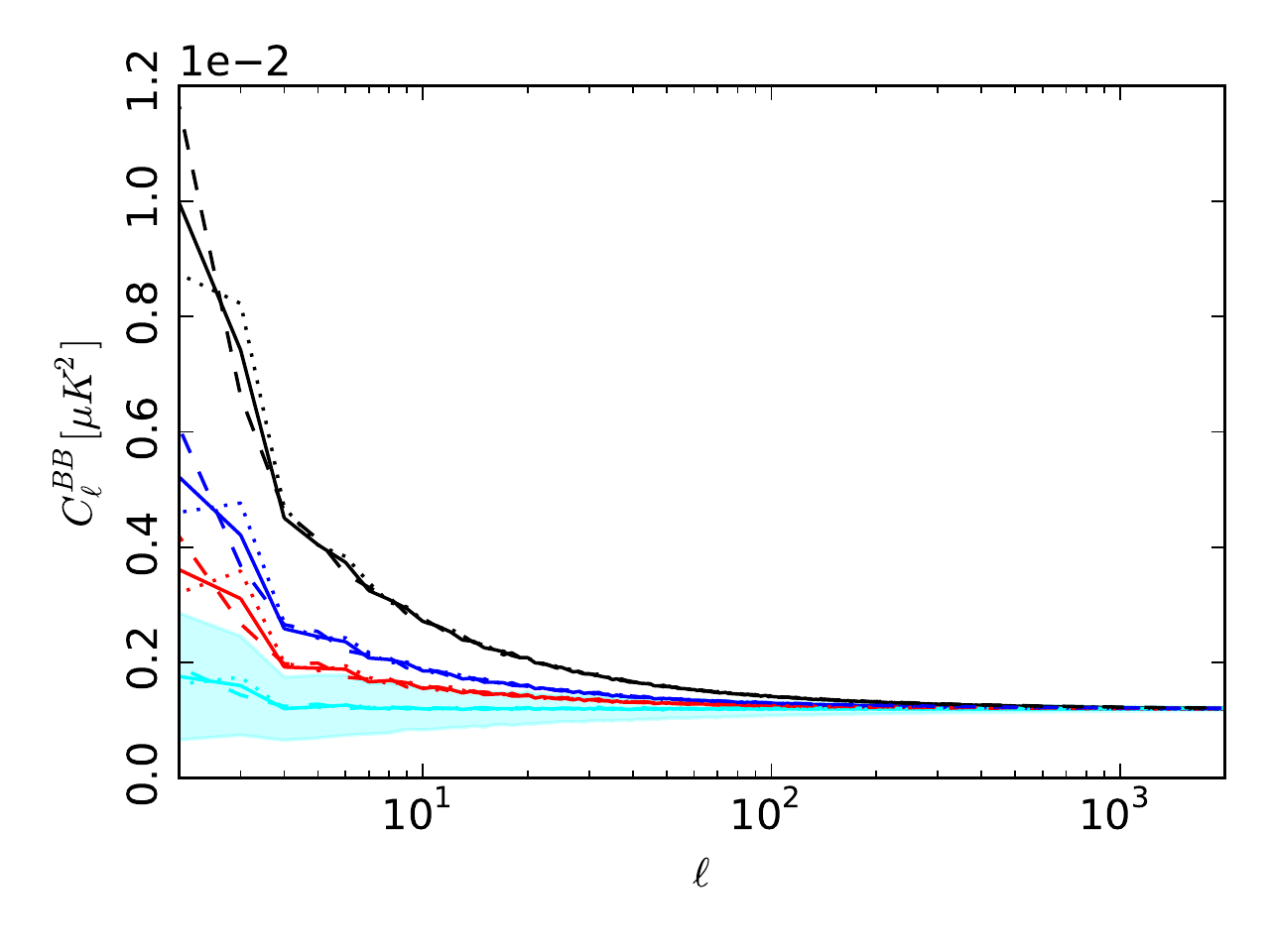}
\caption{\footnotesize{$TT$ (left), $EE$ (centre) and $BB$ (right) APS
    of the baseline simulations for the boresight detectors
    considering several knee frequencies $f_k$. We show the APS from
    the boresight detectors (solid lines), high detectors (dotted
    lines) and low detectors (dashed lines).}}
\label{fig:APS_fknee}
\end{figure}

In Fig. \ref{fig:APS_fknee} we show the average APS from 1000 MC
realizations for the knee frequencies $f_k$ of 10\,mHz (red line),
20\,mHz (blue line), 50\,mHz (black line). They are compared with the
pure white noise case $f_k=0$\,mHz and its $1\,\sigma$ dispersion
(cyan line and shaded region). In the same Figure we show the results
for a pair of low detectors (dashed lines) and high detectors (dotted
lines). As expected, we do not observe any difference between the
position of the detectors in the focal plane.  The effect of the
destriping residuals is a larger amplitude of the noise spectrum at
large scales ($\ell < 100$) and, as expected, the residuals increase
with increasing $f_k$. The 10\,mHz case lies at the edge of the
$1\,\sigma$ dispersion of the white noise MC. Knee frequencies lower
than this value will generate noise maps that cannot be practically
distinguished from pure white noise both in temperature and
polarization. Therefore low frequency noise drifts have negligible
effects if $f_k<10$\,mHz. We notice that a knee frequency of 20\,mHz
is still an acceptable compromise showing an increase in the noise APS
mostly confined to $\ell < 10$.

We now turn to comparing the amplitude of $1/f$ noise residuals with
primordial polarization signals.  In the above analysis we considered
a pair of detectors at 145\,GHz. The proposed configuration of
\CORE\ has 2100 detectors in the frequency range from 60 to 600\,GHz
\citep{ECO.instrument.paper}. We can use the above results to infer
the impact in the APS of a noise map obtained from the entire focal
plane. We consider the six cosmological channels between 130 and
220\,GHz which have the lowest noise, and produce a noise power
spectrum from the combination of these channels by inverse noise
weighting. We then rescale the amplitude of the noise APS derived from
a pair of detectors (shown in Fig. \ref{fig:APS_fknee}) to match this
noise spectrum at $\ell=300$. This approach does not fold in
contributions to the final error budget arising from sources other
than instrumental noise (e.g. foreground separation residuals) and for
this reason we avoid using channels below 130\,GHz and above 220\,GHz,
which still contain useful CMB signal. The same combination of
channels has been considered in the \CORE\ cosmological parameters and
inflationary forecasts discussed in \cite{ECO.Parameters.paper} and
\cite{CORE_inflation}.

\begin{figure}
\centering
\includegraphics[scale=0.5]{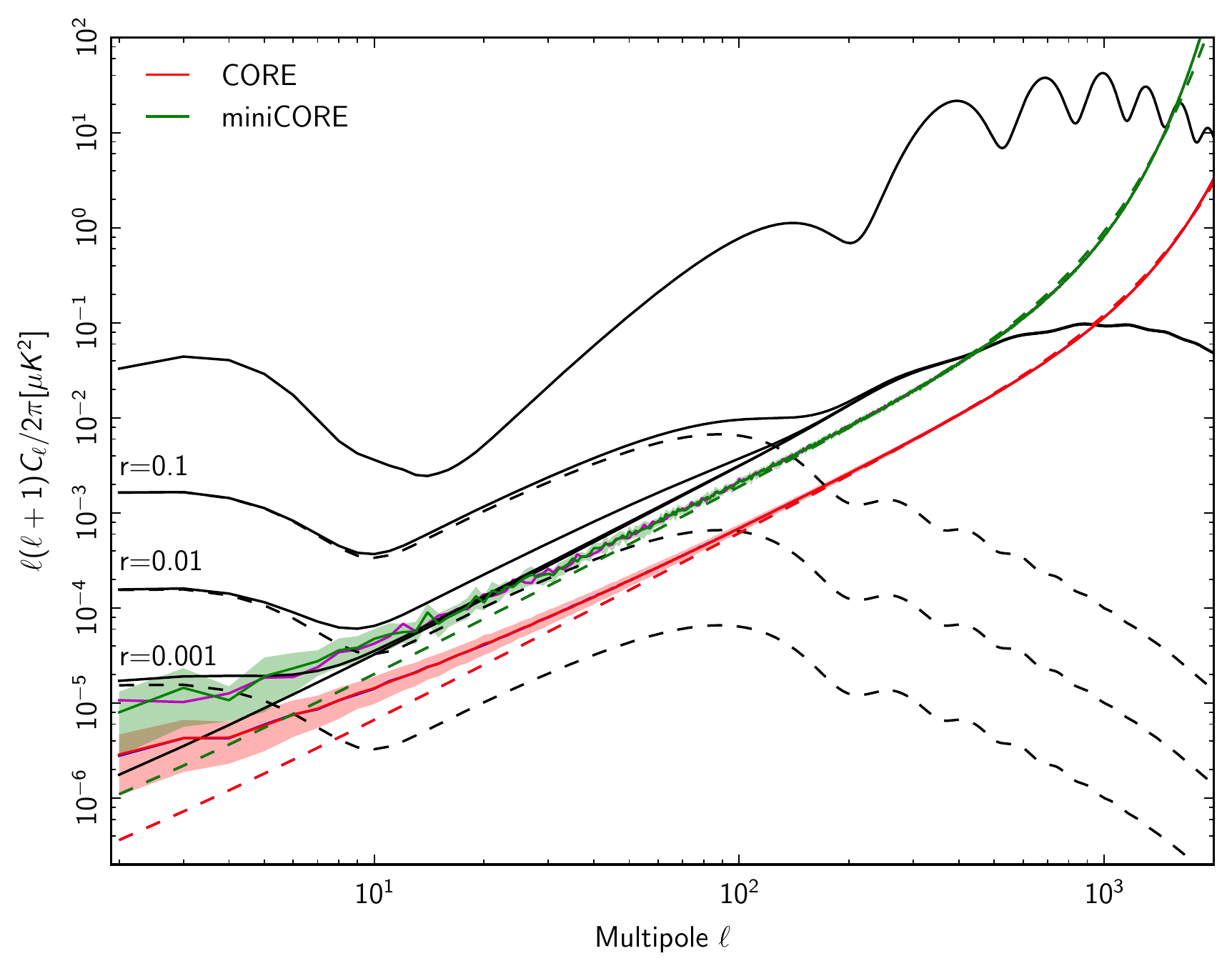}
\caption{\footnotesize{Polarized noise angular power spectra (coloured
    lines) in $EE$ and $BB$ for \CORE\ (red and blue lines
    respectively) and \miniCORE\ (magenta and cyan respectively),
    considering all channels between 130 and 220\,GHz, compared to
    $EE$ and $BB$ CMB theoretical spectra for several values of the
    tensor-to-scalar ratio $r$ (the solid curve includes lensing B
    modes). The shading corresponds to the $1\,\sigma$ uncertainty
    region. See text for details.}}
\label{fig:test_noise}
\end{figure}

Results are shown in Fig.\ \ref{fig:test_noise}, where they are
compared to $EE$ and $BB$ CMB spectra, considering for the latter
several values of the tensor to scalar ratio $r$, while the other
cosmological parameters are kept to the \Planck\ 2015 updated best fit
values \citep{2016A&A...596A.107P}. The red dashed line shows the
noise spectrum from the inverse noise weighting of the six CMB
channels as described above. The solid line shows noise APS of a pair
of detectors as result of the average of 1000 noise realizations
(assuming $f_k=50$\,mHz and a beam of 5 arcmin), rescaled to match the
dashed line at $\ell=300$. We show both $EE$ (blue line) and $BB$ (red
line) spectra although they are almost indistinguishable. We also show
the $1\,\sigma$ dispersion of the realizations as the shaded
region. The effect of $1/f$ noise is noticeable at $\ell \lesssim
100$. However, even in this pessimistic assumption of $f_k=50$\,mHz,
the noise spectrum is well below the $BB$ spectrum for $r=10^{-3}$ for
$\ell \lesssim 10$. We also show in Fig.~\ref{fig:test_noise} the
forecasted noise spectra (magenta for $EE$, cyan for $BB$) for the
so-called \miniCORE\ design, a down-scoped configuration of
\CORE\ (see \cite{ECO.instrument.paper} for more detail). We follow
the same approach described above, but consider the
\miniCORE\ parameters for the beam FHWM (11.9 arcmin at 145\,GHz,
corresponding to a sampling rate of 54.8 Hz) and averaging over 10
noise realizations instead of 1000.  All the other parameters, except
of course the number of detectors per channel, are identical to
\CORE\ (see table \ref{tab:toast_parameters}). Despite the noise level
for \miniCORE\ is now significantly closer to the $BB$ spectrum for
$r=10^{-3}$, this design still allows plenty of margin for an accurate
measurement of tensor modes in view of $1/f$ residual contamination,
especially considering that $f_k=50$\,mHz is taken here as a worst
case scenario.

\section{Cross-correlated noise}
\label{sec:CorrelatedNoise}

Cross-correlated noise contributions between different detectors have
been reported by several CMB experiments \citep[see
  e.g.][]{2006A&A...458..687M,
  2008ApJ...681..708P,2016A&A...596A.107P}. They are a source of
potential concern for the densely packed focal planes of the current
and forthcoming generation of CMB experiments. The presence of these
common modes is usually neglected during analysis, though optimal
treatment can easily be included in the GLS framework discussed in
Sect.~\ref{sec:map-making}, at the cost of increasing the
computational burden \citep{2008ApJ...681..708P,2016A&A...593A..15D},
as detailed in Appendix~B. In short, a solution formally identical to
Eq.~\ref{eq:gls_mm} can be obtained for the the estimated map
$\widetilde{\mathbf{m}}$:
\begin{equation}
 \mathbf{\widetilde m} = \left( \mathbf{A}^T 
\mathbf{N}^{-1} \mathbf{A}\right)^{-1}   
\mathbf{A}^T \mathbf{N}^{-1} \mathbf{d}, 
\end{equation} 
where $\mathbf{d}$ and $\mathbf{A}$ are respectively a multi-detector
timeline and pointing matrix and $\mathbf{N}$ a generalized noise
matrix:
\begin{equation}   
\mathbf{N} \equiv \left\langle \mathbf{n}_t \mathbf{n}_{t^\prime}\right\rangle=   
\left(     
\begin{array}{ccc}       \left\langle   n_t^{(1)}   n_{t^\prime}^{(1)} \right\rangle &        
\cdots &       
\left\langle   n_t^{(1)}    n_{t^\prime}^{(k)} \right\rangle \\       
\vdots & \ddots & \vdots  \\        
\left\langle   n_t^{(k)}    n_{t^\prime}^{(1)} \right\rangle &       
\cdots &       
\left\langle    n_t^{(k)}    n_{t^\prime}^{(k)} \right\rangle      
\end{array}   
\right), 
\end{equation} 
where $\left\langle \cdot \right\rangle$ denotes the expectation value
and $( \cdot )$ labels different channels.  As is customary, we assume
stationary noise, implying that the statistical properties of the
noise do not change over the mission lifetime. To be specific, when
considering the $j$-th and $\ell$-th detectors and the noise elements
$n_t^{(j)}$ and $n_{t^\prime}^{(\ell)}$ at time $t$ and $t^\prime$,
respectively, we assume that $\left\langle n_t^{(j)}
n_{t^\prime}^{(\ell)}\right\rangle $ only depends on
$|t-t^{\prime}|$. Moreover, we consider that the noise statistical
properties are known and completely described by the noise covariance
matrix. We assume that the noise auto- and cross-spectra can be
estimated either by on-ground instrument calibration or directly from
observational data by dark-sky measurements or iterative procedures
\citep[see e.g.][]{2001misk.conf..421P}.

To find $\mathbf{\widetilde m}$, the optimal GLS formula can be solved
iteratively using a Fourier based, preconditioned conjugate gradient
method as outlined in Sect.~\ref{sec:map-making}. Note however that,
given the full noise covariance matrix (whose effective size is
proportional to $N+N_c\times(N_c-1)/2$, where $N$ is the total number
of detectors and $N_c$ is the number of with cross-correlated noise),
a single iteration of the preconditioned conjugate gradient solver
scales linearly with the total number of samples, but quadratically
with $N_c$.

\begin{figure}
\centering
\includegraphics[width=0.49\textwidth]{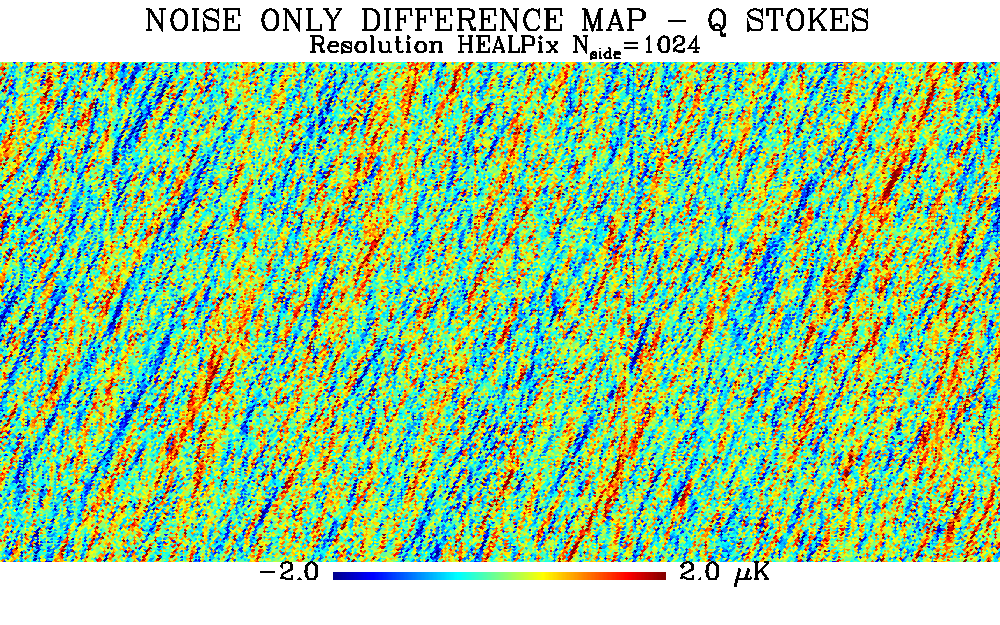}
\includegraphics[width=0.49\textwidth]{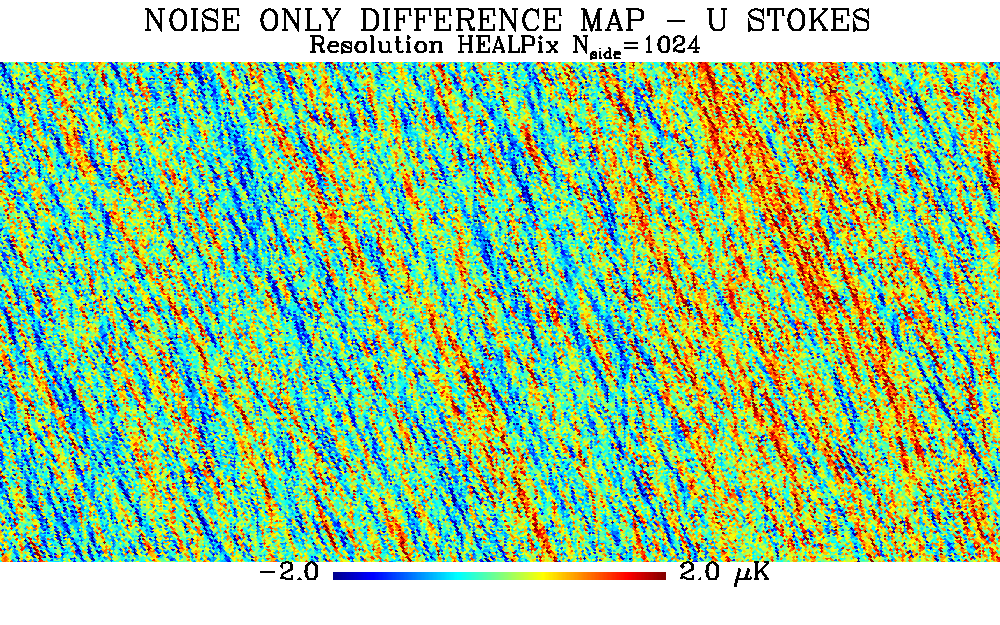}
\caption{\footnotesize{Difference of the noise-only maps of the Stokes
    parameters $Q$ (left) and $U$ (right) between the cases accounting
    for and ignoring cross-correlated noise in the map-making
    (Cartesian projection). Maps have been estimated assuming 2
    detectors and one year of operation. Maps are in Ecliptic
    coordinates, in units of $\mu K$ and at resolution HEALPix
    $N_{side}$=1024 (3.435 arcmin/pixel). The stripes are due to the
    cross-correlated noise, which is mitigated when it is taken into
    account in the map-making algorithm.}}
\label{fig:diff_map}
\end{figure}

\begin{figure}
\centering
\includegraphics[angle=90,width=0.49\textwidth]{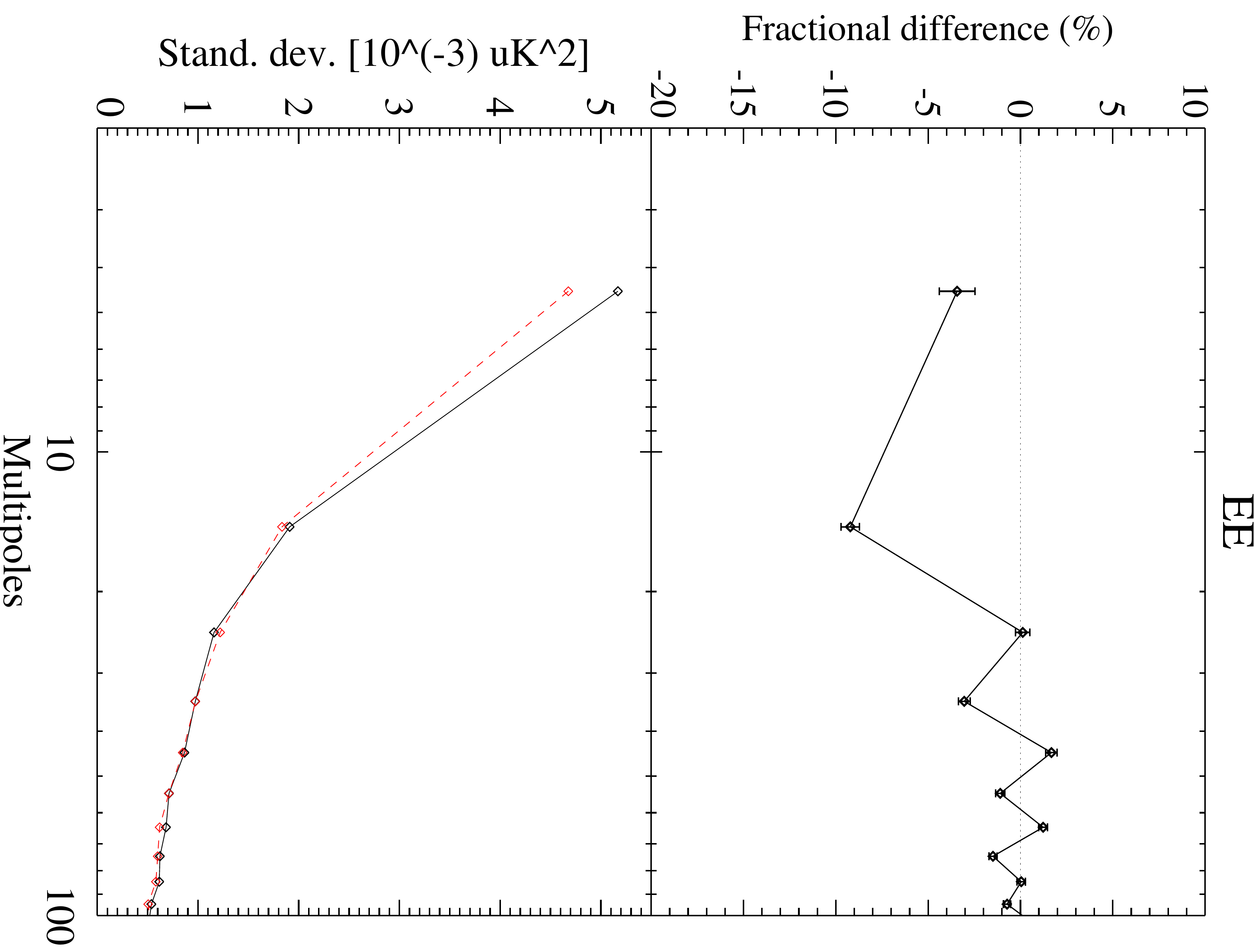}
\includegraphics[angle=90,width=0.49\textwidth]{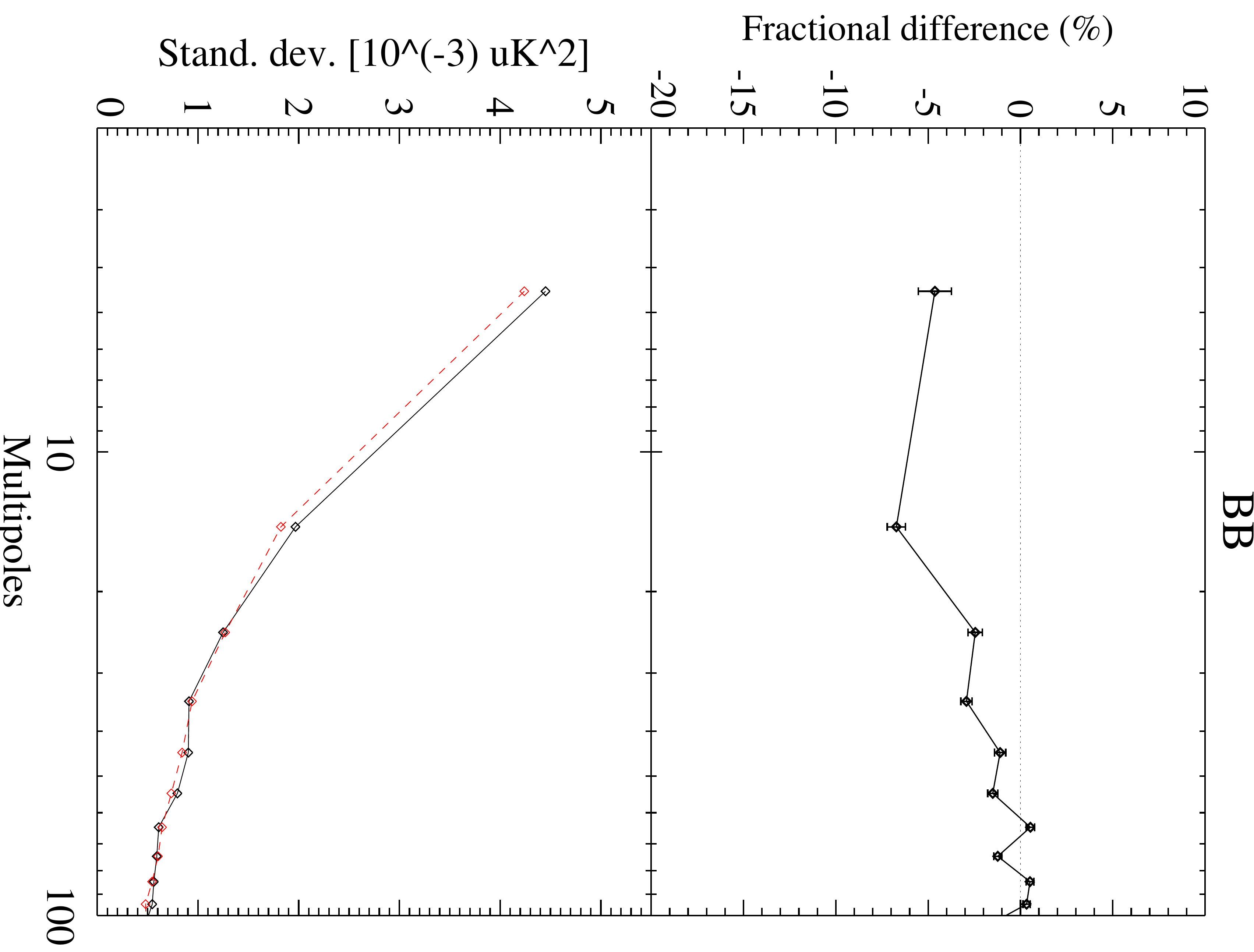}
\caption{\footnotesize{Comparison of $EE$ and $BB$ angular power
    spectra between the cases accounting for and ignoring the
    cross-correlated noise in the map-making algorithm. Spectra have
    been estimated from 20 noise-only Monte Carlo realizations,
    assuming 2 detectors and one year of operation. On the top:
    fractional difference of $EE$ (left) and $BB$ (right) average
    spectra between the two cases. On the bottom: $EE$ (left) and $BB$
    (right) spectrum standard deviations of the cases accounting for
    (dashed red line) and ignoring (solid black line)
    cross-correlation. Standard deviations correspond to the
    dispersion of the simulations.}}
\label{fig:comp_spec}
\end{figure}

We set up a \CORE\ test case by generating a noise timeline with the
TOAST software assuming the \CORE\ scanning strategy and detector
parameter baseline (see Tab.~\ref{tab:toast_parameters}), considering
the cases of 2 and 4 detectors for 1 year of operations. Following the
behaviour of \Planck\ HFI \citep{2016A&A...596A.107P}, we assume that
the noise of the $i$-th detector has the following properties:
\begin{eqnarray}
n_i &=& \tilde{n}_i + n_c\\
P_i(f) \equiv  \left<\tilde{n}_i \tilde{n}_i\right> &=& A \left [\left ( \frac{f_k}{f} \right) + 1 \right] \\
P_{ij}(f) \equiv  \left<n_i n_j \right>  = \left<n_c n_c\right> &=& A \left [\left ( \frac{f_1}{f} \right)^2 + C \right] \mathrm{for\ } i \ne j 
\end{eqnarray}
where $\tilde{n}_i$ and $n_c$ refer to the self- and cross-correlated
noise component, with knee frequencies $f_k = f_1 = 20$\,mHz. We
assume $C=0$ and use the NET from Table~\ref{tab:toast_parameters} for
the amplitude $A$. We then generate noise maps with:
\begin{enumerate}
\item a GLS map-making algorithm that considers only the diagonal
  terms of the noise covariance matrix that correspond to the
  auto-correlated $1/f$ noise (the ROMA MPI-parallel code, see
  \citealt{2001A&A...372..346N} and \citealt{2005A&A...436.1159D});
\item an optimal GLS map-making algorithm extended to account for all
  the off-diagonal terms of the noise covariance matrix that
  correspond to the correlations between detectors (the extended ROMA
  MPI-parallel code, see \citealt{2016A&A...593A..15D}).
\end{enumerate}
For each case above, we generate the noise-only maps of the Stokes
parameters $Q$ and $U$.  In Fig.\ref{fig:diff_map} we show the
difference of the noise-only $Q$ and $U$ maps between the cases with
and without cross-correlated noise in the map-making code (for the 2
detector case). These difference maps show stripy structures,
suggesting that the inclusion of the cross-correlation properties in
the noise model mitigates the residual correlation along the scan.  We
then compute the r.m.s.\ values of the maps, noting very slight
differences between cases 1 and 2 above (of few parts out of
$10^4$). However, it should be emphasized that the map r.m.s.\ is not
the most appropriate figure of merit here, since the effect of
accounting for the noise cross-correlation is limited only to the very
large angular scales. An APS analysis is better suited inasmuch as it
separates the contribution of different angular scales. We therefore
produce 20 Monte Carlo noise-only maps and generate the corresponding
APS.  We then compare the average spectra for cases 1 and 2 above by
calculating their ratio (see Fig.\ref{fig:comp_spec}). We notice that
neglecting the noise cross-correlation results in a larger noise
amplitude at very large angular scales. This excess is suppressed by
accounting for the detector cross-correlation in the map-making: in
fact, we have a decrease of the noise spectrum amplitude up to 5-10\%
at very low multipoles.  Finally, we compare the APS standard errors
as computed from the the dispersion of the simulated Monte Carlo maps
(see Fig.\ref{fig:comp_spec}). We find that the inclusion of the
cross-correlated noise provides smaller error-bars at the largest
angular scales (again, up to 5-10\%), while, as expected, we do not
find any relevant difference at smaller scales.

Results shown in Fig.\ref{fig:comp_spec} correspond to the 2 detector
case. Tests on the 4 detector case have shown the same qualitative
behaviour: we still find a decrease in both noise spectrum amplitude
and standard deviations of 5-10\% after including the noise
cross-correlation information in the map-making process.  We emphasize
that the common-mode noise contribution does not easily integrate down
with increasing the number of detectors, as opposite to the auto noise
component. For this reason, the noise cross-correlation should be
carefully accounted for in any polarization experiment with a large
number of detectors, and the map-making process is the natural place
to deal with it.

We emphasize that the amplitude of this effect is not straightforward
to forecast accurately (as it will crucially depend of the
instrumental set-up), but we highlight the importance of having a
pipeline able to handle it. Indeed, our analysis shows that the effect
of the noise cross-correlation is not negligible. Due to the faintness
of the B-mode polarization signal, the improvement provided by the
extended map-making algorithm may prove crucial for accurate
characterization of such contribution at the largest angular scales:
accurate modelling of noise at low resolution is an important task to
reliably measure the B mode reionization bump, since for some highly
efficient estimators devoted to this task noise misestimation may
induce bias \citep{2015MNRAS.453.3174M}, contrary to what happens for
GLS map-making.

\section{Bandpass mismatch}
\label{sec:BandpassMismatch}

Difference in detector bandpasses induces leakage from intensity to
polarization for any astrophysical component with a non-CMB spectrum
after calibrating the data on the CMB. This arises after differencing
measurement from detectors, with slightly different sensitivity to
component signals, to estimate the polarization signal. This effect is
studied in~\cite{BPMMpaper1}, and it has been shown that the amplitude
of the leakage is strongly coupled to the single detector
cross-linking parameters measuring the uniformity of angle coverage in
each pixel.  This systematic effect was observed in
\Planck\ data~\citep{2016A&A...596A.107P} at the percent level and
required correction to avoid biasing the estimation of CMB
polarization~\citep{planck.2015.08.HFI.data.processing.calibration.and.maps}.
Given the sensitivity of the \CORE\ mission, this effect has to be
studied carefully and correction methods must be designed. In this
section, we study this effect in the specific case of \CORE\ baseline
scanning strategy. We first describe our model of the effect and the
simulations performed. We then introduce a correction approach and
show to which accuracy the effect could be reduced.

\subsection{Model of the bandpass mismatch effect}

As discussed in Sect.~\ref{sec:map-making} (see in particular
Eq.~\ref{eq:signal_model_mm}), a detector observing the sky at the
frequency $\nu_0$ with a polarizer oriented at an angle $\psi_t$ at
time $t$ measures the quantity:
\begin{equation}
  d_t(\nu_0) = I_{p(t)}(\nu_0) + Q_{p(t)}(\nu_0) \cos\left(2\psi_t\right) + U_{p(t)}(\nu_0) \sin\left(2\psi_t\right) + n_t,
\label{pol_obv_model}
\end{equation}
where $I_{p(t)}(\nu_0)$, $Q_{p(t)}(\nu_0)$ and $U_{p(t)}(\nu_0)$ are
the Stokes parameters at the position $p(t)$ on the sky measured in
the local reference frame and $n_t$ is the random instrumental and
photon noise.

For a given sky pointing, each of the Stokes parameters $(I, Q, U)$
receives contributions due to the emission of different components
$c$. For simplicity, let us model the intensity in terms of these
components. After integrating over the detector bandpass, the total
flux par unit steradian received by a detector from the sky can be
written as (see~\cite{BPMMpaper1} for details on this model):
\begin{equation}
{dF\over d\Omega} = \int \sum_{c} g(\nu) f_c(\nu,p) d\nu,
\end{equation}
where $g(\nu)$ is the detector bandpass, $f_c(\nu,p)$ the emission
spectrum of the $c$ component, which can depend on the pixel.  After
calibrating on the CMB, the intensity reads:
\begin{equation}
I(\nu_0) = I_{\rm CMB}(\nu_0) + \sum_{c \, \neq \, {\rm CMB}} \gamma_c(p) I_c(\nu_0),
\end{equation}
where $I_c(\nu_0)$ is the mean intensity of component $c$ at a
reference frequency $\nu_0$ and $\gamma_c(p)$ the relative amplitude
of component $c$ in CMB temperature units, which is slightly pixel
dependent if the component spectra depend on the sky region
considered. Since the signal has been calibrated on the CMB, the
factor $\gamma_{\rm CMB}$ is normalized to unity. The component
amplitude coefficient, defined for the reference frequency $\nu_0$,
can be related to the transmission of the band and the spectrum of the
component by:
\begin{equation}
 \gamma_c = {\int g(\nu) f_c(\nu) f_c(\nu_0)^{-1} d\nu \over \int g(\nu)\left.{dB(\nu,T_0)\over d\nu}\right |_{\nu_0} d\nu}\left.{dB(\nu,T_0)\over d\nu}\right|_{\nu_0},
\label{IntgT}
\end{equation}
where $T_0$ is the mean temperature of the CMB and $B(\nu,T_0)$ is the
blackbody spectrum of the CMB. The quantity $\gamma_c$ is close to 1
for a chosen $\nu_0$ near the center of the band. A similar
relationship applies for the $Q$ and $U$ Stokes parameters.

The expression above describes an ideal situation and does not include
real-world complications such as beam asymmetries and miscalibration.
We follow this approach in order to isolate the bandpass mismatch
effect. Potential couplings with other systematic effects are ignored
at this stage, in the spirit of the discussion in
Sect.~\ref{sec:intro}.

The problem of bandpass mismatch can be understood by calculating the
data model for the sky signal for a set of detectors
$\lbrace(i)\rbrace$. Each detector $(i)$ in the set will have its own
$\gamma_c^{(i)}$ which can be written as
\begin{equation}
\gamma_c^{(i)} = \overline{\gamma}_c + \delta\gamma_c^{(i)}
\end{equation}
where $\overline{\gamma}_c$ is the mean of scaling parameter $\gamma$ for the set $\lbrace(i)\rbrace$ and the component $c$ and $\delta\gamma_c^{(i)}$ is its deviation from this mean. The data model for the sky signal can now be written, using a vector notation in boldface, as
\begin{equation} \label{eq:bandpass-leakage-model}
\vt{d}^{(i)} \, = \,   \sum_c \overline{\gamma}_c \left [ \vt{I}_c + \vt{Q}_c \cos\left(2\psi\right) + \vt{U}_c \sin\left(2\psi\right) \right ]
				 +  \sum_{c \, \neq \, {\rm CMB}} \delta\gamma^{(i)}_c \left [ \vt{I}_c + \vt{Q}_c \cos\left(2\psi\right) + \vt{U}_c \sin\left(2\psi\right) \right ] + \vt{n}.
\end{equation}
The first term on the right hand side is the `ideal' sky signal with
all components including CMB, while the second term is the leakage
term of non-CMB components due to their different bandpasses. Since
the signal for each detector has been calibrated on the CMB, we expect
$\delta \gamma_{CMB}$ to be zero, so it is absent from the second
term.

\subsection{Simulations of the bandpass mismatch effect}

We evaluate the impact of the mismatch on a set of simulations of the
data using a simplified version of the \Planck\ sky model (PSM, see
\citet{2013A&A...553A..96D}), which comprises only two foreground
components (thermal dust and synchrotron) and CMB at $145$\,GHz, at
HEALPix resolution $N_{\mathrm{side}}=1024$, and symmetric Gaussian
beams of FWHM $7.6^\prime$. We use four detectors, with polarization
angles evenly spaced at intervals of $45^\circ$, at the boresight, and
with varying square bandpasses with a 1\% random error on the edge of
the bands. This gives similar variations of the $\gamma$ parameters
than for the most pessimistic cases of \Planck\ filters. Nevertheless,
we expect less variations between future mission detector filters.
The sky component is integrated over the corresponding spectrum
following Eq.~\ref{IntgT}, using a top hat instrumental bandpass. In
simulations, we have included the complexity of component spectra as
modelled in the PSM, and consequently the resulting $\gamma$
parameters are pixel dependent. We have run simulations of pure signal
and of signal plus white noise separately, using the nominal noise
level expected for \CORE, observing for one year using the baseline
\CORE\ scanning strategy as discussed in Sect.~\ref{Analysis_maps}
above.

As described in Sect.~\ref{sec:map-making}, we produce intensity and
polarization maps using the maximum likelihood estimator for the
Stokes parameters, Eq.~\eqref{eq:gls_mm}. In the first step we ignore
the correction for bandpass leakage and use the same pixelization as
the input maps to avoid introducing additional pixelization
effects. Maps of the timeline noise are made separately and subtracted
from total maps. This is justified since the map-making method which
has been used is linear, and so instrumental noise will be purely
additive.

In the absence of bandpass leakage, that is, if all $\gamma_c^{(i)}$,
for each detector $(i)$ are identical, and there is sufficient
modulation of the polarization angle $\psi$, we expect $\widehat{Q_p}
= \overline{Q}_p$ and $\widehat{U_p} = \overline{U}_p$, the actual
signal on the sky with no additional bias, for all pixels given our
noiseless simulations.

By comparing the resulting polarization maps with the input, we
estimate the impact of bandpass leakage on the polarization
measurement for the chosen detector set. We compute the $\widehat{Q}$
and $\widehat{U}$ maps and compute the $EE$ and $BB$ power spectra of
the difference with the input CMB $Q$ and $U$ maps after masking 25\%
of the sky where the Galactic dust emission is the brightest. The
resulting power spectrum of the residual signal (see
Fig.~\ref{bandpass_leakage_correction}) is above the primordial
$B$-mode signal even for $r=0.1$, and is thus completely unacceptable
for measuring $r=0.001$. It is also above the lensing signal for $\ell
< 100$ (see, again, Fig~\ref{bandpass_leakage_correction}). The
prediction for any number of detectors $N_{\rm det}$ can be obtained
by rescaling the power spectrum of a factor $4 / N_{\rm det}$ (since
the result described here is for four detectors), assuming independent
filter variations between detectors. The scaling of the spectrum with
the inverse of the number of detectors is demonstrated
in~\cite{BPMMpaper1}.

\subsection{Correction algorithm}

We now describe an approach to correct the data for bandpass mismatch
that consists in estimating jointly the map $Q$ and $U$ Stokes
parameters as well as the leakage component using as input data
timelines built from the individual detectors at $145$\,GHz bands and
the templates built from $350$\,GHz and $90$\,GHz recovered intensity
maps for thermal dust and synchrotron respectively. Correction methods
specific to the \Planck\ mission were implemented at the map as well
as timeline level and are detailed
in~\cite{planck.2015.08.HFI.data.processing.calibration.and.maps}
and~\cite{2016A&A...596A.107P}. We now develop a more generalized
correction algorithm based on the model of the data introduced in the
last section and describe the simulation employed to validate it.

The baseline focal plane design for the \CORE\ mission uses detectors
with dual polarization sensitivity in one single focal plane pixel, as
well as pairs of single polarization detectors with orthogonal
polarization sensitivity scanning the sky along the same
trajectory. The former can directly be differenced to cancel intensity
and get a polarization signal, while the latter can also be
differenced but after correcting for the appropriate
time-shift. Differencing in this way the timelines of two orthogonal
detectors, we obtain
\begin{align}
\vt{d} = & \, \frac{1}{2} \left( \vt{d}^{(a)} - \vt{d}^{(b)} \right) \nonumber \\
	= &\,  \sum_c \overline{\gamma}_c \left [ \vt{Q}_c \cos\left(2\psi\right) + \vt{U}_c \sin\left(2\psi\right) \right ] 
	+ \frac{1}{2} \left[ \sum_{c \, \neq \, {\rm CMB}} \left( \delta\gamma^{(a)}_c - \delta\gamma^{(b)}_c \right) \vt{I}_c \right]
	+  \left( \vt{n}^{(a)} - \vt{n}^{(b)} \right) \nonumber \\
	= & \, \vt{Q} \cos\left(2\psi\right) + \vt{U} \sin\left(2\psi\right)
	+  \sum_{c \, \neq \, {\rm CMB}} y_c  \, \vt{I}_c +   \vt{n},
	\label{eq:leakage-pair-model}
\end{align}
and we are left with a reduced set of Stokes parameters $[\vt{Q},
  \vt{U}]$ from the CMB and Galactic sources, $\vt{I}_c$ are the
timelines from reference foreground intensity maps with $y_c$ their
corresponding amplitude, each of them given by $\frac{1}{2} \left(
\delta\gamma^{(a)}_c - \delta\gamma^{(b)}_c \right)$, and $\vt{n}$ is
the noise term.

In equation \ref{eq:leakage-pair-model}, the first term on the right
hand side is the term of interest (a linear combination of the
polarization Stokes parameters), while the other two are nuisance
terms, a bandpass leakage term proportional to a sum of foreground
components, and a noise term $\vt{n}$.

By recasting our data set in the form of equation
\ref{eq:leakage-pair-model}, we have isolated a leakage term which is
a sum of bandpass mismatch coefficients times foreground intensity
templates. If the bandpass mismatch coefficients $y_c$ are perfectly
known by calibration, our measurements can be considered as linear
combinations of $\sum \vt{Q}_c$ (sky Stokes $Q$ in channel $c$), $\sum
\vt{U}_c$ (sky Stokes $U$ in channel $c$), and additional foreground
intensity maps $\vt{I}_c$. The system can be inverted in the usual
way. If however the calibration of the bandpasses is not perfect, we
want to solve also for $y_c$.

We propose to correct for bandpass mismatch terms with the following
approximation. Assume that we have at hand measured templates for the
foregrounds. Such measurements can be obtained directly for
\CORE\ intensity data, either at other frequency (i.e. higher
frequency for a dust template, or lower for a synchrotron template),
or at the reference frequency of the channel of interest after
component separation in intensity. Such templates are not perfect,
i.e.
\begin{equation}
\vt{I}_c = k \, \widetilde {\vt{I}}_c +   \delta \vt {I}_c,
\end{equation}
where $k$ is a global scaling factor, and $\delta \vt{I}_c$ the
difference between the scaled template and the real foreground map. By
replacing the (unknown) $\vt{I}_c$ by its expression in terms of the
(known) template $\widehat {\vt{I}}_c$ in
Eq.~\ref{eq:leakage-pair-model}, neglecting the second-order term
proportional to $y_c \delta \vt{I}_c$, and absorbing the global
scaling factor $k$ in $y_c$, we get, in matrix-vector notation
\begin{equation}
\vt{d} = \vt{A} \vt{m} + \vt{T} \vt{y} + \vt{n}, 
\label{eq:model-leakage-vecmat}
\end{equation}
where $\tn{A}$ is a reduced pointing matrix with has two non-zero elements in each row
\begin{equation}
\vt{A}_{tp(t)} = \begin{bmatrix}
\cos\left(2\psi_t\right) & \sin\left(2\psi_t\right) \\ \end{bmatrix},
\end{equation}
$\vt{m}$ is the reduced set of Stokes parameters, containing only
$[\vt{Q}, \vt{U}]$ from both the CMB and the Galactic sources,
$\vt{T}$ is built from the \emph{known} foreground template maps
$\widetilde{\vt{I}}_c$, $\vt{y}$ is the set of amplitudes of the
leakage of the corresponding templates in that difference timeline,
and $\vt{n}$ is the noise term.

We find an unbiased estimator $\widetilde{\vt{m}}$ free (to first
order) from the leakage term with a standard generalized least square
estimator of the form
\begin{equation} \label{eq:direct_estimator_S}
  \widetilde{\vt{m}} = \left( \vt{A}^T \vt{N}^{-1} \vt{F}_{\vt{T}} \vt{A} \right)^{-1} \vt{A} \vt{N}^{-1} \vt{F}_{\vt{T}} \vt{d},
\end{equation}
\begin{equation} \label{eq:direct_estimator_y}
  \widetilde{\vt{y}} = \left( \vt{T}^T \vt{N}^{-1} \vt{F}_{\vt{A}} \vt{T} \right)^{-1} \vt{T} \vt{N}^{-1} \vt{F}_{\vt{A}} \vt{d},
\end{equation}
and 
\begin{equation} \label{eq:filter_S}
  \vt{F}_{\vt{A}} = \left\lbrace \mathbf{1} - \vt{A} \left( \vt{A}^T \vt{N}^{-1} \vt{A} \right)^{-1} \vt{A}^T \vt{N}^{-1} \right\rbrace,
\end{equation}
\begin{equation} \label{eq:filter_y}
  \vt{F}_{\vt{T}} = \left\lbrace \mathbf{1} - \vt{T} \left( \vt{T}^T \vt{N}^{-1} \vt{T} \right)^{-1} \vt{T}^T \vt{N}^{-1} \right\rbrace.
\end{equation}

We identify the terms $\vt{F}_{\vt{A}}$ and $\vt{F}_{\vt{T}}$ as
operators that filter out the component of the signal in the
respective space of $\vt{A}$ and $\vt{T}$. Also, the noise covariance
term $\vt{N}$ for the case of white noise is diagonal given by
$\sigma^2 \mathbf{1}$, where $\sigma$ is the standard deviation of the
white noise, and it cancels out in the previous equations. We thus
freely drop the noise covariance term $\vt{N}$. It is to be noted that
this algorithm can be suited for any number of systematic effects
whose leakage signal can be modelled as a nuisance term proportional
to a temporal template as shown in equation
\eqref{eq:leakage-pair-model}. An analogous approach to subtract
temporal templates was implemented in~\cite{Poletti_2017} for the
Polarbear experiment using the direct estimation of maps as in
equation \eqref{eq:direct_estimator_S}.

To make matters computationally feasible we perform our correction
iteratively by first estimating the amplitudes $\vt{y}$ of the
templates using Eq.~\eqref{eq:direct_estimator_y}, then use these to
perform a simple GLS map-making by subtracting the leakage term from
the timeline by
\begin{equation}
\widetilde{\vt{m}} =  \left( \vt{A}^T \vt{A} \right)^{-1} \vt{A}^T \left( \vt{d} - \vt{T} \, \widetilde{\vt{y}} \, \right)
\end{equation}

\begin{figure}[h]
  \centering
  \includegraphics[width=\hsize]{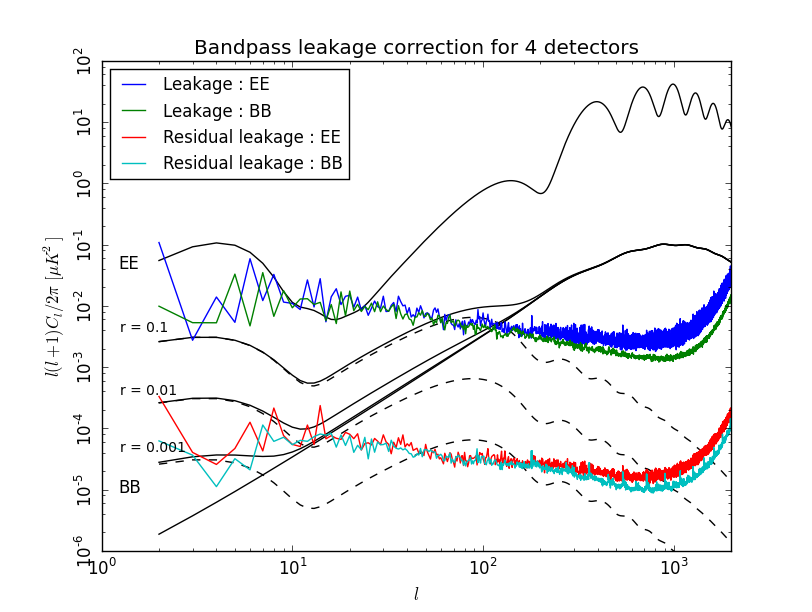}
  \caption{$EE$ and $BB$ power spectra of bandpass mismatch leakage
    for a set of 4 detectors at 145\,GHz before and after implementing
    the correction algorithm described in the text. After subtracting
    the leakage induced by thermal dust and synchrotron to first
    order, the power is reduced by more than two orders of magnitude
    (from blue to red for $EE$ and from green to light blue for
    $BB$).}
  \label{bandpass_leakage_correction}
\end{figure}

We test here this correction algorithm on the simulations described in
the previous subsection. For our templates we simulate intensity maps
for thermal dust at at 350\,GHz and synchrotron at 90\,GHz using the
PSM. The $EE$ and $BB$ power spectra of the leakage maps are shown in
Fig.~\ref{bandpass_leakage_correction}. The algorithm reduces the
leakage by more than two orders of magnitude in power. The residual
after correction for a set of 4 detectors is now comparable to the
primordial B-modes for a level of $r$ in the 0.001-0.01 range, and
below the lensing signal for $\ell \geq 10$.

As already emphasized, the power spectrum of the leakage after
averaging $N$ pairs of detectors will be reduced by $N$. The
\CORE\ 145\,GHz channel comprises 144 detectors (72 pairs). Hence we
estimate that the residual leakage after correction and averaging all
detectors of the channel will be one order of magnitude or more below
the target sensitivity at all angular scales. If necessary, this
approach can be extended to second order to further reduce the
residuals.

\section{Asymmetric beam}
\label{sec:AsymmetricBeam}

\begin{figure}
\includegraphics[width=\textwidth]{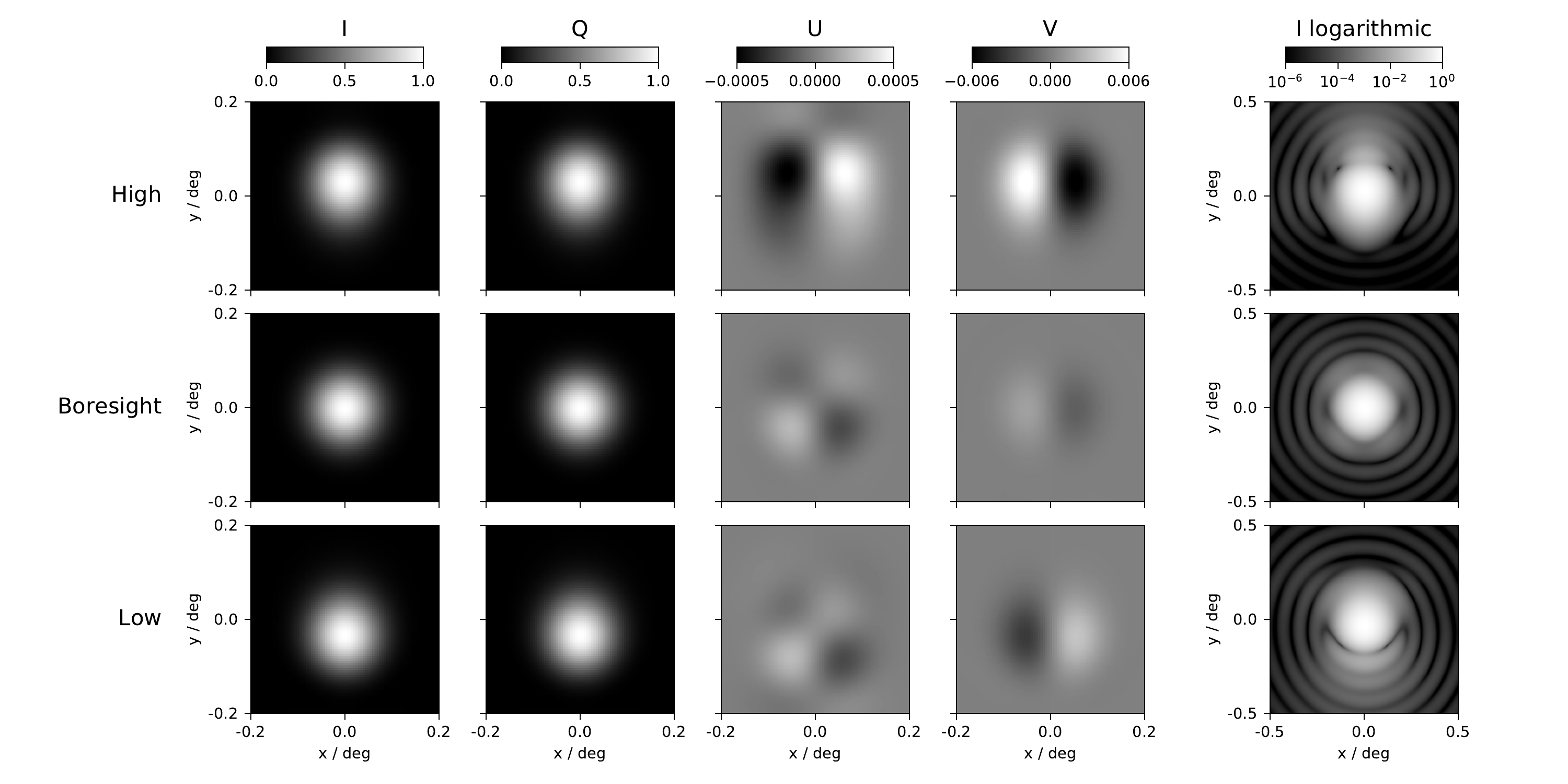}
\caption{Simulated 145\,GHz beams for the cross-Dragonian design of
  the \CORE\ telecope.  The three rows show the beams for the high,
  boresight, and low detectors (see text for details).  The first 4
  columns show the $I$, $Q$, $U$, and $V$ Stokes parameters of the
  beams on a linear scale.  The last column shows a wider view of the
  $I$ Stokes parameter on a logarithmic scale.  The Stokes parameters
  have been normalized to unity at the peak in intensity.}
\label{fig:beam_maps}
\end{figure}

The convolution of the CMB signal with an asymmetric beam will cause
leakage between intensity and and $E$- and $B$-mode polarization when
$I$, $Q$ and $U$ maps are reconstructed using the generalized least
squares solution of equation \ref{eq:gls_mm} without proper measures
to take into account this beam asymmetry.  Given that the primordial
intensity signal is much larger than the polarization signal, the most
important effect is the temperature-to-polarization leakage.

In this section, we both investigate the amplitude of this effect for
representative \CORE\ beams, and propose methods to correct this
effect by data processing.
To this effect, we first generate simulations of the \CORE\ beams at
145\,GHz for the cross-Dragonian design of the telescope.  Beam maps
are computed using GRASP for three locations in the focal plane, the
boresight, a `high' detector 4 degrees from the boresight towards the
spin axis, and a `low' detector 4 degrees from the boresight in the
opposite direction (note the locations of the high and low detectors
considered here differ from those assumed in
Sect.~\ref{Analysis_maps}).  In each location all components of the
beams (Stokes parameters $I$, $Q$, $U$ and $V$) are simulated for two
orthogonal polarizations, $x$ and $y$, where $x$ is aligned with the
scan direction.  This arrangement of detectors does not reflect the
layout of a real focal plane, which would naturally have detectors of
the same frequency grouped together.  Instead it is designed to
explore the variation of the beam shapes with position in the focal
plane using a representative CMB-dominated frequency channel as a test
case.  Figure~\ref{fig:beam_maps} shows the Stokes parameters of the
$x$-polarized beams for the three locations.  The corresponding
$y$-polarized beams are almost identical in shape, and only differ in
the sign of the $Q$ and $U$ Stokes parameters.

These beams are then used to simulate the CMB signal observed by the
\CORE\ instrument.  Two complementary paths have been followed in this
investigation.  The first starts with simulated \CORE\ timelines using
a high-resolution implementation of direct-space convolution.  Maps of
$I$, $Q$, and $U$ are reconstructed using equation \ref{eq:gls_mm},
and beam asymmetry effects are corrected using a re-observation of the
reconstructed map of $I$.  In the second, we directly use a
power-spectrum approach, the QuickPol formalism \citep{Hivon:2016}, to
estimate and correct for biases in the power spectra.  We now describe
simulations and results for these two approaches.

\subsection{Real space convolution and first-order de-projection}

The beams of the proposed \CORE\ mission are relatively small. The
impact of beam asymmetry must be calculated with sufficient numerical
precision. To this effect, we implement a real-space convolution of
the CMB sky by a pixelized beam map, which allows us to simulate a
realistic timeline signal for any beam shape. Deviations of the beam
from a symmetric Gaussian shape project onto the time ordered signal
in a manner that reflects the orientation of the beam along the scan.

The real space beam convolution technique uses a method specifically
designed for a spinning experiment, which breaks up a $N \times N$
pixelized beam map into $N$ rows that are aligned with the scanning
direction. In each row, every beam pixel observes the same signal, up
to a time shift and a global scaling term, that is, the contribution
of the row to the total signal is obtained by one-dimensional
convolution of a signal timeline by the corresponding beam cut (a row
of beam pixels). The data timeline is cut in smaller periods with some
overlap to perform the convolution in practice.  After each timeline
is convolved in Fourier space by the corresponding beam row, the sum
of all the individual rows gives us the full convolved timeline. This
method scales as $N \times log(N)$ instead of $N^{2}$ as would be the
case if the contribution of all $N^{2}$ pixels was computed
independently for each data point.

\begin{figure}[h]
   \centering
   \includegraphics[width=0.7\linewidth]{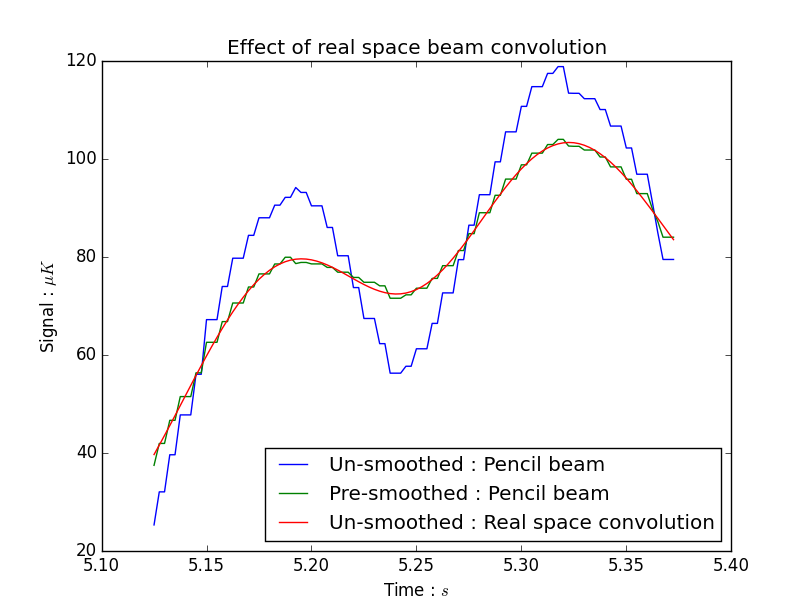}
   \caption{Comparison of signals obtained from various beam
     convolution methods. The signal obtained using a pencil beam to
     scan a map pre-smoothed in harmonic space with an $8^\prime$ beam
     (green) is compared to the signal obtained using the real-space
     convolution technique to scan an un-smoothed map with the same
     pixel size (red). All calculations are performed using HEALPix
     maps at $N_{\mathrm{side}}=4096$, corresponding to a pixel size
     of about 52$^{\prime\prime}$. The green and the red signals
     signals follow each other closely, although we distinguish small
     discontinuities in the former due to pixelization, which are
     absent in the latter.}
   \label{fig:real_space_convolution}
\end{figure}

The real space convolution has the added advantage of reducing
sub-pixel effects, for a more accurate timeline signal that avoids
part of the sky pixelization effects.  This is illustrated in
Fig.~\ref{fig:real_space_convolution}, which compares the timeline
obtained by real-space convolution with an $8^\prime$ Gaussian
symmetric beam, with that obtained by scanning with a pencil beam the
map after convolution with the same beam in harmonic space. The
real-space convolution, which uses beam pixels of size $\simeq
1^\prime$, produces a smoother signal, without the numerical artifacts
due to the discontinuities between the pixels of the HEALPix map.  All
simulations performed here use high resolution maps at
$N_{\mathrm{side}} = 4096$ for adequate sampling of the underlying CMB
signal, and hence good numerical precision.

\begin{figure}[h]
   \centering
   \includegraphics[width=0.9\linewidth]{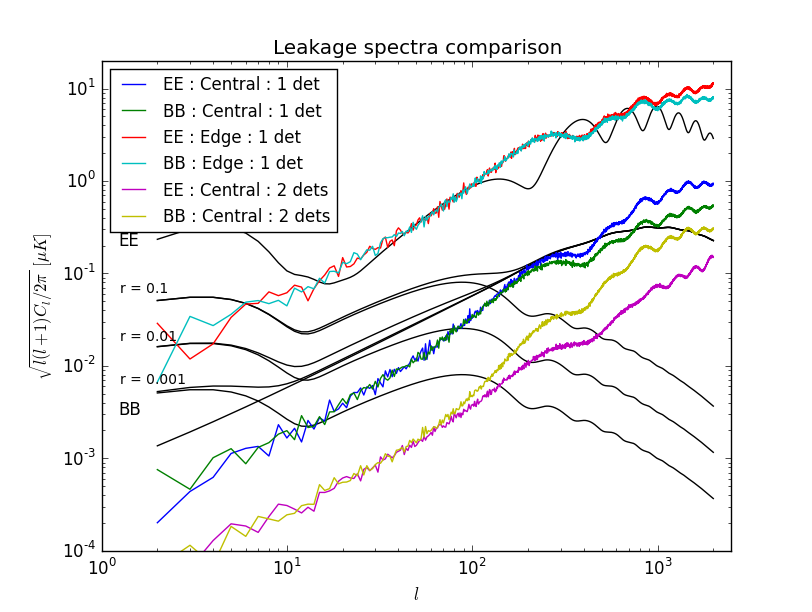}
   \caption{$EE$ and $BB$ spectra of leakage due to beam asymmetry and
     cross-polarization.  The spectra are computed from the difference
     of the output CMB map with the input map. The effect of the
     additional asymmetry of the edge detectors is reflected in the
     increase in the leakage spectra as compared to the boresight
     detector. The leakage is decreased when using a pair of
     orthogonally polarized detectors due to better modulation of
     angles.}
   \label{beam_asymmetry_leakage_spectra}
\end{figure}

Simulations for \CORE\ are performed by scanning an unsmoothed CMB map
using the pixelized beam maps of Fig.~\ref{fig:beam_maps} with the
real space convolution described above.
Maps made from the resulting timelines suffer from the
temperature-to-polarization leakage due to the asymmetric beams and
from sub-dominant cross-polarization terms, projected onto the
reconstructed `observed' map as a function of the scanning
strategy. Reconstructed maps of $Q$ and $U$ (and hence $E$ and $B$),
are thus contaminated with leakage from temperature to
polarization. These effects can be quite large, as illustrated in
Fig.~\ref{beam_asymmetry_leakage_spectra}, in particular for detectors
near the edge of the focal plane, which have more asymmetric beams
than those of detector located at the boresight, for which even the
$E$ modes would be significantly contaminated. We note however that
the temperature-to-polarization leakage is reduced for maps obtained
with two detectors instead of one, when these two detectors are in the
same location, but are sensitive to orthogonal polarizations. This
happens because the $I$ beams of two such detectors are very similar,
which leads to much of the leakage being cancelled in the map-making
stage.

To correct for the temperature-to-polarization leakage in the map due
to the beams, we consider an intensity map that has been observed by
our given set of detectors as a template. This template, however, is
not perfect because it contains noise and errors due to the beam
asymmetry itself. Still, we can use this template to estimate the
intensity-to-polarization leakage to first order. To do this, the
template is deconvolved in harmonic space using an effective average
symmetric beam that best fits the angular resolution of the
channel. In practice, this deconvolution is performed only up to a
certain limit of $\ell$ (see \cite{Banerji-beams} for details). This
deconvolved map is then re-scanned using the same scanning strategy
and the intensity beam of the detector of interest. The resulting
timeline will therefore contain the intensity signal, along with
higher order leakage terms due to the asymmetric beams, and noise, as
well as the initial leakage terms being observed again. When projected
onto $(Q, U)$ maps, this timeline propagates an estimate of the
intensity leakage terms to the polarization map for that particular
detector (or detector set when several timelines are combined). Thus
to first-order, we can use it to estimate the
temperature-to-polarization leakage due to asymmetric beams. The
leakage in the observed polarization maps is then cleaned by
subtracting the estimated leakage maps. At first order, this method
does not correct for cross-polar leakage that mixes $E$ and $B$ modes,
which would require additional modelling terms (see discussion about
systematics in \cite{2017arXiv170604516D} for a general introduction).

\begin{figure}[h]
   \centering
   \includegraphics[width=0.9\linewidth]{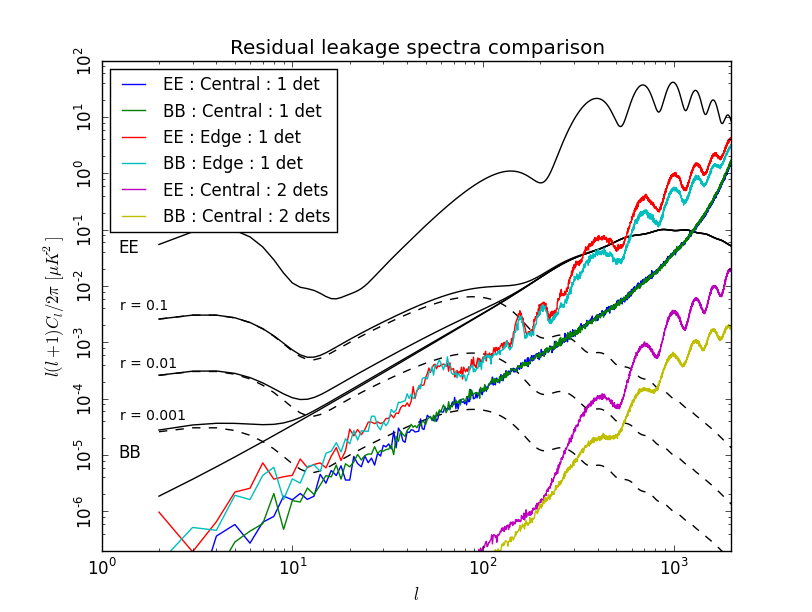}
   \caption{$EE$ and $BB$ spectra of the residual beam leakage after
     correction by the real-space method.  The residual spectra are
     dominated by cross-polar leakage and by pixelization noise due to
     the process of deconvolution and re-observation of the template
     map.}
   \label{beam_asymmetry_correction_spectra}
\end{figure}

After such first-order correction, as seen in
Fig.~\ref{beam_asymmetry_correction_spectra}, the leakage for a single
\textcolor{red}{high/low} detector is significantly reduced, below the
$BB$ lensing signal for $l \leq 300$. For one boresight detector the
residual leakage is at the level of the primordial BB spectra for
$r=0.001$ for the first few $10$'s of multipoles. For the pair of
boresight detectors, the residual leakage is dominated by cross-polar
leakage (which is not corrected for in this analysis) and is well
below the primordial $BB$ spectrum f for all $l \leq 200$, even for
$r=0.001$.

\subsection{Harmonic space}

\begin{figure}
\centering{
  \includegraphics[width=0.8\textwidth]{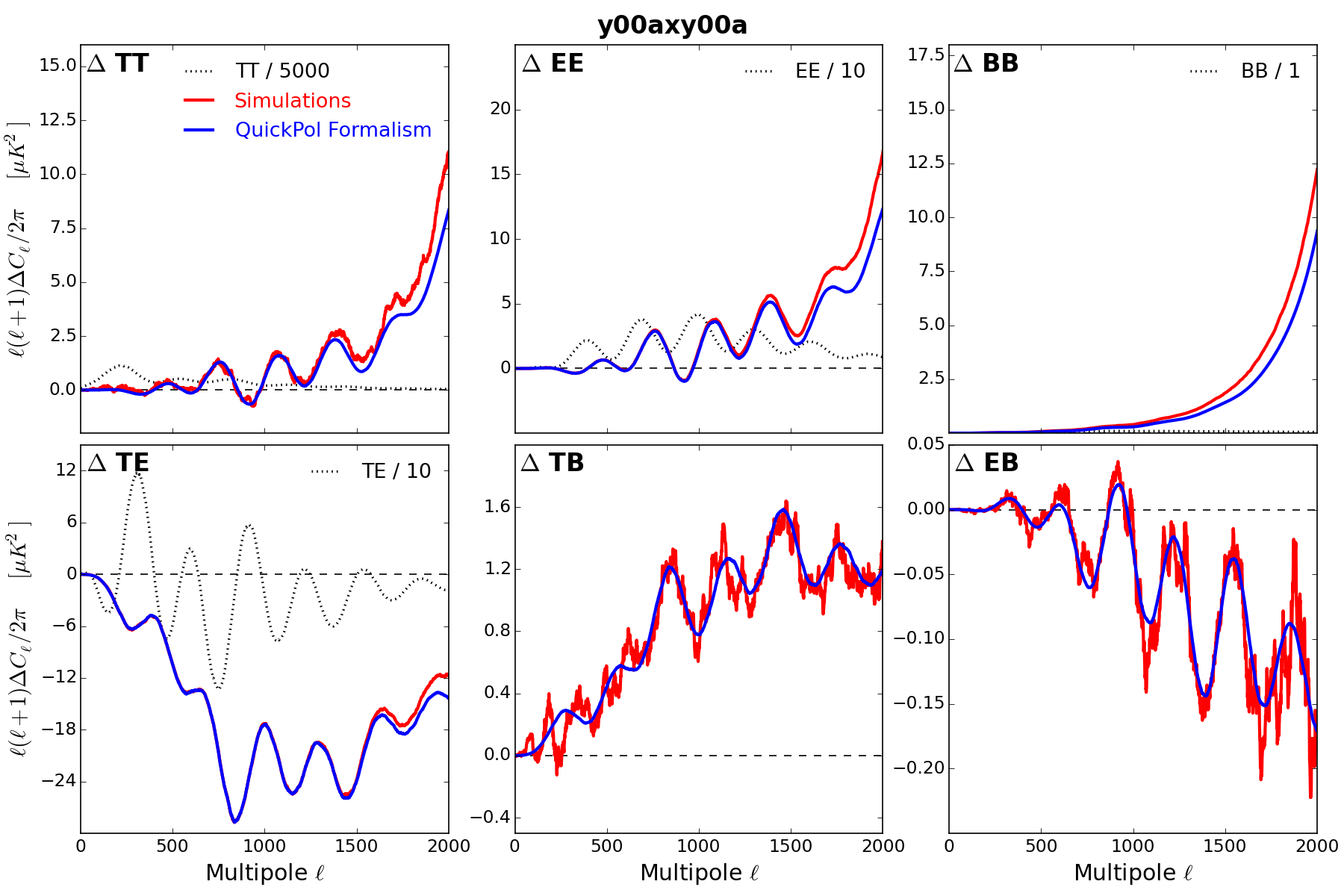}
  \includegraphics[width=0.8\textwidth]{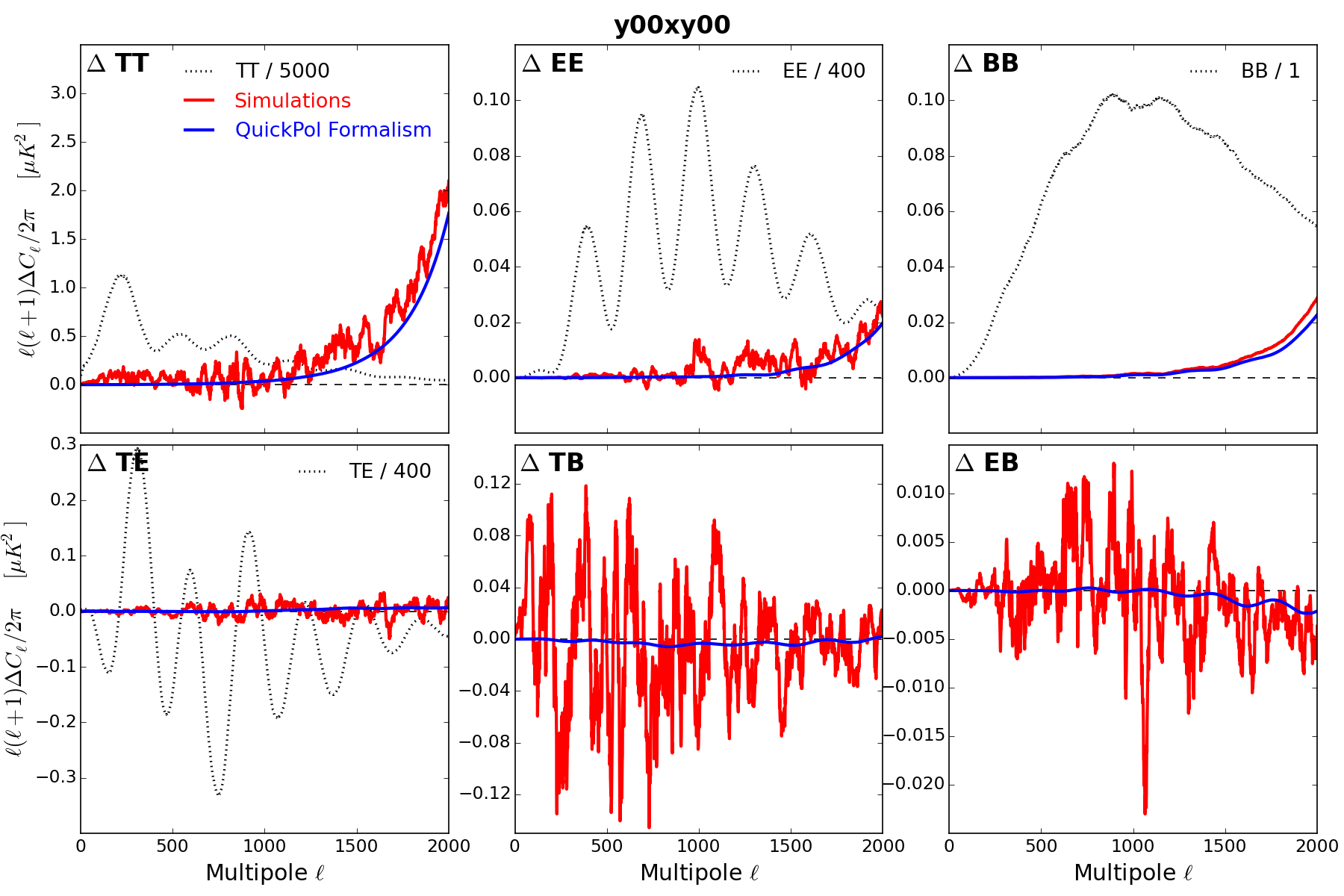}
}
\caption[Beam leakage]{Discrepancy between the measured
  $\ell(\ell+1)C_\ell$ and the input, smoothed with $\Delta \ell=31$,
  for a single \CORE\ detector (6 upper panels) and for a pair of
  orthogonally polarized detectors in the same location (6 lower
  panels) at the boresight.  Results obtained in noiseless simulations
  of \CORE\ observations (red lines) are to be compared with those of
  the QuickPol formalism (blue lines).  In panels where it does not
  vanish, a fraction of the input power spectrum is also shown in
  black dots for comparison.}
\label{fig:beam_leakage}
\end{figure}

An alternative way to account and correct for the effect of the beam
leakage is to work at the angular power spectrum level.  The QuickPol
formalism \citep{Hivon:2016}, allows such a description, as
illustrated in Fig.~\ref{fig:beam_leakage} where its predictions are
compared to simulations of observations of either a single detector or
the pair of detectors at the boresight.  The agreement with the
simulations is very good, including at very large multipoles, where
the sub-pixel effects, discussed in Sec.~\ref{sec:pointing_accuracy},
dominate.

The formalism only needs a description of the detector scanning
strategy, in the form of the statistics of the orientation of the
detectors on the sky in a subset of sky pixels, the spherical harmonic
representation of the beams, and some extra description of the
detectors: orientation in the focal plane, gain calibration,
polarization efficiency.  It returns a beam matrix describing how, at
each multipole $\ell$, the input $TT$, $TE$, $EE$, and $BB$ $C_{\ell}$
power spectra are related to the measured $\widetilde{C}_{\ell}$
spectra, that is
\begin{equation}
  \widetilde{C}^{XY}_{\ell} = \sum_{X'Y'} W^{XY,\; X'Y'}_\ell C^{X'Y'}_{\ell},
  \label{eq:beam_matrix}
\end{equation}
with $X,Y,X',Y' \in \{T,E,B\}$.  Such a $W^{XY,\; X'Y'}_\ell$ beam
matrix can be used in the cosmological analysis of the measured power
spectra, in order to determine the model that best fits the data.
Since the algorithm is very fast, it can be used for a Monte-Carlo
determination of the systematics-related error bars on the final
$C_\ell$. It simply has to be run multiple times while varying the
values of the instrumental parameters known with limited accuracy,
such as the polarization efficiency or even the beam shape.

\subsection{Beam asymmetry conclusions}

In the absence of fast modulation of polarization such as with a
rotating half-wave plate, the non-idealities of the beams could hinder
the precise measurements of primordial $BB$ spectrum by \CORE.
However, as we have shown, we have at our disposal two different and
complementary approaches to deal with such systematics, one map based,
and the other one power spectrum based.  They both allow us to
simulate rapidly and precisely the impact of \CORE\ beam
non-idealities on the maps or the power spectra, and can therefore be
used to respectively clean the maps of most of these systematic
effects, and account precisely for these systematics in the
cosmological analysis of the measured power spectra, even with an
imperfect knowledge of the instrument features.

While these investigations must be pursued in more detail, no
technical limitation has been identified from beam asymmetry effects
for the \CORE\ space mission. Further investigations must also address
the simultaneous deprojection of all potential systematics, and are
postponed to further study.

\section{Calibration}
\label{sec:Calibration}

In this section we discuss the systematic effects that can potentially
affect the quality of the photometric calibration of \CORE's
detectors. By `photometric calibration', we mean the process of
converting the output produced by a detector into the amplitude of the
signal entering \CORE's optical system. Assuming that the output of
the detectors depends linearly on the input signal entering the
optical system, the result of a photometric calibration is a timeline
of multiplicative calibration factors that convert the output of a
detector into thermodynamic temperature. In the case of \Planck\ HFI,
the stability of the detectors was good enough to allow the use of a
constant calibration factor over the 2.5 years of the nominal mission
\citep{planck.2013.08.HFI.calibration.and.mapmaking,planck.2015.08.HFI.data.processing.calibration.and.maps}. In
this work, we aim to determine how well we can detect changes in the
calibration factors of \CORE\ detectors that unfold over timescales of
the order of a few hours to a few days.

The ability to track changes in the value of the calibration factors
is affected by the strength of the signal used for the calibration. As
was the case with \WMAP\ and \Planck, we plan to use the dipolar
signal induced by the motion of the Sun with respect to the CMB rest
frame as a calibration signal. We call this signal the `dipole' for
brevity. For simplicity, in this analysis we neglect the contribution
of the time-varying orbital dipole, caused by the yearly motion of the
\CORE\ spacecraft with respect to the Solar System. This
component, called the \emph{orbital dipole}, is usually $\sim10\,\%$
of the \emph{solar dipole}, the component due to the motion of the Sun
with respect to the CMB rest frame
\citep{planck.2015.05.LFI.calibration}. Including this
component would have improved the estimation of the calibration
factors at the expense of a more complex analysis pipeline.

We have developed an improved version of the calibration code used for
\Planck\ LFI \citep{planck.2015.05.LFI.calibration}, because code has
the ability to track changes of the calibration with
time\footnote{Unlike HFI, which used stable bolometric detectors based
  on transition edge sensors (TES), LFI implemented square-law
  detectors based on high electron mobility transistors (HEMT), which
  have shown non-negligible changes (of the order of several percent)
  in the gain constants over the 4 years of the extended
  mission.}. Using an approach similar to the one described by
\citet{Tristram_2011}, the LFI calibration algorithm, named DaCapo, is
an hybrid calibration and map-making tool which is able to estimate
the calibration factors, the $1/f$ noise timelines, and the sky map
simultaneously, by means of a modified destriping algorithm. We have
rewritten the code from scratch in Python\ 3 and have incorporated it
in TOAST. DaCapo uses a maximum-likelihood approach which is similar
to destriping techniques employed to solve map-making problems, and it
shares with them a number of concepts (See Sect.~\ref{sec:map-making} above). We
have modified the original algorithm in order to allow the timescale
on which the calibration is assumed to be constant to vary. In the
original version for the \Planck\ LFI, this timescale was assumed to
be one pointing period of 40--60\,min.

We present a brief overview of the algorithm here; for additional
details, the reader can refer to the source code of the programs we
have used in this analysis \citep[]{2016ascl.soft12007T}.

The DaCapo algorithm models the output of a detector scanning the sky
using the following approximation:
\begin{equation}
\label{eq:calDaCapoRadiometerModel}
s (t_i) = G_k \bigl(T_i + D_i\bigr) + b_n + n_i,
\end{equation}
where $s (t_i)$ is the signal measured by the detector at time $t_i$,
$T_i$ is the temperature of the sky (including Galactic signals, CMB,
point sources, etc.)\footnote{We do not attempt to model polarized
  signals in Eq.~\eqref{eq:calDaCapoRadiometerModel}, as we have found
  that the systematic effect induced by this approximation is far
  smaller than other systematics discussed later.}, $D_i$ is the
amplitude of the solar and orbital components of the Doppler dipole
(due to the motion of the spacecraft on its orbit in the solar system,
and of the solar system with respect to the CMB: as previously mentioned, we neglect the orbital contribution in the following analysis), $n_i$ is a
white-noise term, $G_k$ is the gain, and $b_n$ is an offset which
keeps track of slow fluctuations due to $1/f$ noise. The DaCapo
algorithm obtains a maximum-likelihood estimate of $G_k$, $T_i$, and
$b_n$ given $s$ and $D$. The last quantity is given by:
\begin{equation}
D(\mathbf{x}, t) = T_\mathrm{CMB} \left(\frac1{\gamma\, \bigl(1 - \mathbf{\beta}\cdot \mathbf{x}\bigr)} - 1\right),
\end{equation}
where $T_\mathrm{CMB}$ is the mean temperature of the CMB,
$\mathbf{\beta} = \mathbf{v}/c$ is the velocity of the spacecraft with
respect to the CMB rest frame, $\gamma = (1 - \beta)^{1/2}$, and
$\mathbf{x}$ is the direction of the detector's main beam axis at time
$t$. The values for $T_\mathrm{CMB}$ and $\mathbf{v}$ have been taken
from \citet{mather1999} and \citet{planck.2015.01.Overview},
respectively.

The quantities $G$ and $b$ in Eq.~\eqref{eq:calDaCapoRadiometerModel}
are indexed by $k$ and $n$ instead of $i$. This relates to the fact
that the sampling rate of these two quantities must be smaller than
the sampling rate of $s$: in this way, the number of unknowns of the
calibration problem is smaller than the number of variables. We call
the inverse of the sampling rate of $G_k$ the `calibration period';
its fluctuations unfold on periods of the order of several hours; the
inverse of the sampling rate of $b_n$ is the `offset period', and it
must be shorter than the knee frequency of the $1/f$ noise
component. Unlike the code used in
\citet{planck.2015.05.LFI.calibration}, we allow the calibration
period to be different from the offset period; in this way, we have an
additional free parameter that can be tuned to optimize the quality of
the calibration. Specifically, this change allows us to make the
calibration period longer than the offset period, thus potentially
improving the detection of the dipole signal thanks to a better S/N
without degrading too much the estimation of the $1/f$ component. This
change is important in the context of the \CORE\ proposal, as we
expect \CORE\ detectors to be more stable than those of \Planck\ LFI.

We solve for the unknowns in Eq.~\eqref{eq:calDaCapoRadiometerModel} by minimizing the quantity
\begin{equation}
\label{eq:calChiSquare}
\chi^2 = \sum_i \frac{\bigl(s_i - s(t_i)\bigr)^2}{\sigma_i^2},
\end{equation}
where $s_i$ is the sample acquired by the detector, and $s(t_i)$ is
the model in Eq.~\eqref{eq:calDaCapoRadiometerModel}. To make the
problem well-posed, \citet{planck.2015.05.LFI.calibration} adds the
constraint that the map $\mathbf{m}$ of the sky temperature $T$ be
orthogonal to the dipole map $D$, and that the monopole of map $T$ be
zero. Using Lagrange multipliers, the coupling between the
minimization of Eq.~\eqref{eq:calChiSquare} and these additional
constraints leads to the following solution\footnote{In the following
  equations, we adopt the convention of using \textbf{bold} to
  indicate vectors and matrices, and \textit{italic} to indicate
  scalar values. The $ij$ coefficient of matrix $\mathbf{m}$ is
  therefore indicated as $m_{ij}$.}:
\begin{equation}
\label{eq:calDaCapoSolution}
\mathbf{F}^T \mathbf{N}_w^{-1} \mathbf{Z} \mathbf{F} \mathbf{a} = \mathbf{F}^T \mathbf{N}_w^{-1} \mathbf{Z} \mathbf{s}, 
\end{equation}
where $\mathbf{N}_w$ is the covariance matrix of the white noise
component $n_i$ in Eq.~\eqref{eq:calDaCapoRadiometerModel},
$\mathbf{a}$ is a vector containing all the values of $b_n$ and $G_k$,
$\mathbf{F}$ is defined by the following relation
\begin{equation}
\sum_j F_{ij} a_j \equiv G_k \left(D_i + \sum_p P_{ip} m_p\right) + b_n
\end{equation}
and
\begin{align}
\mathbf{Z} &= \mathbf{I} - \mathbf{\tilde P} \bigl(\mathbf{M} + \mathbf{N}_w^{-1} \bigr)^{-1} \mathbf{\tilde P}^T \mathbf{N}_w^{-1},\\
\mathbf{M} &= \mathbf{\tilde P}^T \mathbf{N}_w^{-1} \mathbf{\tilde P},\\
\bigl(\mathbf{M} + \mathbf{N}_w^{-1}\bigr) &= \mathbf{M}^{-1} - \mathbf{M}^{-1} \mathbf{m}_c \bigl(\mathbf{m}_c^T \mathbf{M}^{-1} \mathbf{m}_c\bigr)^{-1} \mathbf{m}_c^{-1} \mathbf{M}^{-1},
\end{align}
where $\mathbf{I}$ is the identity matrix, $\mathbf{m}_c$ is a
two-column matrix containing the map of the dipole signal $D$ and a
constant map with each pixel set to 1, and $\mathbf{\tilde P}$ is the
pointing matrix $\mathbf{P}$ where each non-zero element has been
multiplied by the corresponding gain $G_k$.

In the next paragraphs, we discuss how this model can be applied to
timelines of signals acquired by detectors scanning the sky using
\CORE's scanning strategy, the kind of statistical and systematics
errors we should expect using this approach, and the estimated quality
of the results.

\subsection{Time dependence of the dipole signal}
\label{sec:calDipoleTimeDependence}

\begin{figure}[tb]
  \centering	
  \includegraphics[width=0.4\columnwidth]{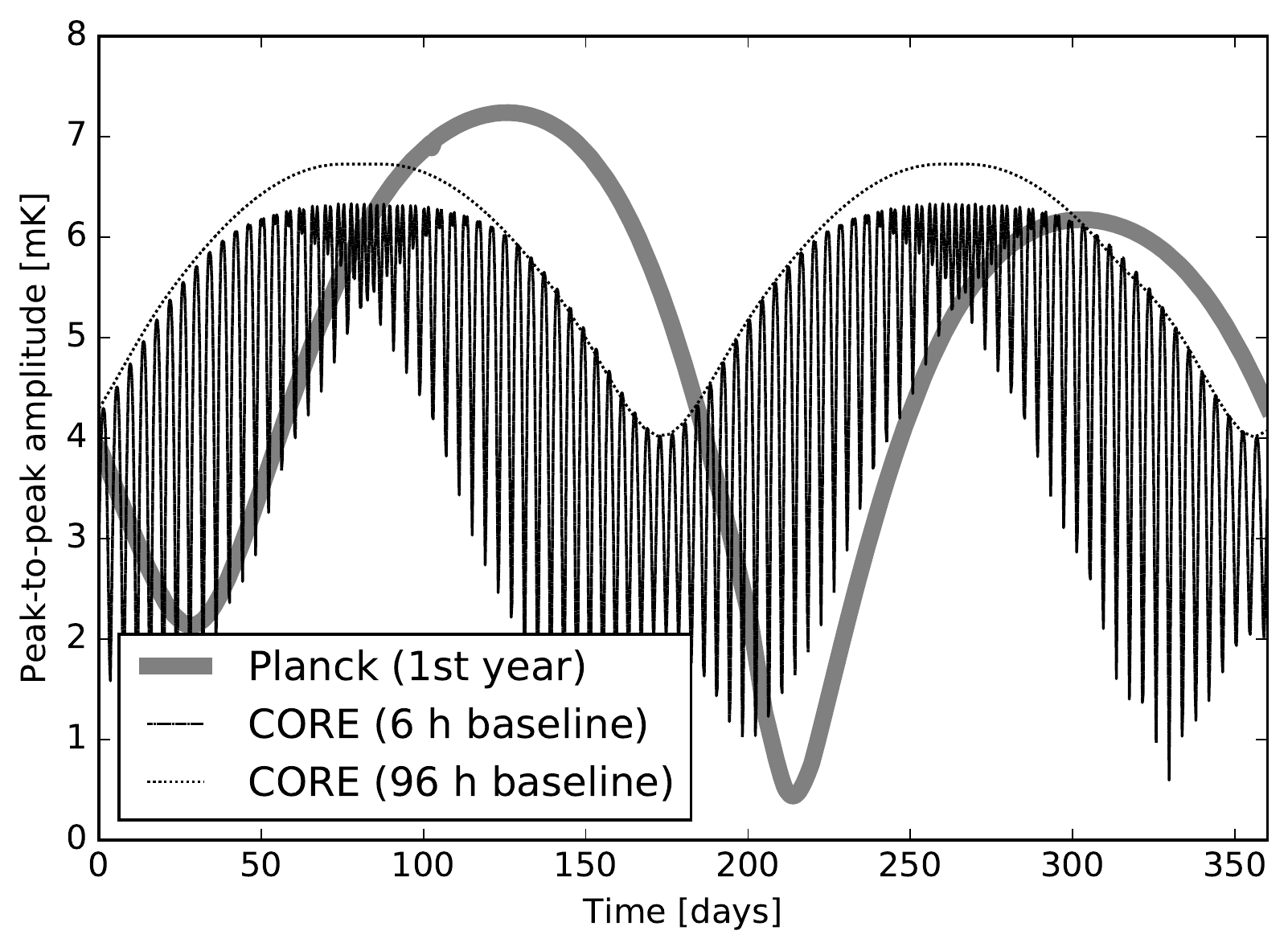}
  \includegraphics[width=0.45\columnwidth]{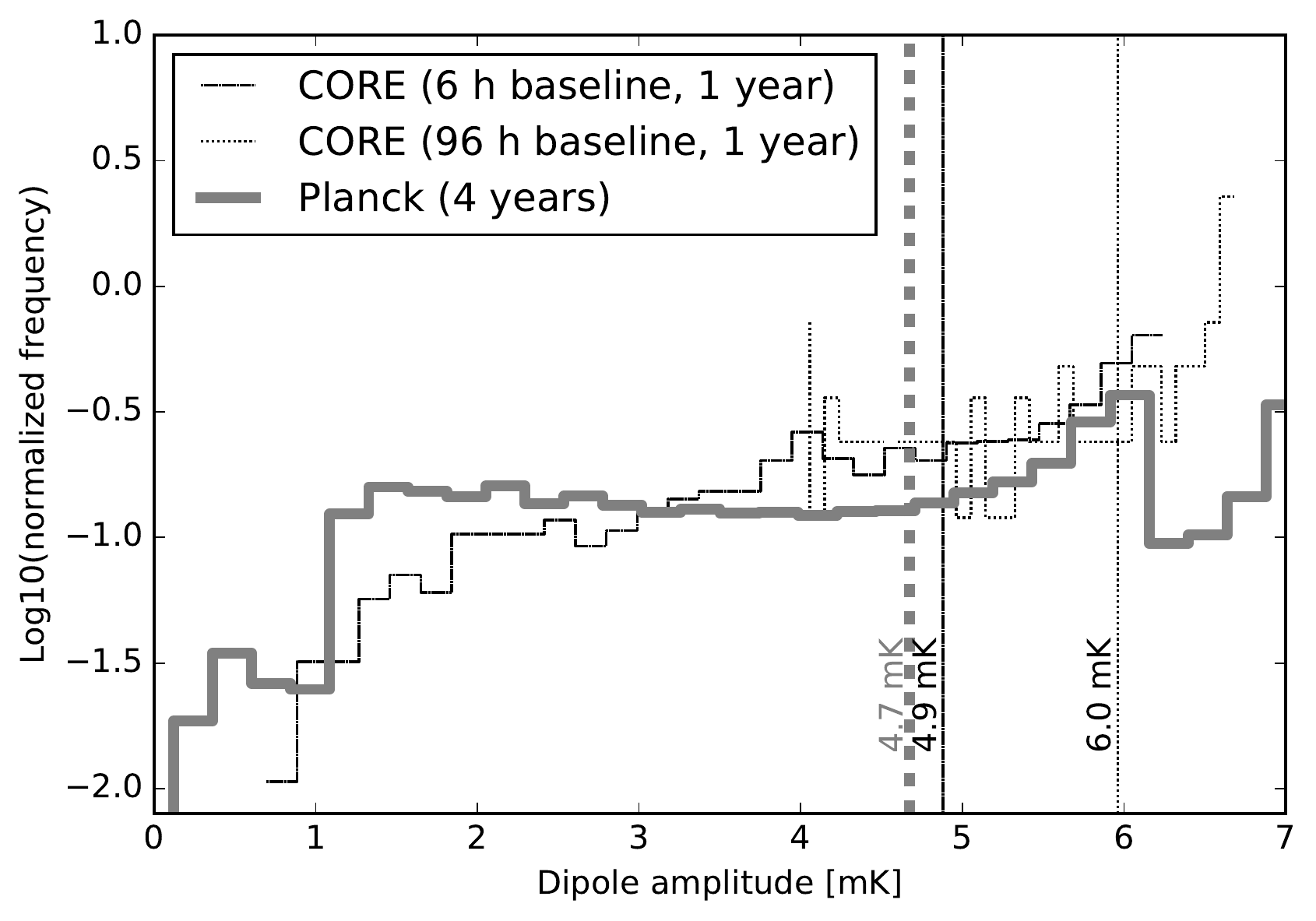}    
  \caption{\textit{Left:} Peak-to-peak amplitude of the hourly dipole
    signal as seen by \CORE\ (thin black line for 6 hour baseline and
    thin dotted line for 96 hour baseline) and \Planck\ (thick grey
    line). \CORE's faster variations will allow to pinpoint the
    calibration constant more accurately than was possible with
    \Planck\ data. \textit{Right:} Distribution of the amplitudes
    shown in the left plot. Although the maximum amplitude of the
    hourly dipole signal seen by \CORE\ is smaller than those seen by
    \Planck, the tail at low amplitudes is considerably smaller. Thus,
    the median amplitude seen during each hour of \CORE's sky
    observation is 5.8\,mK, instead of \Planck's value of 4.7\,mK
    ($\sim 20\,\%$ larger).}
  \label{fig:calDipoleAmplitudeVsPlanck} 
\end{figure}

The quality of the estimates for $G_k$
(Eq.~\ref{eq:calDaCapoRadiometerModel}) depends on the peak-to-peak
variation in the amplitude of the dipole $D$ during each calibration
period. We have therefore studied the variation in time of the dipole
signal $D$ as observed by a typical \CORE\ detector. 

Figure~\ref{fig:calDipoleAmplitudeVsPlanck} shows the expected
peak-to-peak variation of $D$ within one calibration period, for
different values of this period. We have considered periods going from
6\,hours to 4\,days; the latter case is interesting, since this is the
precession period used in \CORE's baseline scanning
strategy. Therefore, using calibration periods of this length produces
large coverages within each period. We compare our estimates with the
peak-to-peak variation of $D$ as seen by a \Planck\ detector
LFI27M-00), whose calibration periods lasted 1 hour: \CORE's scanning
strategy will lead to larger peak-to-peak variations in the
calibration signal $D$, thus potentially improving the quality of the
calibration.

In the next sections we show the results of a number of simulations
run under different assumptions, and quantify the quality of \CORE's
calibration more precisely.

\subsection{Systematics}
\label{sec:calSystematics}

\begin{figure}
  \centering
  \includegraphics[width=\textwidth]{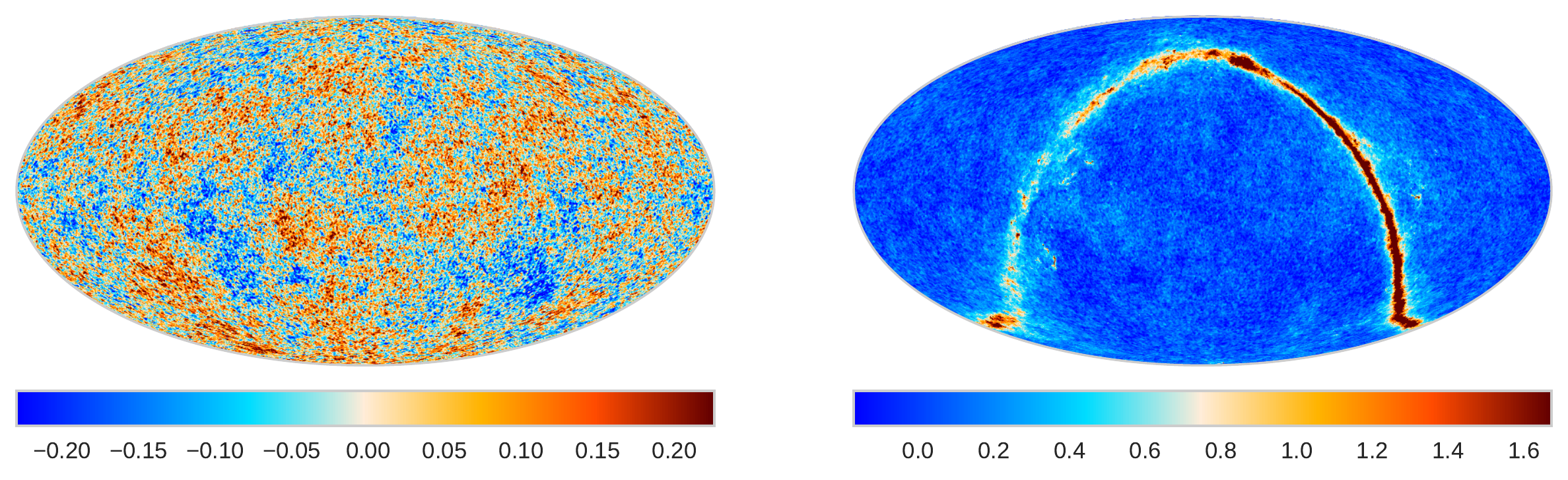}
  \caption{Foreground maps used to estimate the quality of the
    calibration of \CORE\ detectors. Temperatures are expressed in
    mK. \textit{Left:} The CMB map produced by \Planck in the Ecliptic
    coordinates. \textit{Right:} The \Planck\ HFI 143\,GHz temperature
    map in the Ecliptic coordinates.}
\label{fig:calForegroundMaps} 
\end{figure}

We run a number of simulations to estimate the impact of different
systematics on the calibration of the \CORE\ detectors. Our
simulations are generated using the following approach:
\begin{enumerate}
\item Created timelines by scanning sky maps under the assumption of
  \CORE's proposed scanning strategy\footnote{The code used to produce
    the timelines is based on TOAST. It is freely available at
    \url{https://github.com/ziotom78/create_timelines}.}. The signal
  in the timelines is
  \begin{equation}
    \label{eq:calTodSimulation}
    s (t_i) = G \times \bigl(I_i + Q_i\cos2\psi + U_i\sin2\psi\bigr) + \tilde n_i,
  \end{equation}
  where $G$ is the gain, constant throughout the whole simulation,
  $I_i$, $Q_i$, and $U_i$ are the Stokes parameters of the pixel being
  observed at time $t_i$, and $\tilde n_i$ is a $1/f$ plus white noise
  realization. The knee frequency of the $1/f$ noise has been set to
  20\,mHz. We produce two sets of timelines using two sky maps: the
  first one being the \Planck\ 2015 CMB map, the second one being
  \Planck\ 2015 143\,GHz map.

\item We run DaCapo on the simulated timelines using different values
  of the calibration period, and compare the estimates of the gains
  $G_k$ with the input gain $G$ used in
  Eq.~\eqref{eq:calTodSimulation}.
\end{enumerate}

\begin{figure}
  \centering
  \includegraphics[width=\textwidth]{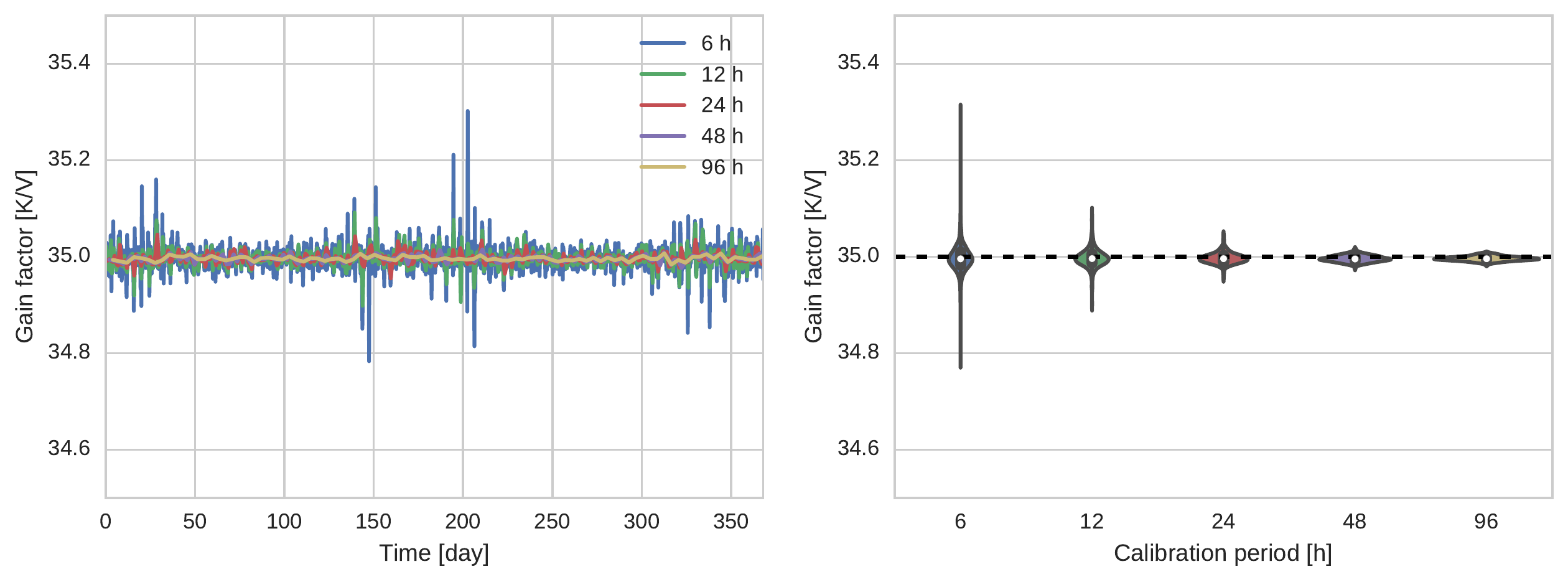}
  \caption{\textit{Left:} Variation of the gains estimated in the case
    of a CMB-only sky. The gain $G$ used in
    Eq.~\protect\eqref{eq:calTodSimulation} to produce the simulated
    timelines was 35.0\,K/V. \textit{Right:} Violin plots of the gain
    estimates shown on the left. Increasing the length of a
    calibration period reduces the dispersion of the gain estimates.}
  \label{fig:calGainTimelinesCmbOnly}
\end{figure}

We consider 5 values for the calibration period $\Delta t_G$:
6\,hours, 12\,hours, 1\,day, 2\,days, and 4\,days. The value actually
used in the analysis of the data acquired by \CORE\ will depend on the
overall stability of the detectors; the experience acquired with HFI
shows that bolometers are extremely stable, and intrinsic variations
are usually smaller than gain uncertainties
\citep{planck.2015.08.HFI.data.processing.calibration.and.maps}.

The results of our simulations are shown in
Fig.~\ref{fig:calGainTimelinesCmbOnly}. The amplitude of the
fluctuations is reduced for large values of the calibration period, as
expected. For the best case ($\Delta t_G = 4\,\text{days}$), the
estimated gain is $G = 34.996_{-0.008}^{+0.010}\,\text{K/V}$ (95\,\%
C.L.), in agreement with the input value $G = 35\,\text{K/V}$ within
$0.01\,\%$.

\begin{figure}
  \centering
  \includegraphics[width=0.6\textwidth]{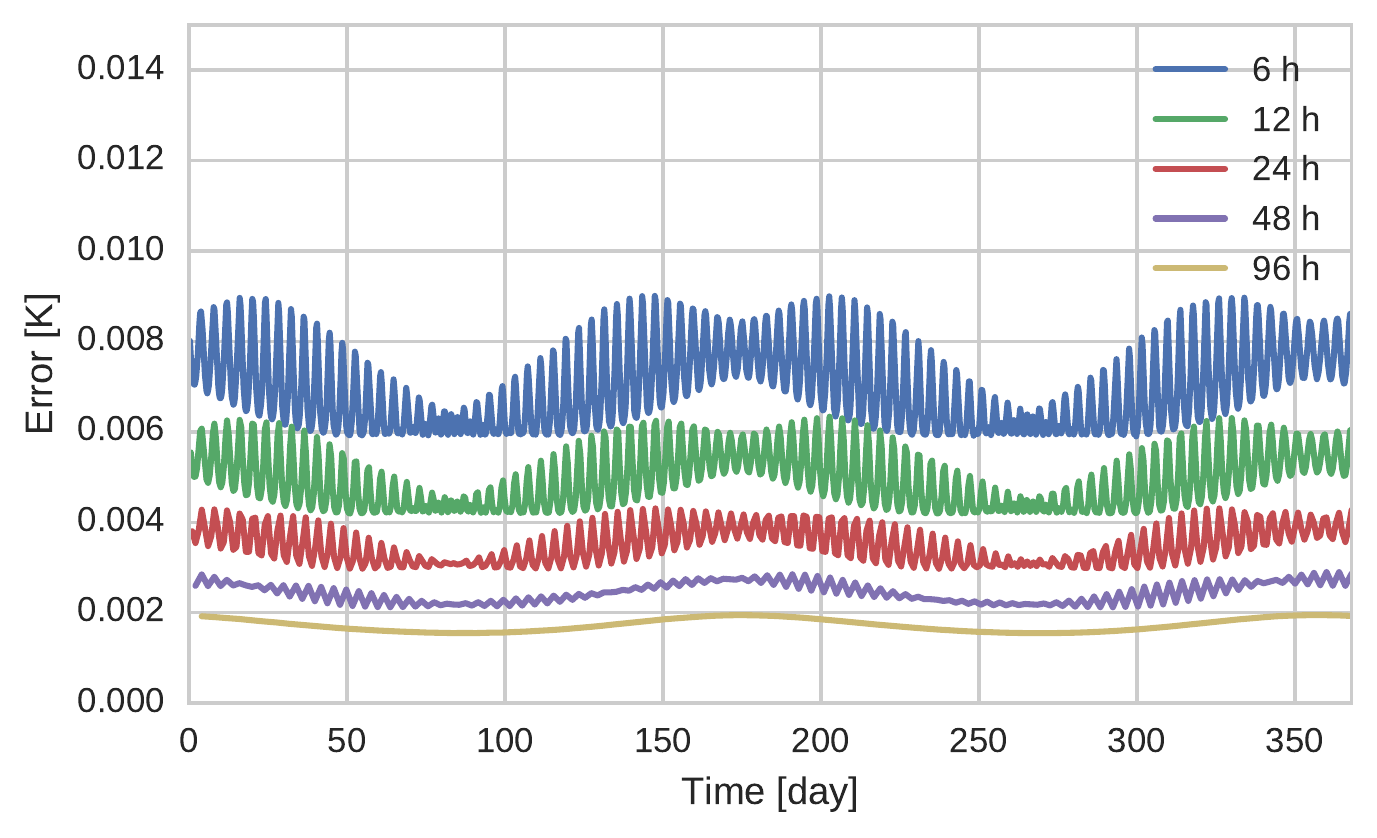}
  \caption{The uncorrelated statistical errors for the gains shown in
    the plot on the left of
    Fig.~\protect\ref{fig:calGainTimelinesCmbOnly} in the case of a
    CMB-only sky for different lengths of the calibration period. The
    errors have been computed assuming diagonality for the gain
    covariance matrix, and they are therefore a lower bound. Both the
    short- and long-period fluctuations in the error are
    anticorrelated with the amplitude of the dipole signal; see the
    plot on the left of
    Fig.~\protect\ref{fig:calDipoleAmplitudeVsPlanck}.}
  \label{fig:calGainErrors}
\end{figure}

The algorithm is able to estimate the error of each gain estimate,
under the following assumptions:
\begin{enumerate}
\item Only the statistical noise is considered (that is, due to the
  presence of the $n_i$ term in
  Eq.~\eqref{eq:calDaCapoRadiometerModel});
\item No correlation is assumed between the gains of the same
  timeline.
\end{enumerate}
This estimate is a lower bound on the true statistical error, since
the latter will also include the effect of imperfect $1/f$ cleaning
and the correlation between pairs of gains. The behaviour of the
errors is shown in Fig.~\ref{fig:calGainErrors}: their time dependence
is clearly anticorrelated with the amplitude of the dipole variations
shown in Fig.~\ref{fig:calDipoleAmplitudeVsPlanck}.

\subsection{Systematics due to the Galaxy}
\label{sec:calGalacticSystematics}

The presence of a Galactic signal in the temperature measured by
\CORE\ detectors is more problematic than the presence of the CMB
signal considered in Sect.~\ref{sec:calSystematics}, because the
Galaxy shows large scale features that can be easily mistaken for a
component of the dipole signal, thus causing leakage from $T$ to $D$
in Eq.~\eqref{eq:calDaCapoRadiometerModel}. There are two effectsthat
are caused by this leakage\footnote{Obviously, the sum of two dipolar
  signal on the sky is still a dipolar signal, as it can be readily
  proven by adding their harmonic coefficients.}; they depend on the
relative orientation of the Galactic dipole axis with respect to the
CMB dipole axis:
\begin{enumerate}
\item The axis of the total dipole is tilted with respect to the CMB
  dipole axis; this induces a time-dependent bias whose exact shape
  depends on the scanning strategy, but in any case it is likely to
  produce large-scale stripes in the sky maps;
\item The amplitude of the dipole is either increased or decreased,
  depending on whether the angular separation between the two axes is
  smaller or larger than 90$^\circ$, respectively; this effect leads
  to a constant offset of the gain estimates $G_k$, as it does not
  depend on the scanning strategy. The effect of this systematic
  effect is to change the amplitude of the CMB signal.
\end{enumerate}
Both effects are potentially dangerous and should be properly
characterized. One possible approach to reducing the impact of
Galactic contamination would be to use foreground map templates of the
form $\epsilon_k \textbf{F}_k$ in
Eq.~\eqref{eq:calDaCapoRadiometerModel}, where $k$ indexes different
foregrounds (synchrotron, CO lines, dust, etc.), and solve for the
unknown scalar factors $\epsilon_k$; this is the approach used by
\Planck\ to characterize bandpass mismatches among HFI bolometers
\citep{planck.2015.08.HFI.data.processing.calibration.and.maps}. A
possibly better approach would be to do the calibration in tandem with
a component separation algorithm, using the following iterative
procedure:
\begin{enumerate}
\item \label{enum:calCleaningEstimateG} Estimate the gains $G_k$
  ignoring the presence of the Galaxy, and produce maps of the sky
  signal at different frequencies;
\item Run a component-separation algorithm on the maps and estimate
  the Galactic signal;
\item \label{enum:calCleaningDecalibration} Scan the map of the
  Galactic signal into a timeline, and use the gains estimated in
  step~\ref{enum:calCleaningEstimateG} to decalibrate this timeline;
\item Clean the data timeline using the one estimated in
  step~\ref{enum:calCleaningDecalibration} and repeat the process from
  the beginning.
\end{enumerate}

\begin{figure}
  \centering
  \includegraphics[width=\textwidth]{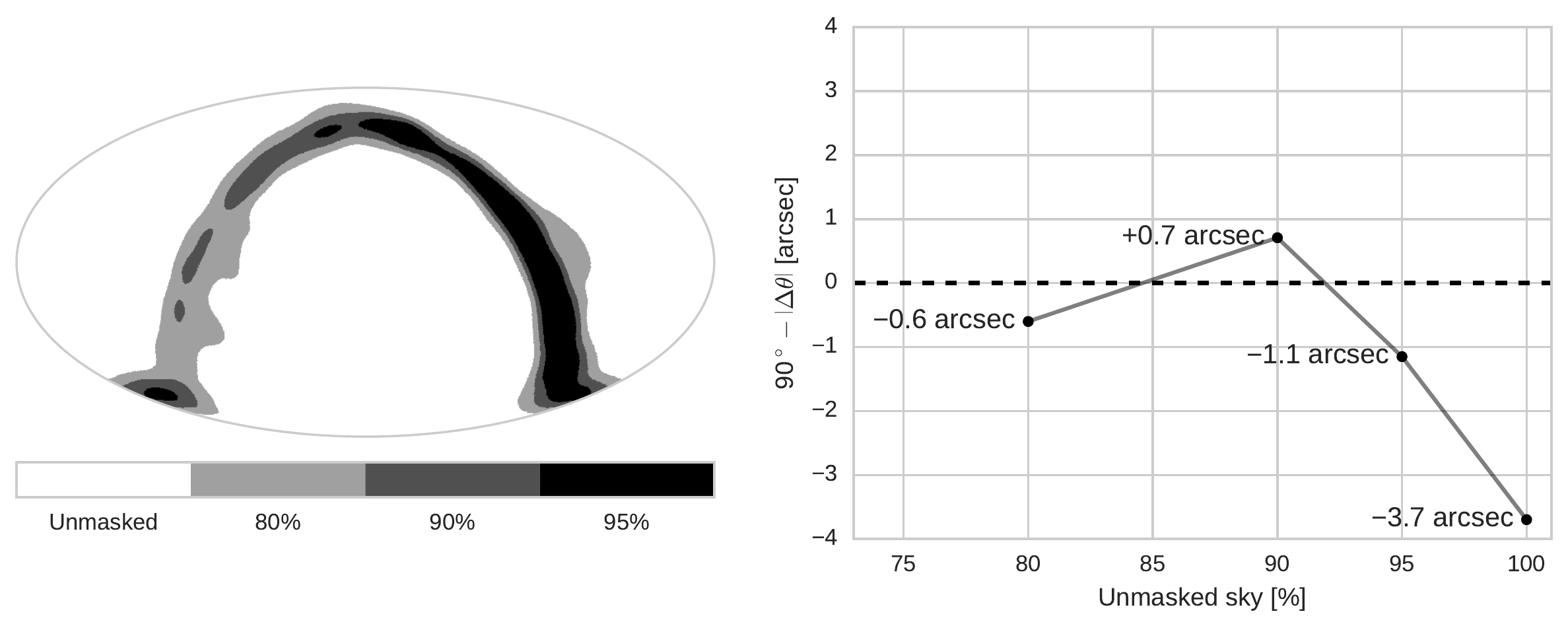}	
  \caption{\label{fig:calDipoleAxesSeparation} \textit{Left:}
    Mollweide projection of the three masks used to remove the
    contribution of the Galactic dust from the data used in the
    calibration simulations. The masks have been created using the
    \Planck\ 353\,GHz temperature maps. They are shown here in
    Ecliptic coordinates. \textit{Right:} Angular separation between
    the axis of the Galactic dipole in the \Planck\ HFI 143\,GHz
    temperature map and the axis of the CMB solar dipole, as a
    function of the Galactic mask. For every mask, the two axes are
    nearly perpendicular; however, in the case of the 90\,\% mask the
    separation is slightly less than 90$^\circ$. This has implications
    for the calibration, as shown in
    Fig.~\protect\ref{fig:calGainTimelinesGalaxy}.}
\end{figure}

For the purpose of this work, we adopt the simpler approach of
applying a mask which removes the brightest parts of the Galaxy from
the computation, as was done in
\citet{planck.2015.05.LFI.calibration}. To produce the timelines, we
simulate the observation using the 143\,GHz map from the \Planck\ 2015
data release. The masks have been created using the 353\,GHz map from
the same data release to remove those sky regions where the dust
signal is strongest. In Fig.~\ref{fig:calDipoleAxesSeparation} we show
the three sky masks we use and the angular separation between the
Galactic dipole axis and the CMB dipole axis as a function of the sky
mask. Since the axis of the Galactic dipole in the \Planck\ sky maps
is nearly perpendicular to the CMB dipole axis\footnote{This is true
  for the sky maps produced by \WMAP\ and \Planck\ LFI and HFI, as the
  angle between the two dipole axes always differs from $90^\circ$ by
  only a few arcseconds.}, slightly varying the masked region can make
the separation between the two axes smaller or greater than
$90^\circ$, thus changing the sign of the overall bias induced by the
Galaxy. As the right-hand side of
Fig.~\ref{fig:calDipoleAxesSeparation} shows, in the case of the
143\,GHz map, we expect the bias to be positive for the 90\,\% mask,
and negative in the other cases.

\begin{figure}
  \centering
  \includegraphics[width=\textwidth]{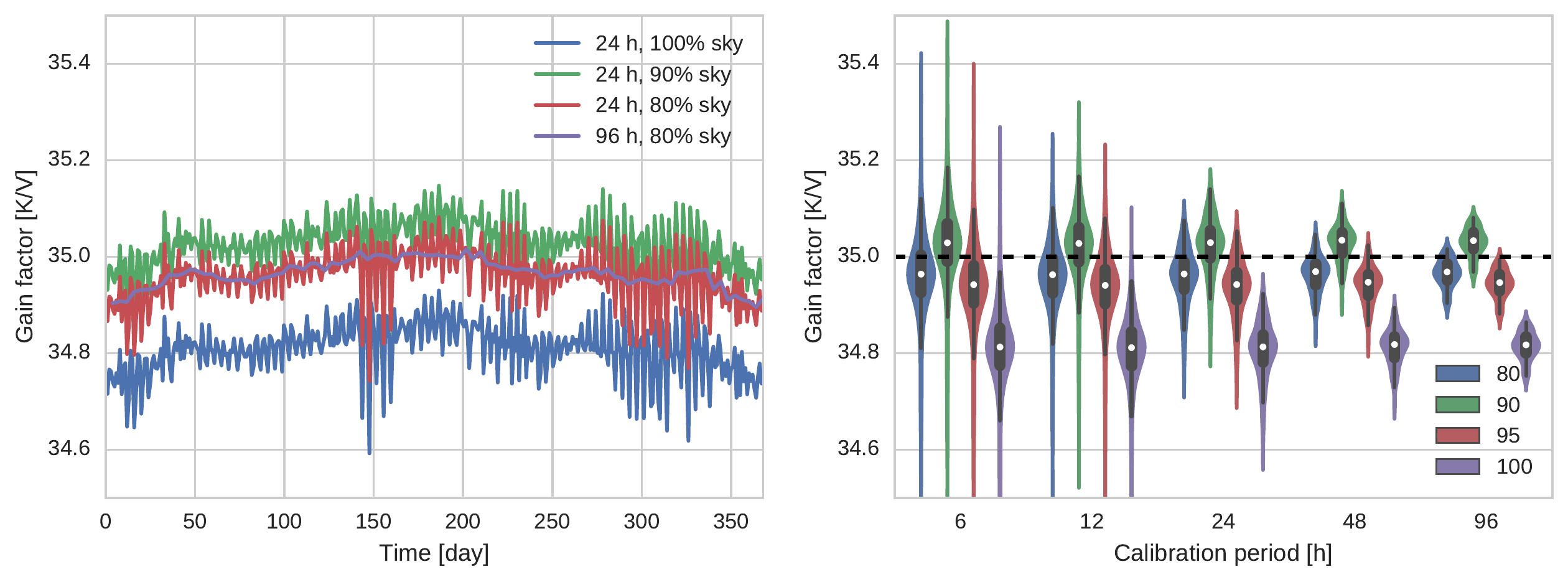}
  \caption{Variation of the estimated gains with time in the case of a
    realistic sky (\Planck\ HFI 143\,GHz map). As in
    Fig.~\protect\ref{fig:calGainTimelinesCmbOnly}, $G =
    35.0\,\text{K/V}$. The presence of the Galaxy induces a systematic
    offset, mainly due to the additional dipolar component which
    biases the calibration. Applying masks helps in reducing this
    effect, as the cases with a 80\,\% mask show. The sign of the bias
    is always negative except in the case of the 90\,\% mask: this
    depends on the relative orientation of the Galactic dipole axis
    with respect to the CMB solar dipole, as explained in the
    text. (See also Fig.~\protect\ref{fig:calDipoleAxesSeparation}.)}
  \label{fig:calGainTimelinesGalaxy}
\end{figure}

Figure~\ref{fig:calGainTimelinesGalaxy} shows the results of the
simulations. Unlike Fig.~\ref{fig:calGainTimelinesCmbOnly}, the two
systematic effects we expected are now clearly evident: (1) a
systematic bias in the overall level of the gains, which is positive
when the 90\,\% mask is used and negative otherwise, and (2)
time-dependent fluctuations which are not reduced if longer
calibration periods are used. If a calibration period of 4\,days is
used, and 80\,\% of the sky is used, then the gain estimate is $G =
34.97_{-0.06}^{+0.04}\,\text{K/V}$, which should again be compared to
the reference value $G = 35\,\text{K/V}$ used in
Eq.~\eqref{eq:calTodSimulation}. The results are still consistent with
the reference value, with a $0.1\,\%$ error.

\begin{figure}
  \centering
  \includegraphics[width=\textwidth]{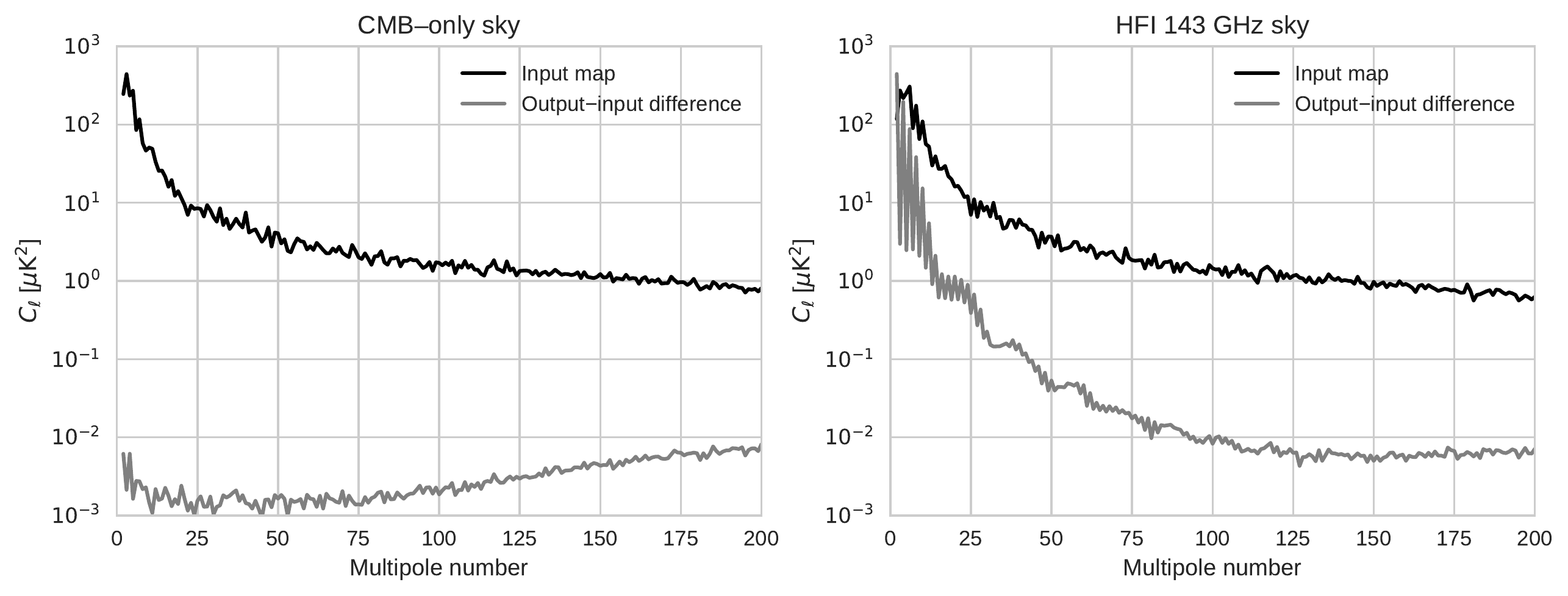}
  \caption{\textit{Left:} Comparison of the power spectrum of the CMB
    map with the spectrum of the difference between the same CMB map
    and the output map produced by DaCapo. \textit{Right:} The same
    comparison for the HFI 143\,GHz map.}
  \label{fig:calAPS}
\end{figure}

Our results show that residual Galactic emission outside the mask
produces a systematic effect in the determination of $G$. We have
tested this and found that this effect is larger than the systematic
error due to neglecting polarization in the map (see
Eq.~\ref{eq:calDaCapoRadiometerModel}), as the latter is at least one
order of magnitude smaller. To assess the impact of this kind of
systematic error on the scientific outcome of \CORE, a more detailed
set of simulations need to be carried out. However, we expect that the
impact of such errors on polarization measurements will be smaller
than 0.1\,\%, as the error on $G$ is highly correlated between
detectors. Therefore, it should cancel when differencing data from
detector pairs like the one used in our simulations (that is, two
detectors oriented at $-22.5^\circ$ and $67.5^\circ$ with respect to
the scan direction of motion). In fact, for a given detector pair the
systematic contamination due to the Galaxy will cancel out to first
order when differencing the data. When averaging $N$ such pairs,
oriented the same way, residual noise will be dominated by
uncorrelated contributions, thus scaling down as $1/\sqrt{N}$.

Figure~\ref{fig:calAPS} shows a comparison between the power spectrum
of the input map and the that of the residual map between the input
and the output map produced by DaCapo. In the case of a CMB sky, the
residuals are more than two orders of magnitude smaller than the map
itself. In the case of a realistic sky with Galactic dust emission,
the residuals become of the same order of magnitude as the signal
itself at large angular scales ($\ell < 10$). To address these
discrepancies, we expect to use the calibration pipeline in tandem
with high quality models of the sky and component-separation
methods. As a result, the overall calibration accuracy should fall
between the optimistic case presented in
Sect.~\ref{sec:calSystematics} (0.01\,\%) and the pessimistic case
discussed in Sect.~\ref{sec:calGalacticSystematics} (0.1\,\%).

\section{Pointing accuracy and reconstruction uncertainty}
\label{sec:pointing_accuracy}

\begin{figure}[!tbp]
  \centering
  \includegraphics[width=0.75\textwidth]{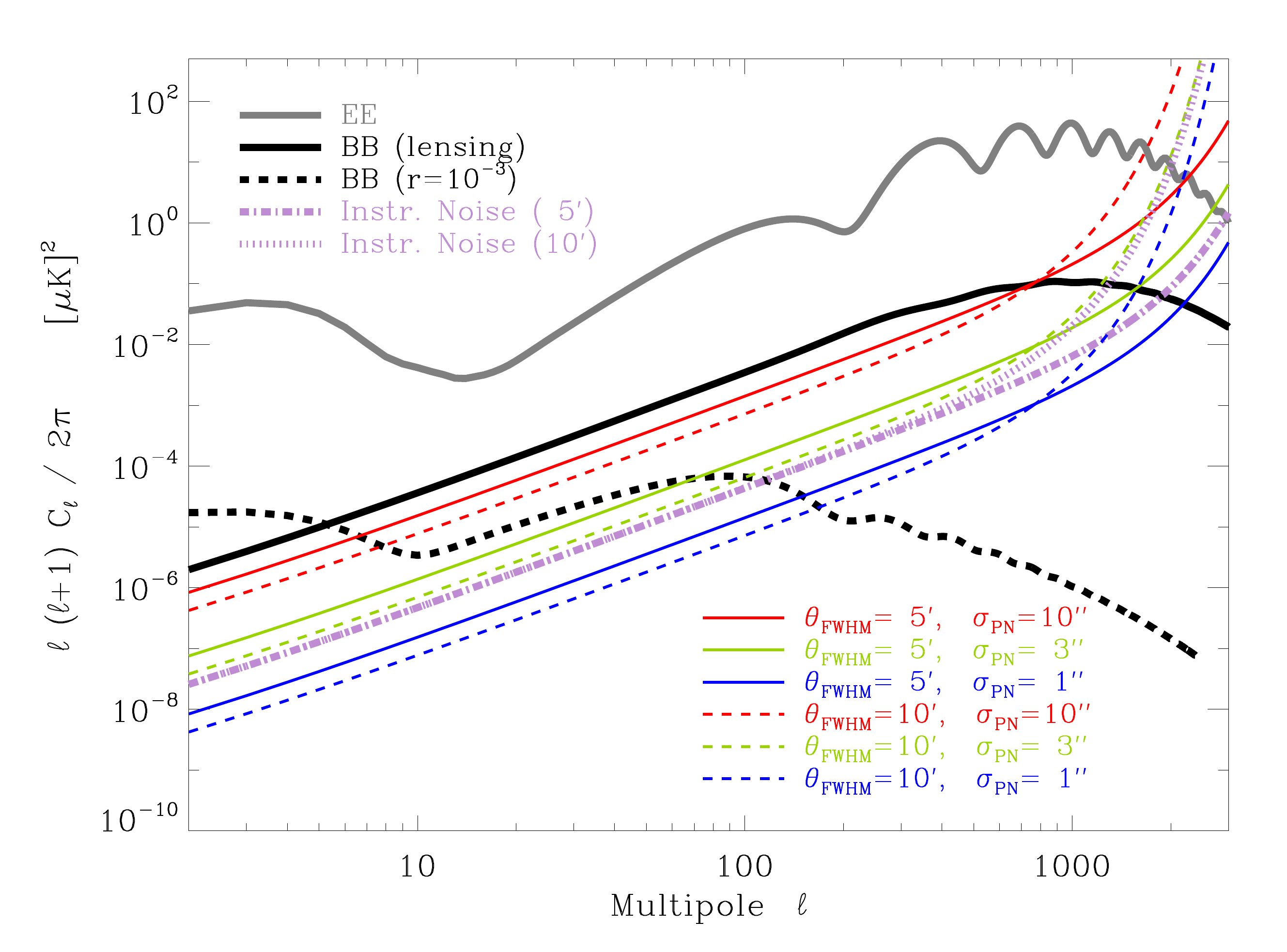}
  \caption[Pointing noise spectrum]{Angular power spectrum of the
    noise induced by pointing inaccuracy (red, green and blue curves),
    after correction of the beam window functions, compared to the
    $EE$ and $BB$ (either pure lensing or primordial with $r=10^{-3}$)
    spectra (grey and black curves). The instrumental noise (purple
    curves) is assumed to be 2$\mu$K$\cdot$arcmin, the expected
    \CORE\ sensitivity to CMB polarization.}
\label{fig:pointing_noise_Cl}
\end{figure}

The variation of the temperature signal at scales smaller than the
beam size may induce spurious polarized signals in two different, but
closely related, ways.  (1) Sub-pixel effects due to the variation of
the sky signal within a map pixel can generate artefacts, that depend
mainly on the distance between the nominal pixel centre and the
average position of the samples in the pixel, multiplied by the local
signal gradient.  If the pointing were perfectly known, and once a
first estimate of the signal is available, they could be corrected
iteratively at the map or at the power spectrum level
\citep{Hivon:2016}.  This is, however, limited by (2) the accuracy of
the pointing reconstruction (for example, a few arcsec in the case of
\Planck), which will create a pointing noise contaminating all power
spectra at a comparable amplitude.  If the pointing error in a pixel,
integrated over all the samples and detectors contributing to the map,
is assumed to be weakly correlated with the adjacent pixels, then the
resulting pointing induced noise will also be white, with an amplitude
determined by the pointing error variance and the variance of the
temperature gradient.  Figure \ref{fig:pointing_noise_Cl} shows how
this pointing-induced noise compares to the angular power spectra of
the polarized signal and instrumental noise, for residual pointing
error per pixel of 1, 3, and 10\,arcsec \emph{rms} and beam FWHM of 5
and 10\,arcmin. This shows that a pointing error of a few arcsec or
less is enough to allow a good measurement of the $BB$ power spectrum.
Furthermore, assuming that the pointing error variance is measured,
correction schemes can be implemented, as proven by several CMB
experiments, including \Planck\ \citep{2016A&A...594A...7P} and
\Spider.

If the pointing error is correlated between pixels, and presents some
long term drifts or correlations with external factors, then specific
tools would have to be developed to treat it.

\section{Conclusions}
\label{sec:conclusions}

In the context of the proposed \CORE\ satellite mission to map CMB
polarization, we have discussed the impact of the main systematic
effects we expect to affect the observations, with emphasis on their
projected impact on frequency maps. We have chosen to deal with one
effect at a time, for the sake of carefully assessing its impact in
isolation from the others. In a real world application, however,
systematic effects will interact one another, implying that their
combined effect cannot always be deduce by linear superposition. Their
treatment will thus require a global processing pipeline. An accurate
assessment by means of complete end-to-end simulations is needed, but
will have to wait until we will possess detailed knowledge of the
individual effects. We therefore defer this analysis to future work.

We have employed the TOAST simulation pipeline to generate timelines
of realistic instrumental noise and beam-convolved sky signals. We
have also employed the flexible generalized destriping code MADAM to
produce maps with statistically optimal noise properties, having tuned
the algorithm to the specific \CORE\ design and scanning strategy.

We have explored in detail the properties of the \CORE\ noise maps,
focusing for the sake of simplicity on a pair of detectors at
145\,GHz, at positions in the centre and at the edges of the focal
plane, to produce a triplet of $(I,Q,U)$ maps. We find that for all
cases this set-up achieves full sky coverage and is able to cleanly
separate the Stokes parameters, in the sense that the map-making
equations are well-conditioned as measured by the reciprocal condition
number of the covariance matrix of Stokes parameters in each
pixel. This is a reassuring test, since an ill-conditioned map-making
system is vulnerable to several systematic effects. Its outcome was
not obvious beforehand: in the absence of an active hardware modulator
such as a half-wave plate, the \CORE\ instrument only relies on the
spacecraft's scanning strategy to modulate the polarization signal.
We have also verified that the pixel covariance matrices exhibit low
intensity-to-polarization couplings, a desirable property that helps
to keep intensity-to-polarization leakage low either by design, or
under control with proper analysis tools. At the same time, residual
$QU$ couplings are not negligible and need to be accounted for during
analysis. We have also attempted to optimize the scanning strategy, in
particular the telescope offset and precession angles, finding that
any advantage over the baseline configuration is minimal within a
range compatible with reasonable assumptions about the spacecraft's
design and operational constraints.

As a consequence of the very well interconnected scanning strategy,
the \CORE\ map-making achieves excellent levels of suppression of the
correlated (`$1/f$' type) noise component. We find that the residual
correlation in the noise maps are in fact negligible for a detector
knee frequency of the order of 10\,mHz and could easily be handled
even for values a few times higher. We hence do not expect $1/f$ noise
to be a driver in the design and performance of the
mission. Foreground contamination, which may leave hard to minimize
residuals, is more of a concern on the largest angular scales,
although it appears that the frequency coverage of \CORE\ is adequate
to guarantee a detection of $r \simeq 0.001$ even in the presence of
complex foregrounds \citep{ECO.foregrounds.paper}.

We have also taken in consideration the case of noise correlations
between detectors, such as those generated by a `common mode' of
thermal or other origin. Assuming a realistic model for this component
(derived from \Planck\ results), we find that a dedicated map-making
scheme that takes such cross-correlations into account within the
noise correction model would be beneficial for \CORE\ (at the cost,
however, of significantly increasing the computational burden of the
problem, whose cost grows quadratically with the number of correlated
detectors).

Since the \CORE\ analysis uses multi-detector map-making (using a
minimal set of two or four detectors as explored in this paper),
bandpass mismatch leakage is a potential source of concern. Since
there is no reason for the bandpass leakage systematic to be
correlated between detectors, the effect is predicted to average out
when increasing the number of detectors. In addition, we have
implemented and demonstrated a correction scheme, which reduces the
amount of leakage to a level well below the bounds on primordial $BB$
spectrum allowed by the \CORE\ error budget.

Leakage from intensity to polarization arising due to beam asymmetries
is a potential concern for high-precision CMB polarization
experiment. The case of \CORE\ requires even greater attention, due to
the lack of active modulation in the optical path to regularize the
beam shape independently of the scan. We have demonstrated two
complementary approaches, in real space and harmonic space, that both
allow to model accurately and quickly the impact of \CORE\ beam
non-idealities on the maps or the power spectra. When these
simulations are employed to clean the \CORE\ data, the uncorrected
level of residuals is well within the \CORE\ science requirements,
even when accounting for imperfect knowledge of the instrument.

In an effort to move away from ideal simulations at the stage before
map-making, we have implemented a prototype calibration pipeline for
\CORE, based on that used for \Planck\ LFI. We have discussed its
robustness to several non-idealities, arising due to the instrument
and the sky, concluding that the calibration requirements of
\CORE\ can be already met with existing knowledge and algorithms.

Without doubt, an all-sky experiment to map the CMB polarization to
cosmic variance level is an ambitious effort. Systematic contributions
are expected to dominate the error budget and, if the mission is
selected, a considerable effort will be necessary in the years ahead
to build an analysis pipeline that accurately deals with them. While
such a task will only be completed after critical information about
the instrument will become available (that is, during the study phase
and later), it would make sense to start building the necessary
infrastructure as soon as possible. At the same time, it is reassuring
that simple yet realistic assumptions about the main contaminants,
implemented within quick and robust simulation and correction
algorithms -- all of which owe much to the heritage of \Planck\ --
demonstrate that our requirement to keep \CORE\ systematics under
control rests on solid grounds.


\acknowledgments

This research used resources of the National Energy Research
Scientific Computing Center, which is supported by the Office of
Science of the U.S. Department of Energy under Contract
No. DE-AC02-05CH11231. We also thank CSC -- IT Center for Science
(Finland) for computational resources.  We acknowledge financial
support by ASI Grant 2016-24-H.0 and Academy of Finland grant
295113. This work has received funding from the European Union's
Horizon 2020 research and innovation programme under grant agreement
number 687312.  CJM is supported by an FCT Research Professorship,
contract reference IF/00064/2012, funded by FCT/MCTES (Portugal) and
POPH/FSE (EC). JGN acknowledges financial support from the Spanish
MINECO for a `Ramon y Cajal' fellowship (RYC-2013-13256) and the I+D
2015 project AYA2015-65887-P (MINECO/FEDER). GR acknowledges support
from the National Research Foundation of Korea (NRF) through NRF-SGER
2014055950 funded by the Korean Ministry of Education, Science and
Technology (MoEST), and from the faculty research fund of Sejong
University in 2016. We thank Jean Kaplan for useful comments on the
manuscript.


\section*{Appendix A: Monte Carlo simulations}

In order to run TOAST to simulate one year of observations, we need a
supercomputer. During the initial development of the \CORE\ pipeline,
making single runs and small sets of Monte Carlo realizations (10 at
maximum), we used the high performance computing (HPC) resources at
the National Energy Research Scientific Computing
Center\footnote{http://www.nersc.gov} (NERSC) in the USA. In
particular we used the Edison machine, a Cray XC30 supercomputer with
134,064 compute cores for a peak performance of 2.57 petaflop/s.

For the larger MC runs needed to derive the results shown in
Sect.~\ref{Analysis_maps} and in particular the ones for the
destriping tests in Sect.~\ref{fknee}, we used the Centre for
Scientific Computing (CSC) Sisu
supercomputer\footnote{http://www.csc.fi} in Finland. It is a Cray
XC40 supercomputer with a total theoretical peak performance of 1.69
petaflop/s.

We considered 12 cases, generating 1000 realizations for each
case. For each run we used 960 cores (40 nodes). We used four values
of the knee frequency, $f_{k} = (0, 10, 20, 50)\,\mathrm{mHz}$, and
for each value of $f_{k}$ we generated maps for the `boresight',
`high', and `low' detector positions (see
Table~\ref{tab:toast_parameters}). In
Table~\ref{tab:noisemc_resources} we report the CPU and memory
resources needed for each case. We saved the simulated noise,
destriped and not, for all 1000 realizations, and the hit-map, the
white noise covariance matrix and its inverse from the first
realization (these are identical for all the realizations). The total
size of these files is 3.4 TB. The total computational cost of these
simulations was 162\,100 CPU-hours.

\begin{table}[t]\footnotesize
\begin{center}
\begin{tabular}{|c|c|c|}
\hline
$f_{k}$ [mHz] & CPU time [CPUh] & Memory footprint [GB] \\
\hline
\multicolumn{3}{|c|}{Boresight}    \\
\hline
 $0$ & $18\,320$  & $586$ \\
 $10$ & $11\,580$  & $586$ \\
 $20$ & $12\,280$  & $586$ \\
 $50$ & $11\,730$  & $586$ \\
\hline
\multicolumn{3}{|c|}{High}    \\
\hline
 $0$ & $18\,270$  & $570$ \\
 $10$ & $11\,480$  & $570$ \\
 $20$ & $11\,260$  & $570$ \\
 $50$ & $11\,950$  & $570$ \\
\hline
\multicolumn{3}{|c|}{Low}    \\
\hline
 $0$ & $18\,610$  & $600$ \\
 $10$ & $11\,650$  & $600$ \\
 $20$ & $11\,690$  & $600$ \\
 $50$ & $13\,280$  & $600$ \\
\hline
\end{tabular}
\label{tab:noisemc_resources}
\caption{Resource consumption of the noise Monte Carlo cases}
\end{center} 
\end{table}\mbox{}

\section*{Appendix B: Algebra for cross correlation map-making}
\label{Appendix:B}

In this Appendix we briefly review the algebra of the map-making
procedure and derive the formalism adopted in
Section~\ref{sec:CorrelatedNoise}. We suggest the interested reader
refers to the references given in the text for a detailed discussion.

We model the observed data as follows:
\begin{equation}
\mathbf{d} = \mathbf{A} \mathbf{m} + \mathbf{n},
\end{equation}
where the timelines of the $k$ detectors are combined:
\begin{equation}
\mathbf{d} \equiv
\left(
\begin{array}{c}
\mathbf{d}^{(1)} \\
\vdots \\
\mathbf{d}^{(k)}
\end{array}
\right),
\end{equation}
and the generalized $k N_d \times 3 N_p$ pointing matrix becomes:
\begin{equation}
\mathbf{A} \equiv
\left(
\begin{array}{ccccc}
A_{tp}^{(1)} &&A_{tp}^{(1)} \cos 2 \psi_t^{(1)} && A_{tp}^{(1)} \sin 2 \psi_t^{(1)} \\
\vdots  && \vdots && \vdots \\
A_{tp}^{(k)} && A_{tp}^{(k)} \cos 2 \psi_t^{(k)}&& A_{tp}^{(k)} \sin 2 \psi_t^{(k)} \\
\end{array}
\right).
\end{equation}
Here the $N_d\times N_p$ matrix $A^{(j)}_{tp}$ is the pointing matrix
of the $j$-th detector, with elements equal to unity if the pixel $p$
is observed at time $t$ and zero otherwise.  Each row of the pointing
matrix is multiplied either by the cosine or the sine of $2\psi_t$,
where $\psi_t$ is the angle defining the polarimeter orientation at
time $t$ with respect to the chosen reference frame. Similarly, the
sky signal can be expressed as a $3 N_p$ vector:
\begin{equation}
\mathbf{m} \equiv
\left(
\begin{array}{c}
\mathbf{I}  \\
\mathbf{Q} \\
\mathbf{U}
\end{array}
\right),
\end{equation}
where $\mathbf{I}$, $ \mathbf{Q}$ and $ \mathbf{U}$ are $N_p$ Stokes
parameter vectors of the pixelized CMB sky. The noise timeline is:
\begin{equation}
\mathbf{n} \equiv
\left(
\begin{array}{c}
\mathbf{n}^{(1)} \\
\vdots \\
\mathbf{n}^{(k)}
\end{array}
\right).
\end{equation}
where $\mathbf{n}^{(j)}$ is the $N_{d}$ element noise vector of the
$j$-th detector, accounting for instrumental noise, atmospheric and
temperature fluctuations, cosmic-ray hits and any other random
systematic effect.

In the case of white noise, the generalized least squared (GLS)
approach yields the following optimal estimator
$\widetilde{\mathbf{m}}$ of ${\mathbf{m}}$:
\begin{equation}
  \mathbf{\widetilde m} = \left( \mathbf{A}^T
 \mathbf{A}\right)^{-1}
\mathbf{A}^T  \mathbf{d}.
\end{equation}
This simply means to bin the samples in the map pixels. In the
presence of correlated $1/f$ noise, this approach, usually known as
`naive' map-making, leaves stripy structures in the map.  Thus, the
above formula is extended like follows:
\begin{equation}
  \mathbf{\widetilde m} = \left( \mathbf{A}^T
\mathbf{N}^{-1} \mathbf{A}\right)^{-1}
\mathbf{A}^T \mathbf{N}^{-1} \mathbf{d},
\end{equation}
where $\mathbf{N}$ is the noise covariance matrix
$\left\langle\mathbf{n}\mathbf{n}^T\right\rangle$. The matrix
$\mathbf{N}$ is block-diagonal with respect to the detector index and
each block can be inverted independently. Including also the
cross-correlated noise among different detectors (i.e. the
off-diagonal terms), the most general matrix $\mathbf{N}$ is given by:
\begin{equation}
\mathbf{N} \equiv \left\langle \mathbf{n}_t \mathbf{n}_{t^\prime}\right\rangle=
\left(
\begin{array}{ccc}       \left\langle   n_t^{(1)}   n_{t^\prime}^{(1)} \right\rangle &
\cdots &
\left\langle   n_t^{(1)}    n_{t^\prime}^{(k)} \right\rangle \\
\vdots & \ddots & \vdots  \\
\left\langle   n_t^{(k)}    n_{t^\prime}^{(1)} \right\rangle &
\cdots &
\left\langle    n_t^{(k)}    n_{t^\prime}^{(k)} \right\rangle
\end{array}
\right),
\end{equation}
where $\left\langle \cdot \right\rangle$ denotes the expectation
value. Some assumptions are made on the noise, in particular we assume
that its statistical properties do not change over the mission life
time (stationarity).

The stationarity property implies that $\mathbf{N}$ is a
block-circulant matrix\footnote{Strictly speaking, stationarity implies that this matrix is block-Toeplitz, not block-circulant. Assuming it is circulant produces undesirable correlations between the end and the beginning of each  block. This effect can be avoided by carefully zero padding the blocks before they are Fourier transformed.} In this way, the inversion of $\mathbf{N}$ is
much easier, since the Fourier counterpart of a block-circulant matrix
is block-diagonal. Let us define the multichannel Fourier operator
$\bar{F}$ such that:
\begin{equation}
\bar{\mathbf{n}}=\bar{F}{\mathbf{n}},
\end{equation}
where:
\begin{equation}
\bar{\mathbf{n}} \equiv
\left(
\begin{array}{c}
F{\mathbf{n}}^{(1)} \\
\vdots \\
F{\mathbf{n}}^{(k)}
\end{array}
\right),
\end{equation}
contains end to end the Fourier transforms, $F{\mathbf{n}}^{(j)}$, of
each segment of ${\mathbf{n}}$. Thus, in the case of a number of
cross-correlated detectors, the inverse of $\mathbf{N}$ is given by:
\begin{equation}
{\mathbf{N}}^{-1}=\bar{F}^{T}{\mathbf{R}}^{-1}\bar{F},
\end{equation}
where, under the assumption that the $N_d\times N_d$ matrix $
\left\langle n_t^{(i)} n_{t^\prime}^{(j)} \right\rangle $ is
circulant, $R^{(ij)}$ is a diagonal matrix with elements given by the
noise cross-power spectrum between detectors $i$ and $j$ at frequency
$f$:
\begin{equation}
{R^{(ij)}}_{ff^\prime}=P^{(ij)}(f) \delta^{f^\prime}_f,
\end{equation}
To find $ \mathbf{\widetilde m}$, the optimal GLS formula can be
solved iteratively by the use of a Fourier-based, preconditioned
conjugate gradient method.

To achieve a good convergence speed it is of paramount importance to
provide a good preconditioner for the matrix $\left(
\mathbf{A}^T\mathbf{N}^{-1} \mathbf{A}\right)$. Our choice is to
approximate the matrix $\mathbf{N}^{-1}$ with its diagonal part and
our preconditioner, $\mathbf{H}$, will be:
\begin{equation}
 \mathbf{H} = \left[\mathbf{A}^T diag \left(\mathbf{N}^{-1}\right) \mathbf{A}\right]^{-1}.
\end{equation}
It can be shown that the preconditioning operator $\mathbf{H}$ is a
$3\times 3$ block diagonal matrix, where each block is the linear
operator that solves for the three Stokes parameters of the given
pixel (assuming white noise).

We define the inverse pixel condition number $\mathrm{R_{cond}}$ as
the ratio of the absolute values of the smallest and largest
eigenvalue of each block of $\mathbf{H}$. The condition number is a
useful tool to trace the errors in the Stokes parameter estimation due
to an inadequate polarization angle coverage on the given pixel. In
particular $0 \leq \mathrm{R_{cond}} \leq 0.5$, assuming its lower
value in the worst case and $\mathrm{R_{cond}} = 0.5$ in the limit of
uniform angle coverage.


\bibliographystyle{plainnat}
\bibliography{ECO_Systematics}

\end{document}